\pdfoutput=1
\documentclass[aps,prb,reprint,citeautoscript,superscriptaddress,showpacs]{revtex4-1}

\usepackage{graphicx,epsfig,float}
\usepackage{amssymb} 
\usepackage{amsmath}

\begin{document}
\title{Stacking in incommensurate graphene/hexagonal-boron-nitride heterostructures based on \textit{ab initio} study of interlayer interaction}

\author{Alexander V. Lebedev}
\email{allexandrleb@gmail.com}
\affiliation{Kintech Lab Ltd., 3rd Khoroshevskaya Street 12, Moscow 123298, Russia}
\author{Irina V. Lebedeva}
\email{liv\_ira@hotmail.com}
\affiliation{Nano-Bio Spectroscopy Group and ETSF, Universidad del Pa\'is Vasco, CFM CSIC-UPV/EHU, 20018 San Sebastian, Spain}
\author{Andrey M. Popov}
\email{popov-isan@mail.ru}
\affiliation{Institute for Spectroscopy of Russian Academy of Sciences, Troitsk, Moscow 108840, Russia}
\author{Andrey A. Knizhnik}
\email{kniznik@kintechlab.com}
\affiliation{Kintech Lab Ltd., 3rd Khoroshevskaya Street 12, Moscow 123298, Russia}

\begin{abstract}
The interlayer interaction in graphene/boron-nitride heterostructures is studied using density functional theory calculations with the correction for van der Waals interactions. It is shown that the use of the experimental interlayer distance allows to describe the potential energy surface at the level of more accurate but expensive computational methods. On the other hand, it is also demonstrated that the dependence of the interlayer interaction energy on the relative in-plane position of the layers can be fitted with high accuracy by a simple expression determined by the system symmetry. The use of only two independent parameters in such an approximation suggests that various physical properties of flat graphene/boron-nitride systems are interrelated and can be expressed through these two parameters. Here we estimate some of the corresponding physical properties that can be accessed experimentally, including the correction to the period of the Moir\'{e} superstructure for the highly incommensurate ground state of graphene/boron-nitride bilayer coming from the interlayer interaction, width of stacking dislocations in slightly incommensurate systems of boron nitride on stretched graphene and shear mode frequencies for commensurate graphene/boron-nitride systems, such as a flake on a layer. We propose that the commensurate-incommensurate phase transition can be observed in boron nitride on stretched graphene and experimental measurements of the corresponding critical strain can be also used to get an insight into graphene/boron-nitride interactions.
\end{abstract}

%\pacs{}
\maketitle

\section{Introduction}

Following the discovery of 2D analogues of graphene a new research field has emerged on layer-by-layer design of van der Waals heterostructures \cite{Geim2013,Novoselov2012} opening fresh possibilties to observe unusual properties and physical phenomena. Absolute leaders among such heterostructures are those based on graphene and hexagonal boron nitride (h-BN). Chemically inert, flat and wide gap boron nitride substrate  allows to avoid graphene rippling, supress the charge inhomogeniety and improve the carrier mobility providing a quality comparable to suspended graphene \cite{Dean2010, Xue2011, Zomer2011, Decker2011}. The advanced performance of graphene/boron-nitride interfaces has been already utilized in a number of nanoelectronic devices where few-layer and monolayer boron nitride serves as a gate dielectic \cite{Dean2012,Ponomarenko2011,Kim2011,Jain2013} or tunnel barrier \cite{Lee2011,Britnell2012a}, including novel field-effect tunneling transistors \cite{Britnell2012}. Nanocapacitors based on graphene/hexagonal-boron-nitride heterostructures have been proposed theoretically \cite{Ozcelik2013} and  fabricated \cite{Shi2014} recently. Multi-layer graphene intercalated by boron nitride has been also suggested for the use in ultra-scaled interconnects of integrated circuits \cite{Li2012}.

Double-layer graphene with an ultrathin boron nitride spacer has been considered to study strong Coulomb drag \cite{Gorbachev2012}  and tunable metal-insulator transition \cite{Ponomarenko2011}. It has been also predicted that high-temperature superfluidity can be observed in such a heterostructure \cite{Perali2013}.
Though hexagonal boron nitride and graphene have very close lattice constants, the small lattice mismatch of 1.8\% turns out sufficient for creation of the Moir\'{e} superstructure even when the layers are perfectly aligned \cite{Woods2014, Tang2013, Yang2013}. A weak periodic potential of this superstructure perturbs the electronic spectrum of graphene and leads to emergence of superlattice minibands manifested through the Hofstadter's butterfly  and Dirac points near the edges of the superlattice Brillouin zone \cite{Yankowitz2012,Dean2013,Hunt2013,Ponomarenko2013, Chen2014}. The electronic and optical properties of graphene \cite{Mucha2016,Kumar2015,Slotman2015,Eckmann2013,Jung2015,Sachs2011} on boron nitride modified due to the formation of the Moir\'{e} superstructure can serve as a basis for development of new technologies. 

While the interlayer interaction causes significant adjustment of aligned graphene and boron nitride layers and commensurate domains arise \cite{Woods2014, Argentero2017, Tang2013, Yang2013}, relative rotation of the layers brings the system to the fully incommensurate state in which the interlayer interaction landscape is extremely smooth. Drastic changes in the structure, electronic and optical properties of the system are observed upon such a rotation \cite{Woods2014, Eckmann2013}. A new phenomenon of macroscopic self-reorientation of graphene towards crystallographic
directions on the underlying boron nitride crystal has been recently demonstrated \cite{Woods2016}. On the other hand, robust superlubricity that can be used to reduce the friction in nanoelectromechanical systems has been predicted for interfaces between graphite and boron nitride bulk \cite{Leven2013}. As long as energetic characteristics of interlayer interaction in graphene/boron-nitride heterostructures have not been measured yet experimentally, theoretical studies hold the key to understanding the properties related to the interlayer interaction and elaboration of nanodevices based on these properties. Here we present an approach that allows accurate and detailed calculations of the dependence of interlayer interaction energy on the relative position of the layers in graphene/boron-nitride heterostructures and consider a wide set of experimentally measurable physical properties related to this interaction.  

To analyze the interaction of graphene and boron nitride layers responsible for the phenomena mentioned above first-principles studies have been performed \cite{Fan2011, Li2012,Giovannetti2007,Leven2016,Zhao2014,Zhou2015,Sachs2011,Argentero2017}. Though density functional theory (DFT) calculations \cite{Fan2011,Zhao2014,Zhou2015,Argentero2017}  predict the qualitatively correct dependence of the interlayer interaction energy on stacking of the layers, they give wrong magnitudes of relative energies of states with different stacking as can be deduced from comparison with the more accurate but computationally expensive random phase approximation (RPA) approach \cite{Zhou2015,Sachs2011}.  
This failure of the DFT method is associated with its inability to describe the equilibium interlayer distance since, as known from publications for bilayer graphene \cite{Lebedeva2011, Reguzzoni2012}, the relative energies of states with different stacking decrease exponentially in magnitude upon increasing the interlayer distance. In the present paper we perform DFT calculations of the potential surface of interlayer interaction energy in graphene/boron-nitride heterostructures, i.e. the dependence of the interlayer interaction energy on the relative in-plane displacement of the layers, with the help of the vdW-DF2 functional \cite{Lee2010}. We show that the use of the experimental interlayer distance instead of the optimized one allows to get the results close to the data obtained in the more accurate RPA approach \cite{Zhou2015,Sachs2011} at a lower computational cost. 

Since first-principles methods are restricted to small simulation times and length scales, semiemirical models able to reproduce the potential surface of interlayer interaction energy are invoked for efficient modeling of   
phenomena taking place in heterostructures. Several such models have been proposed for the interaction between graphene and boron nitride layers including the registry-dependent \cite{Leven2016} and Morse-type \cite{Argentero2017} interatomic potentials, registry index model \cite{Leven2013} and approximations based on the first spatial Fourier harmonics \cite{Zhou2015, Jung2015}. Atomistic models using semiempirical potentials were employed to analyze structure \cite{Leven2016, Wijk2014, Argentero2017} and van der Waals interactions \cite{Neek-Amal2014} in Moir\'{e} patterns of graphene/boron-nitride systems. Superlubricity of heterostructures was studied qualitatively on the basis of  the simple registry index model \cite{Leven2013, Zhao2014}. The approximations of the dependence of the interlayer interaction energy on the relative in-plane position of graphene and boron nitride layers by the first spatial Fourier harmonics were applied to analyze structural relaxation, tribological behavior and band gap landscapes\cite{Zhou2015, Jung2015,Kumar2015}. However, the adequacy of such approximations has not been addressed. Here we examine the deviation of such an expression determined by the system symmetry from the potential energy surface obtained by the DFT calculations at the experimental interlayer distance.

The approximation of the potential energy surface at the fixed interlayer distance by the first spatial Fourier harmonics includes only two independent parameters and thus a number of properties of flat graphene/boron-nitride systems related to the interlayer interaction are determined to a large extent by these two parameters. A similar observation has been made previously for pure graphene and boron nitride systems \cite{Popov2012, Lebedev2016, Lebedeva2011} and the barrier to relative rotation of the layers, shear mode frequency, width of stacking dislocations and critical strain for the commensurate-incommensurate phase transition have been estimated for such materials. Here, in addition to these properties, we use the approximation of the potential energy surface to analyze the Moir\'{e} superstructure of  the incommensurate ground-state graphene/boron-nitride heterostructure and to study how the interlayer interaction affects the superstructure period. The possibility of experimental measurements of the calculated properties is discussed.

The paper is organized in the following way. In section II we present the results of our DFT calculations of the potential energy surface of graphene/boron-nitride heterostructures. In section III this surface is approximated using the first spatial Fourier harmonics and the accuracy of this approximation is studied. In section IV the approximation is applied to evaluate physical properties related to the interlayer interaction for flat heterostructures with a different degree of incommensurability including the barrier to relative rotation of the layers, shear mode frequency, period of the Moir\'{e} pattern, width of stacking dislocations and critical strain for the commensurate-incommensurate phase transition. Finally conclusions are summarized.

\section{DFT calculations}
The DFT calculations are performed using the non-local vdW-DF2 functional \cite{Lee2010} taking into account van der Waals interactions as implemeted in the VASP code \cite{Kresse1996}. The projector augmented-wave method (PAW) \cite{Kresse1999} is applied. The rectangular unit cell including 4 atoms of each layer and having height of 20~\AA~is considered under periodic boundary conditions. Integration over the Brillouin zone is performed using the Monkhorst-Pack method \cite{Monkhorst1976} with the $28\times 36 \times 1$ k-point grid (in the armchair and zigzag directions, respectively). The maximum kinetic energy of plane waves is 600 eV. The convergence threshold of the self-consistent field is $10^{-7}$ eV. These parameters provide well-converged values of the barriers to relative sliding of graphene layers \cite{Lebedeva2011}.  

The calculations of the optimal size of the commensurate unit cell of the graphene/boron-nitride heterostructure and potential energy surface are performed at a fixed interlayer distance $d=3.33$~\AA, which is close to the experimentally measured interlayer distances in bulk boron nitride \cite{Pease1950, Pease1952, Lynch1966, Solozhenko1995, Solozhenko1997, Solozhenko2001, Paszkowicz2002, Bosak2006, Fuchizaki2008} and graphite \cite{Bernal1924, Baskin1955, Lynch1966, Ludsteck1972, Trucano1975, Zhao1989, Bosak2007}. 
It has been shown previously \cite{Lebedeva2017,Lebedev2016} that our appoach based on the combination of the vdW-DF2 functional \cite{Lee2010} and the use of the experimental interlayer distance allows to describe adequately such properties of purely graphene or boron nitride systems related to in-plane relative motion of the layers as the shear mode frequency, shear modulus, barrier to relative sliding of the layers and width of stacking dislocations. The DFT calculations using the experimental interlayer distance were found to be much more accurate than the ones using the optimized interlayer distance, while at the experimental interlayer distance the vdW-DF2 functional was demonstrated to perform better than the PBE-D2, PBE-D3, PBE-D3(BJ), PBE-TS, PBE-TS/HI, PBE-TS+SCS, optPBE-vdW functionals \cite{Lebedeva2017}. 

\begin{figure*}
	\centering
	\includegraphics[width=\textwidth]{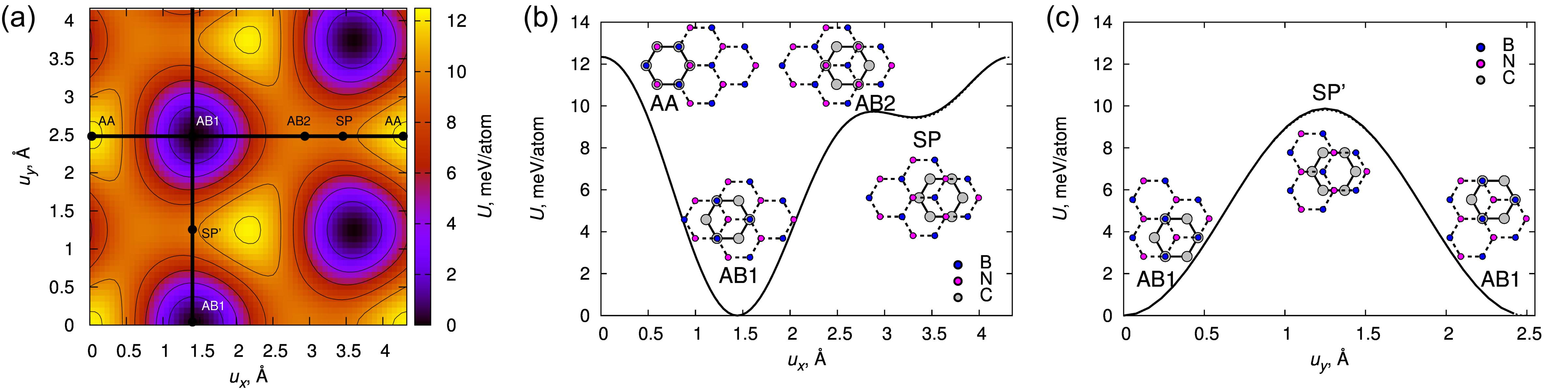}
	\caption{(Color online) Interaction energy of graphene and boron nitride layers $U$ (in meV per carbon atom) obtained by DFT calculations as a function of relative displacement of the layers in the armchair ($u_x$, in \AA) and zigzag ($u_y$, in \AA) directions at the interlayer distance of $d=$3.33~\AA. The energy is given relative to the AB1 stacking. (b,c) The dependences of interlayer interaction energy $U$ on displacements $u_x$ and $u_y$ in the armchair (b) and  zigzag (c) directions obtained by DFT calculations (solid lines) along the thick black lines indicated in figure (a). The curves corresponding to the approximation according to Eq.~(\ref{eq_2}) are shown by dashed lines and are virtually the same as the DFT results. Structures of the symmetric stackings are indicated. Boron, nitrogen and carbon atoms are coloured in blue/dark gray, magenta/medium gray and light gray, respectively.}
	\label{fig:pes}
\end{figure*}

The calculations show that the lowest energy stacking for the commensurate unit cell of the graphene/boron-nitride heterostructure is AB1 in which boron atoms are located on top of carbon atoms and nitrogen atoms are on top of centers of hexagons, in agreement with previous findings \cite{Fan2011,Li2012,Giovannetti2007,Zhao2014,Leven2016,Zhou2015,Sachs2011,Argentero2017} (Fig.~\ref{fig:pes}). For this stacking, the optimized lattice constant of the heterostructure is $a=2.498$~\AA, close to the previously calculated values of 2.48--2.50~\AA~(Ref. \onlinecite{Zhou2015}). The lattice constants of single layers of graphene and hexagonal boron nitride optimized within the same computational approach are $a_{\mathrm{C}}= 2.477$~\AA~and $a_{\mathrm{BN}}=2.521$~\AA~(Ref. \onlinecite{Lebedeva2016}), respectively. Therefore, the graphene and boron nitride layers in the heterostructure are stretched and compressed by 0.86\% and 0.91\%, respectively. It should be also metioned that the calculated lattice constants for graphene and boron nitride are in agreement with the experimental data for graphite of 2.4614$\pm$0.0001~\AA~(Ref. \onlinecite{Pease1952}), 2.45$\pm$0.03~\AA~(Ref. \onlinecite{Bernal1924}), 2.4589$\pm$0.0005~\AA~(Ref. \onlinecite{Baskin1955}), 2.4612$\pm$0.0001~\AA~(Ref. \onlinecite{Ludsteck1972}), 2.464$\pm$0.002~\AA~(Ref. \onlinecite{Trucano1975}), 2.462 \AA~(Ref. \onlinecite{Zhao1989}) and 2.463 \AA~(Ref. \onlinecite{Bosak2007}) and boron nitride bulk of 2.5038$\pm$0.0001~\AA~(Ref. \onlinecite{Pease1950}),  2.50399$\pm$0.00005~\AA~(Ref. \onlinecite{Pease1952}), 2.504$\pm$0.002~\AA~(Ref. \onlinecite{Solozhenko2001, Solozhenko1995}), 2.505$\pm$0.002~\AA~(Ref. \onlinecite{Solozhenko1997}), 2.5047 $\pm$0.0002~\AA~(Ref. \onlinecite{Paszkowicz2002}), 2.506~\AA~(Ref. \onlinecite{Bosak2006}),   2.524$\pm$0.020~\AA~(Ref. \onlinecite{Yoo1997}) and 2.5038$\pm$0.0003~\AA~(Ref. \onlinecite{Yamamura1997}).

The potential energy surface of the graphene/boron-nitride heterostructure calculated for the optimized commensurate unit cell (Fig.~\ref{fig:pes}a) is in qualitative agreement with the previous studies \cite{Fan2011,Zhao2014,Leven2016,Zhou2015,Sachs2011,Argentero2017}. Two types of maxima on the potential energy surface correspond to the AA and AB2 stackings (Fig.~\ref{fig:pes}b). In the first of these stackings, all boron and nitrogen atoms are on top of carbon atoms and the energy of this stacking relative the ground-state AB1 stacking is 12.35 meV/atom, which is within the range of the values obtained at the optimized interlayer distance using different methods of 5.63 meV/atom (DFT-TS) \cite{Zhao2014}, $\sim$5.4 meV/atom (vdW-DF2) \cite{Argentero2017}, 10.86 meV/atom (HSE+MBD) \cite{Leven2016}, 15.68 meV/atom (DFT-D2), 5.35 meV/atom (vdW-DF2), 9.92 meV/atom (RPA) \cite{Zhou2015} and 10.5 meV/atom (RPA) \cite{Sachs2011} (note that all energies in the present paper are given in meV per atom in the upper (adsorbed) layer). The second smaller maximum with the relative energy 9.73 meV/atom corresponds to the AB2 stacking in which nitrogen atoms are located on top of carbon atoms and boron atoms are on top of centers of hexagons. For comparison, the literature values for the relative energy of the AB2 stacking are 4.5 meV/atom (DFT-TS) \cite{Zhao2014}, $\sim$4.6 meV/atom (vdW-DF2) \cite{Argentero2017},  12.96 meV/atom (DFT-D2), 4.68 meV/atom (vdW-DF2), 8.67 meV/atom (RPA) \cite{Zhou2015} and 9 meV/atom (RPA) \cite{Sachs2011}.  

The saddle point (SP) for the transition between adjacent energy minima (Fig.~\ref{fig:pes}a) lies on the straight line in the armchair direction connecting the AA and AB2 stackings on the potential energy surface, 0.427~\AA~away from the AB2 stacking and 1.015~\AA~away from the AA stacking (Fig.~\ref{fig:pes}b). The relative energy of this SP stacking corresponding to the barrier to relative in-plane motion of the graphene and boron nitride layers is 9.46 meV/atom, i.e. only 0.27 meV/atom smaller than the relative energy of the AB2 stacking. The previously reported values of the barrier obtained at the optimized interlayer distance include 3.9 meV/atom (DFT-TS) \cite{Zhao2014}, $\sim$4.4 meV/atom (vdW-DF2) \cite{Argentero2017}, $\sim$13 meV/atom (DFT-D2), $\sim$4.5 meV/atom (vdW-DF2) and $\sim$8.5 meV/atom (RPA) \cite{Zhou2015}. 

It should be noted that the calculated barrier for the graphene/boron-nitride heterostructure is several times greater than the barriers to relative sliding of two graphene layers or two boron nitride layers. The reported barriers to relative sliding of graphene layers range from 0.5 to 2.4~meV/atom (Refs. \onlinecite{Kolmogorov2005, Aoki2007, Ershova2010, Lebedeva2011, Reguzzoni2012, Zhou2015,Lebedeva2016, Lebedeva2017}). There are also estimates of this barrier from the experimental measurements of the shear mode frequency and width of dislocations in few-layer graphene of 1.7 meV/atom~(Ref. \onlinecite{Popov2012}) and 2.4 meV/atom~(Ref. \onlinecite{Alden2013}), respectively. The published values of the barriers for boron nitride layers range from 2.3~meV/atom to 4.3 meV/atom (Refs. \onlinecite{Constantinescu2013,Zhou2015, Lebedev2016, Lebedeva2017}). The same approach as used in the present paper gives the barriers for bilayer graphene and boron nitride of 1.6 meV/atom (Ref. \onlinecite{Lebedeva2016, Lebedeva2017}) and 3.9 meV/atom (Ref. \onlinecite{Lebedev2016, Lebedeva2017}), respectively, i.e. six and two and a half times smaller than the value for the graphene/boron-nitride heterostructure.  

The comparison of the relative energies of symmetric stackings for the graphene/boron-nitride heterostructure obtained here and in previous papers \cite{Zhao2014,Zhou2015,Sachs2011} shows that the approach applied in the present paper, i.e. the combination of the vdW-DF2 functional  with the use of the experimental interlayer distance in DFT calculations, overestimates the characteristics of the potential surface of interlayer interaction energy by 10\% -- 20\% with respect to more accurate RPA calculations \cite{Zhou2015,Sachs2011}. This error can be partially attributed to the fixed interlayer distance in our study since according to the RPA calculations \cite{Zhou2015,Sachs2011}, the optimal interlayer distance varies by 0.2 \AA~for different stackings. Nevertheless, this error is small compared, for example, to the scatter of the experimental data on the total binding energy of graphene layers \cite{Liu2012, Zacharia2004, Benedict1998, Girifalco1956} or more than 50\% deviation of the relative energies of symmetric stackings following from DFT calculations at the optimized interlayer distance \cite{Zhao2014,Zhou2015} from the RPA data. These observations are in agreement with the previous comparison of performance of different methods for pure graphene and boron nitride \cite{Lebedeva2017}. Therefore, the combination of the vdW-DF2 functional with the use of the experimental interlayer distance in standard DFT calculations is a computationally cheap but sufficiently accurate alternative to expensive methods, such as RPA. 

\begin{figure*}
	\centering
	\includegraphics[width=\columnwidth]{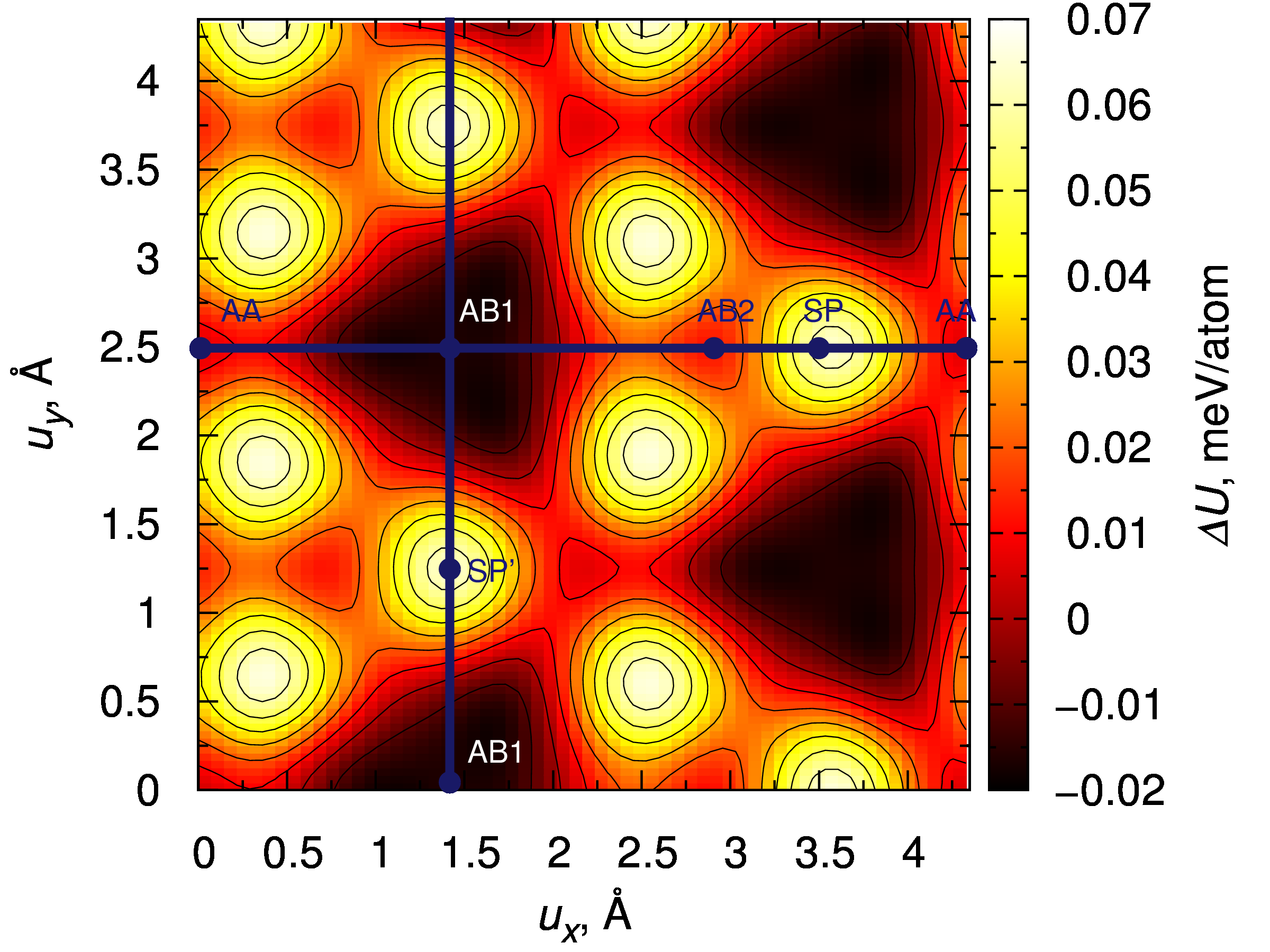}
	\caption{(Color online) Deviation $\Delta U$ (in meV per carbon atom)  of the potential surface of interaction energy of graphene and boron nitride layers approximated according to Eq.~(\ref{eq_2}) from the results of DFT calculations as a function of relative displacement of the layers in the armchair ($u_x$, in \AA) and zigzag ($u_y$, in \AA) directions at the interlayer distance of $d=$3.33~\AA.}
	\label{fig:diff}
\end{figure*}

\section{Approximation of potential energy surface}
The possibility to accurately approximate potential energy surfaces by expressions
containing only the first spatial Fourier harmonics determined by the system symmetry has been previously demonstrated for various types of interacting layers, including purely graphene \cite{Ershova2010,Lebedeva2011,Popov2012,Lebedeva2012,Zhou2015,Reguzzoni2012} and boron nitride \cite{Lebedev2016,Zhou2015} systems as well as carbon nanotube walls in the case when the walls are infinite and commensurate \cite{Belikov2004, Bichoutskaia2005, Bichoutskaia2009, Popov2009, Popov2012a} and when the corrugations of the potential energy surface are determined by edges \cite{Popov2013} or defects \cite{Belikov2004}. An expression for the potential surface of interaction energy of graphene and boron nitride layers including 5 fitting parameters has been also proposed \cite{Zhou2015}. In the present paper we show that this potential energy surface can be approximated using only two independent fitting parameters. A similar approximation of the potential energy surface with the first spatial Fourier components has been used for studies of strain distribution and band gap opening in a graphene layer on the boron nitride crystal \cite{Jung2015,Kumar2015}. However, the adequacy of this approximation has not been addressed. 

We consider the graphene layer as adsorbed on the
boron nitride one and use that the potential energy surface of an atom adsorbed on a 2D trigonal lattice can be approximated by the first Fourier harmonics as \cite{Verhoeven2004}
\begin{equation} \label{eq_1}
	\begin{split}
		U_\mathrm{at} = &\frac{1}{2}U_0 +\\
          &\frac{1}{2}U_1\bigg(2\cos{(k_x u_x)}\cos{(k_y u_y)} +\cos{(2k_x u_x)}+\frac{3}{2}\bigg),
	\end{split}
\end{equation}
where $x$ and $y$ axes are chosen in the armchair and zigzag directions, respectively, $k_x = 2\pi/(\sqrt{3}a)$,  $k_y = 2\pi/a$, $a$ is the lattice constant, $\vec{u}$ describes the relative position of the atom with respect to the 2D lattice and point $\vec{u} = 0$ corresponds to the case when the atom is located on top of one of the lattice atoms. In the case of graphene/boron-nitride heterostructures, we
should sum up interactions of carbon atoms with the boron (CB) and nitrogen (CN) lattices. Thus, we get
\begin{equation} \label{eq_2}
	\begin{split}
		&U = U_{0,\mathrm{tot}} + 3 (U_{1CB}+U_{1CN})+\\
			&U_{1CB}\bigg(2\cos{\bigg(k_x u_x - \frac{\pi}{3}\bigg)}\cos{(k_y u_y)} - \cos{\bigg(2k_x u_x- \frac{2\pi}{3}\bigg)}\bigg)+\\
			&U_{1CN}\bigg(2\cos{\bigg(k_x u_x +\frac{\pi}{3}\bigg)}\cos{(k_y u_y)} - \cos{\bigg(2k_x u_x+\frac{2\pi}{3}\bigg)}\bigg),
	\end{split}
\end{equation}
where $\vec{u}$ describes the relative position of the graphene and boron nitride layers, point $\vec{u} = 0$ corresponds to the AA stacking and $U_{0,\mathrm{tot}} = 2 (U_{0CB}+U_{0CN}) $.  It is seen that corrugations of the potential energy surface in flat graphene/boron-nitride heterostructures are described by two parameters $U_{1CB}$ and $U_{1CN}$, which determine all physical properties related to the interlayer interaction, such as the barrier to relative sliding of the layers, shear modulus and shear mode frequency \cite{Popov2012,Lebedeva2012,Lebedev2016,Lebedeva2016,Lebedeva2017}.

The parameters of the approximation can be found from the relative energies of the symmetric stackings AA, AB1 and AB2 as
\begin{equation} \label{eq_3}
	\begin{split}
		& U_{1CN} = \frac{2}{9}(E_\mathrm{AA}- E_\mathrm{AB1}),\\
  & U_{1CB} = \frac{2}{9}(E_\mathrm{AA}- E_\mathrm{AB2}).
	\end{split}
\end{equation}
The values obtained are $U_{1CB} = 0.58056$~meV/atom and $U_{1CN} = 2.7436$~meV/atom. The greater value of $U_{1CN}$ compared to $U_{1CB}$ can be explained by the stronger repulsion between carbon atoms and negatively charged nitrogen ions compared to positively charged boron ions. 
The standard deviation of the approximated potential energy surface from the one obtained by the DFT calculations is only 0.0305~meV/atom (Fig.~\ref{fig:pes}b,c and \ref{fig:diff}), which is within 0.25\% of the energy difference between the AA and AB1 stackings. The largest deviations of the approximation from the DFT results of up to 0.07 meV/atom are observed for the regions around the saddle-point stacking SP (Fig.~\ref{fig:diff}).

\section{Properties of graphene/boron-nitride heterostructures}
In this section we consider graphene/boron-nitride systems characterized by different stacking patterns and degree of incommensurability. Using the approach introduced by Porovski\u{i}  and Talapov in their pioneer paper \cite{Pokrovsky1978} on the commensurate-incommensurate phase transition and used for description of such a transition in multilayer films on surfaces (see, for example, Ref. \onlinecite{Chung1987}) we distinguish the following states: (1) commensurate system in which the layers have the same lattice constant and are in the AB1 stacking corresponding to the minimal interlayer interaction energy, (2) slightly incommensurate system in which there is a small lattice mismatch of the layers leading to formation of large commensurate domains separated by narrow incommensurate boundaries corresponding to stacking dislocations, (3) highly incommensurate system in which there is a considerable mismatch in the lattice constants of the layers but the positions of atoms in the layers are perturbed by the interlayer interaction and (4) fully incommensurate system in which there is a considerable mismatch in the lattice constants of the layers and the interlayer interaction virturally does not disturb the positions of atoms in the layers.  

The ground state of the graphene/boron-nitride bilayer with the aligned layers is the highly incommensurate system (3) with the period of the Moir\'{e} pattern comparable to the width of boundaries between commensurate domains \cite{Woods2014, Argentero2017, Tang2013,  Yang2013}. The relative rotation of the graphene and boron nitride layers leads to the transition to state (4), in which the potential energy surface is extremely smooth and no deformations of the layers due to the interlayer interactions are observed \cite{Woods2014}. A commensurate system (1) can be realized for a flake on a large substrate layer when the flake size is smaller than the period of the Moir\'{e} pattern or in the case when the graphene layer is stretched so that its lattice constant reaches the one for boron nitride. Biaxial stretching is required for bilayer geometry, while uniaxial stretching is sufficient for a ribbon of the width smaller than the period of the Moir\'{e} pattern on a wide substrate layer or on another ribbon. A slightly incommensurate system (2) can be achieved in this case when the lattice constant of graphene is slightly different from the one for boron nitride.

Our model of the interlayer interaction energy described by Eq. \ref{eq_2} does not take into account changes in the interlayer distance and thus we limit our consideration to the case of flat systems. The experimental measurements gave the height variation for supported graphene/boron-nitride systems of about 0.4 \AA~(Ref. \onlinecite{Yang2013}). This is comparable to the 0.2 \AA~variation of the optimal interlayer distance for different stackings according to the RPA calculations \cite{Zhou2015,Sachs2011}. Therefore, the deviation of 10\% -- 20\% for the relative energies of symmetric stackings obtained here at the fixed interlayer distance from the RPA data \cite{Zhou2015,Sachs2011} for the variable interlayer distance  (see section II) provides an estimate of the accuracy of our model for supported graphene/boron-nitride bilayers. The variation in the interlayer distance should be also insignificant for commensurate or fully incommensurate systems. On the other hand, very strong surface corrugation with the characteristic amplitude of 8 \AA~was predicted for free-standing graphene/boron-nitride bilayers on the basis of atomistic simulations  \cite{Argentero2017, Leven2016}. To be able to describe quantitatively the properties of such corrugated structures the model should be supplemented by the terms describing the dependence of the interlayer interaction energy on the interlayer distance similar to approximations of the potential energy surface proposed for graphene/boron-nitride heterostructures \cite{Zhou2015, Jung2015}  and graphene bilayer \cite{Reguzzoni2012} and interlayer atomic potentials \cite{Kolmogorov2005, Lebedeva2011} for graphene. Nevertheless, even in the present form the model can still provide a qualitative insight into the phenomena coming from the interlayer interaction and serve as an important step towards development of more complicated models with account of variations in the interlayer distance.

The possibility to describe the potential energy surface of graphene/boron-nitride heterostructures with only two independent parameters within Eq.~\ref{eq_2} means that a large number of properties of such systems are interrelated and expressed through these two parameters, similar to the cases of pure graphene and boron nitride systems \cite{Popov2012, Lebedev2016, Lebedeva2011}. Below we estimate the following physical properties of systems of different incommensurability. First the barrier to relative rotation of graphene and boron nitride layers, which corresponds to the transition from state (1) to state (4), is obtained for commensurate flakes on a large substrate layer.  The shear mode frequencies are evaluated for various commensurate systems (1). The period of the Moir\'{e} pattern is analyzed for the highly incommensurate state (3) corresponding to the ground state of the graphene/boron-nitride bilayer with the aligned layers. The phase transition from the commensurate state (1) to the slightly incommensurate state (2) with a low density of stacking dislocations and characteristics of these dislocations are studied in the last  part of this section.

\subsection{Barrier to rotation}
We start consideration of physical properties of graphene/boron-nitride systems from the barrier to relative rotation of the layers from the commensurate state. As mentioned above, though the ground state of the graphene/boron-nitride bilayer is incommensurate and it is difficult to realize the rotation in macroscopic systems, such a rotation is relevant for commensurate flakes on a large substrate layer. From studies for graphene it is known that superlubric behavior \cite{Dienwiebel2004,Dienwiebel2005,Filippov2008,Xu2013} and diffusion \cite{Lebedeva2010,Lebedeva2011a} of flakes on a periodic substrate occurs via rotation, which brings the system to the fully incommensurate state \cite{Woods2014} with an extremely smooth potential energy surface. Correspondingly, the diffusion coefficient of flakes is determined by the barrier to rotation from the commensurate state to the fully incommensurate one \cite{Lebedeva2010,Lebedeva2011a}. The potential energy of the latter can be found as the average interlayer interaction energy given by Eq.~\ref{eq_2}
\begin{equation} \label{eq_4}
	\begin{split}
		\bar{U}  =U_{0,\mathrm{tot}} + 3 (U_{1CB}+U_{1CN}).
	\end{split}
\end{equation}

Therefore, the barrier to relative rotation of the graphene and boron nitride layers from the  commensurate state with the AB1 stacking can be estimated as $U_{\mathrm{in}}= \bar{U} - U (\mathrm{AB1}) = -3 U_{1CB}/2 + 3 U_{1CN} = 7.36$ meV/atom. This value is in agreement with the estimate of the relative energy of the fully incommensurate state from the RPA calculations\cite{Sachs2011} of $\sim$7 meV/atom and is somewhat greater than the reported barriers to relative rotation of two graphene layers of $\sim$5
meV/atom (Ref.~\onlinecite{Popov2012}) and 4 meV/atom (Ref.~\onlinecite{Lebedeva2010, Lebedeva2011a}) and of two boron nitride layers of 6.3 meV/atom (Ref.~\onlinecite{Lebedev2016}). 

\subsection{Shear mode frequency}
One of the experimentally measurable quantities that can be used to probe the potential surface of interlayer interaction energy is the frequency of the shear mode in which commensurate layers rigidly slide with respect to each other parallel to the plane\cite{Lebedeva2011, Lebedeva2012, Popov2012, Lebedev2016, Lebedeva2017}. To estimate shear mode frequencies for commensurate systems of the graphene/boron-nitride bilayer, a graphene flake on a boron nitride layer and a boron nitride flake on a graphene layer the calculations of the curvature $\partial^2 U/\partial x^2$ of the potential energy surface around the AB1 minimum are performed for the commensurate unit cells with different lattice constants corresponding to the optimized lattice constants of the bilayer and single graphene and boron nitride layers.
The graphene and boron nitride layers are rigidly displaced within 0.05~\AA~in the armchair direction and the obtained energy curves are approximated by parabolas. 

The calculations show that the curvatures of the potential energy surface $\partial^2 U/\partial x^2$  for the lattice constants $a_{BN}$ and $a_{C}$ corresponding to the isolated boron nitride and graphene layers differ only by 1.9\% and 2.6\%, respectively, from the value for the lattice constant $a$ of the bilayer. The curvature of the potential energy surface for the bilayer can be also estimated from Eq. \ref{eq_2} as $\partial^2 U/\partial x^2 =4\pi^2(2U_{1CN}-U_{1CB})/a^2= 0.031$ eV/\AA. This estimate is only 1.3\% greater than the value obtained directly from the DFT calculations.

The shear mode frequencies $f_E$ are then found as~\cite{Popov2012, Lebedev2016, Lebedeva2017}
\begin{equation} \label{eq_freq}
	\begin{split}
		f_E = \frac{1}{2\pi}\sqrt{\frac{1}{\mu}\frac{\partial^2 U}{\partial x^2}}.
	\end{split}
\end{equation}
Here for the graphene flake on the boron nitride layer  $\mu = m_C$, for the boron nitride flake on the graphene layer $\mu = m_{BN} =(m_{B} + m_{N})/2$ and for the bilayer  $\mu = m_C m_{BN}/(m_C+m_{BN})$, where $m_C$, $m_{B}$ and $m_{N}$ are masses of carbon, boron and nitrogen atoms, respectively. The shear mode frequencies calculated in this way are listed in Table~\ref{table:freq}. As seen from this table, the results for the graphene flake on the boron nitride layer and the boron nitride flake on the graphene layer are virtually indistinguishable due to the close reduced masses and curvatures of the potential energy surface. 

It should be also noted that the estimated shear mode frequency for the graphene/boron-nitride bilayer of 37 cm$^{-1}$ is somewhat higher than the frequencies for bilayer graphene \cite{Lebedeva2017} of 29 cm$^{-1}$ and boron nitride \cite{Lebedev2016, Lebedeva2017} of 34 cm$^{-1}$ obtained within the same computational approach. This value is also greater than the results of experimental measurements for bilayer graphene of 28 $\pm$ 3 cm$^{-1}$ (Ref. \onlinecite{Boschetto2013}) and 32 cm$^{-1}$ (Ref. \onlinecite{Tan2012}).

\begin{table}[h]
	\caption{Calculated shear mode frequencies $f_E$ of commensurate systems of a graphene/boron-nitride bilayer, a graphene flake on a boron nitride layer and a boron nitride flake on a graphene layer with the different lattice constant $a$.  }
	\begin{tabular}{p{4.5cm}p{1.5cm}p{1.5cm}}
		\hline
		structure & $a$ (\AA) &  $f_E$  (cm$^{-1}$)\\\hline
		graphene flake/boron nitride  & 2.5207 & 26.59 \\\hline
	       bilayer & 2.4978 & 36.95 \\\hline
		boron nitride flake/graphene & 2.4767 & 25.58    \\\hline\hline
	\end{tabular}\label{table:freq}
\end{table}

\subsection{Moir\'{e} pattern}
Let us now obtain the correction to the period of the Moir\'{e} pattern in the incommensurate ground state of graphene/boron-nitride heterostructures induced by the interlayer interaction. First we consider the relationship between the elastic and interlayer interaction energies in the fully incommensurate and commensurate states of the aligned graphene and boron nitride layers. Maintaining the structure of the layers the same as in the absence of the interlayer interaction is related to excess in the average density of the interlayer interaction energy compared to the commensurate state of $w_\mathrm{int} = U_{\mathrm{in}}/\sigma = 44$~mJ/m$^2$, where $\sigma = \sqrt{3}a^2/4 = 2.7$~\AA$^2$ is the area per one carbon atom. Transformation of the incommensurate structure to the commensurate one with the uniformly stretched graphene layer and compressed boron nitride layer, on the other hand, is associated with the penalty in the elastic energy of $w_\mathrm{el} = k\delta^2/(1-\nu)$, where $\delta = (a_{\mathrm{BN}}-a_{\mathrm{C}})/a=1.78$\% is the relative lattice mismatch, $k$ is the reduced elastic constant of the layers and $\nu$ is the Poisson ratio. In the present paper we use the values of the elastic constants of graphene and boron nitride layers $k_{\mathrm{C}} = 331$~J/m$^2$ and $k_{\mathrm{BN}} = 273$~J/m$^2$ calculated previously\cite{Lebedeva2016} using the same computational approach. Although slightly different values of the Poisson ratio were obtained for graphene and boron nitride, $\nu_{\mathrm{C}} = 0.174$ and $\nu_{\mathrm{BN}}= 0.201$, in this section we assume that the Poisson ratios of the layers are close and approximately equal to that for graphene $\nu = \nu_{\mathrm{C}}$ (in section IV D the difference in the Poisson ratios of the graphene and boron nitride layers is taken into account). In the case when both of the graphene and boron nitride layers are free to relax in the plane (this can be achieved by using an incommensurate substrate such as a rotated graphene or boron nitride layer), the reduced elastic constant is $k= k_{\mathrm{C}}k_{\mathrm{BN}}/(k_{\mathrm{C}}+k_{\mathrm{BN}}) = 150$~J/m$^2$ and the density of the elastic energy in the commensurate state $w_\mathrm{el} = 56$~mJ/m$^2$. Comparison of energies $w_\mathrm{el}$ and $w_\mathrm{int}$ suggests that in the ground state the heterostructure should be incommensurate. For a free graphene layer on the boron nitride crystal (similar to the experimental studies \cite{Woods2014,Tang2013, Yang2013}), the interface boron nitride layer can be considered as nearly rigid \cite{Sachs2011,Wijk2014}. In this case the reduced elastic constant of the layers $k \approx k_{\mathrm{C}}$ is approximately twice greater, the density of the elastic energy in the commensurate state is $w_\mathrm{el} = 124$~mJ/m$^2$ and the incommensurate state is even more preferred. The same behavior with $w_\mathrm{el} = 102$~mJ/m$^2$ should also be observed for a free boron nitride layer on the fixed graphite substrate.

Incommensurability of the graphene/boron-nitride systems is manifested through the formation of Moir\'{e} superstructures \cite{Woods2014, Argentero2017, Tang2013,  Yang2013}. In the case when the change in the bond lengths of the layers due to the interlayer interaction is neglected and the layers are completely aligned we can estimate that the period of the Moir\'{e} pattern is $L_0=a/\delta = 14.2$ nm, in agreement with the experimental measurements of $\sim$14 nm \cite{Woods2014,Tang2013} and $15\pm1$ nm \cite{Yang2013} and the values of 14.1 nm (Ref. \onlinecite{Leven2016}) and 13.6 nm (Ref. \onlinecite{Wijk2014}) observed in atomistic simulations. A similar estimate of $\sim$14 nm can be also obtained using typical values of the lattice constants of 2.461 \AA~and 2.504 \AA~from the experimental data for graphite \cite{Pease1952,Baskin1955,Ludsteck1972,Trucano1975,Zhao1989,Bosak2007} and boron nitride \cite{Paszkowicz2002,Yamamura1997,Pease1950,Bosak2006,Solozhenko1997,Solozhenko2001, Solozhenko1995,Pease1952}. Nevertheless, since the characteristic interlayer interlayer interaction energy $w_\mathrm{int}$ is comparable to the characteristic elastic energy $w_\mathrm{el}$, significant modulation of positions of atoms in the interacting layers is possible. Such a modulation has been previously observed in atomistic simulations \cite{Wijk2014, Leven2016, Argentero2017} and calculations using continuum models \cite{Jung2015,Kumar2015}. However, the correction to the period of the Moir\'{e} pattern induced by the interlayer interaction has not been considered explicitly. Below we quantify this correction for aligned graphene and boron nitride layers.

In heterostructures formed by uniformly deformed graphene and boron nitride layers (with the constant in-plane strains) the relative displacement of atoms in the layers varies uniformly with the absolute position of atoms within the layers $u_{0,x} = xa/L$, $u_{0,y} = ya/L$, where $L$ is the period of the Moir\'{e} pattern, axes $x$ and $y$ are chosen in the armchair and zigzag directions and the point $x=0$, $y = 0$ corresponds to the AA stacking following the notations from sections II and III. The Moir\'{e} pattern has the same hexagonal symmetry as the potential energy surface, which is imposed by the geometry of the layers, with the only difference that the unit cell of the Moir\'{e} pattern is scaled by a factor of $L/a$ as compared to the unit cell of the potential energy surface. Obviously this symmetry is not changed by the interlayer interaction. However, the interlayer interaction modifies the period of the Moir\'{e} pattern $L = L_0 + \Delta L$ and the relative displacement of atoms in the layers $\vec{u} = \vec{u}_0 + \Delta\vec{u}$. Here we assume that the perturbations  $\Delta L$ and $\Delta\vec{u}$ are small so that $\Delta L \ll L_0$  and $\Delta u \ll a$. Since the symmetry of the pattern is not affected by the interlayer interaction, the correction $\Delta\vec{u}$ to the relative displacement of atoms should vanish at the symmetry planes. This is automatically satisfied for the correction of the form
\begin{equation} \label{eq_5c}
	\begin{split}
     \Delta\vec{u} =\alpha \nabla U(\vec{u}_0),
	\end{split}
\end{equation}
i.e. in the case when the correction is proportional to the force of interlayer interaction.

The interlayer interaction energy can then be expanded as 
\begin{equation} \label{eq_5}
	\begin{split}
		U (x,y) \approx & \bar{U} +  \alpha \frac{L}{a}\left\{ \left(\frac{\partial U}{\partial u_x}\right)^2 + \left(\frac{\partial U}{\partial u_y}\right)^2\right\} .
	\end{split}
\end{equation}
It is sufficient to stop here at the first-order terms in $\alpha$ as it can be easily checked that the second-order terms vanish upon the integration over the unit cell of the Moir\'{e} superstructure and do not contribute to the total energy. Integrating Eq. \ref{eq_5} over the unit cell of the Moir\'{e} superstructure, we get the average interlayer interaction energy per unit area
\begin{equation} \label{eq_5b}
	\begin{split}
	 \langle U \rangle \approx & \bar{U} + \alpha  \frac{8\pi^2U_1^2}{aL},
	\end{split}
\end{equation}
where 
\begin{equation} \label{eq_5a}
	\begin{split}
         U_1 =\left(U_{1CB}^2+U_{1CN}^2-U_{1CB} U_{1CN}\right)^{1/2}
  \end{split}
\end{equation}
and this parameter corresponds to the characteristic corrugation of the potential energy surface.

Tensile ($\epsilon$) and shear ($\tau$) strains associated with the corrections to the period of the Moir\'{e} pattern and relative displacement of the layers are given by
\begin{equation} \label{eq_6}
	\begin{split}
		 &\epsilon_{x} = \bar{\epsilon}  + \frac{\partial \Delta u_x}{\partial x},  \quad \epsilon_{y} = \bar{\epsilon}  + \frac{\partial \Delta u_y}{\partial y}, \\
  & \bar{\epsilon} = a \left(\frac{1}{L} - \frac{1}{L_0}\right), \quad \tau = \frac{\partial \Delta u_x}{\partial y} + \frac{\partial \Delta u_y}{\partial x}.
	\end{split}
\end{equation}

Using notation for the total strain
\begin{equation} \label{eq_6a}
	\begin{split}
\epsilon (x,y)=\left(\epsilon_{x}^2 + \epsilon_y^2 + 2 \nu  \epsilon_x \epsilon_y +\frac{ (1-\nu) \tau^2}{2} \right)^{1/2},
	\end{split}
\end{equation}
the elastic energy of the heterostructure can be written as 
\begin{equation} \label{eq_7}
	\begin{split}
		W_{el} (x,y) = \frac{k\epsilon^2}{2(1-\nu^2)}. 
	\end{split}
\end{equation}
The average elastic energy per unit area is therefore given by
\begin{equation} \label{eq_8}
	\begin{split}
		 \langle  W_{el} \rangle =\alpha^2 \frac{k U_1^2(4-\nu)}{3(1-\nu^2)}  \left(\frac{2\pi}{L}\right)^4   + \frac{k\bar{\epsilon} ^2}{(1-\nu)}
	\end{split}
\end{equation}

The total energy of the heterostructure $\Delta w = \langle  W_{el} \rangle + \langle U \rangle - \bar{U}$ relative to the incommensurate state with no in-plane deformation of the layers due to the interlayer interaction is minimal for  
\begin{equation} \label{eq_9}
	\begin{split}
		\alpha =-\frac{3(1-\nu^2)}{k(4-\nu)}\left(\frac{L}{2\pi}\right)^2  \frac{L}{a}. 
	\end{split}
\end{equation}

Let us now check that the resulting correction to the relative displacement of atoms in the layers is actually much smaller than the lattice constant $\Delta u  \ll a$ and thus the perturbation approach applied here is valid. From Eqs. \ref{eq_5c} and \ref{eq_9} and relation $L \sim a/\delta$ it can be estimated that the characteristic magnitude of the correction to the relative displacement of atoms of the graphene and boron nitride layers is 

\begin{equation} \label{eq_9a}
	\begin{split}
		\frac{\Delta u}{a} \sim \frac{\alpha U_1}{aL} \sim \frac{3}{8\pi^2}  \frac{U_1}{k\delta^2}.
	\end{split}
\end{equation}

This means that the approach used here makes sense as long as the characteristic magnitude of corrugation of the potential energy surface $U_1$ is less or comparable to the characteristic elastic energy $k\delta^2$. In our case $U_1 = 15$~mJ/m$^2$, $k\delta^2 =46$~mJ/m$^2$ when both of the graphene and boron nitride layers are free to relax in the plane,  103~mJ/m$^2$ for the graphene layer on the fixed boron nitride substrate and 85~mJ/m$^2$ for the boron nitride layer on the fixed graphite substrate, which ensures $\Delta u  \ll a$ and adequacy of the approach used. 

\begin{figure*}
	\centering
	\includegraphics[width=\textwidth]{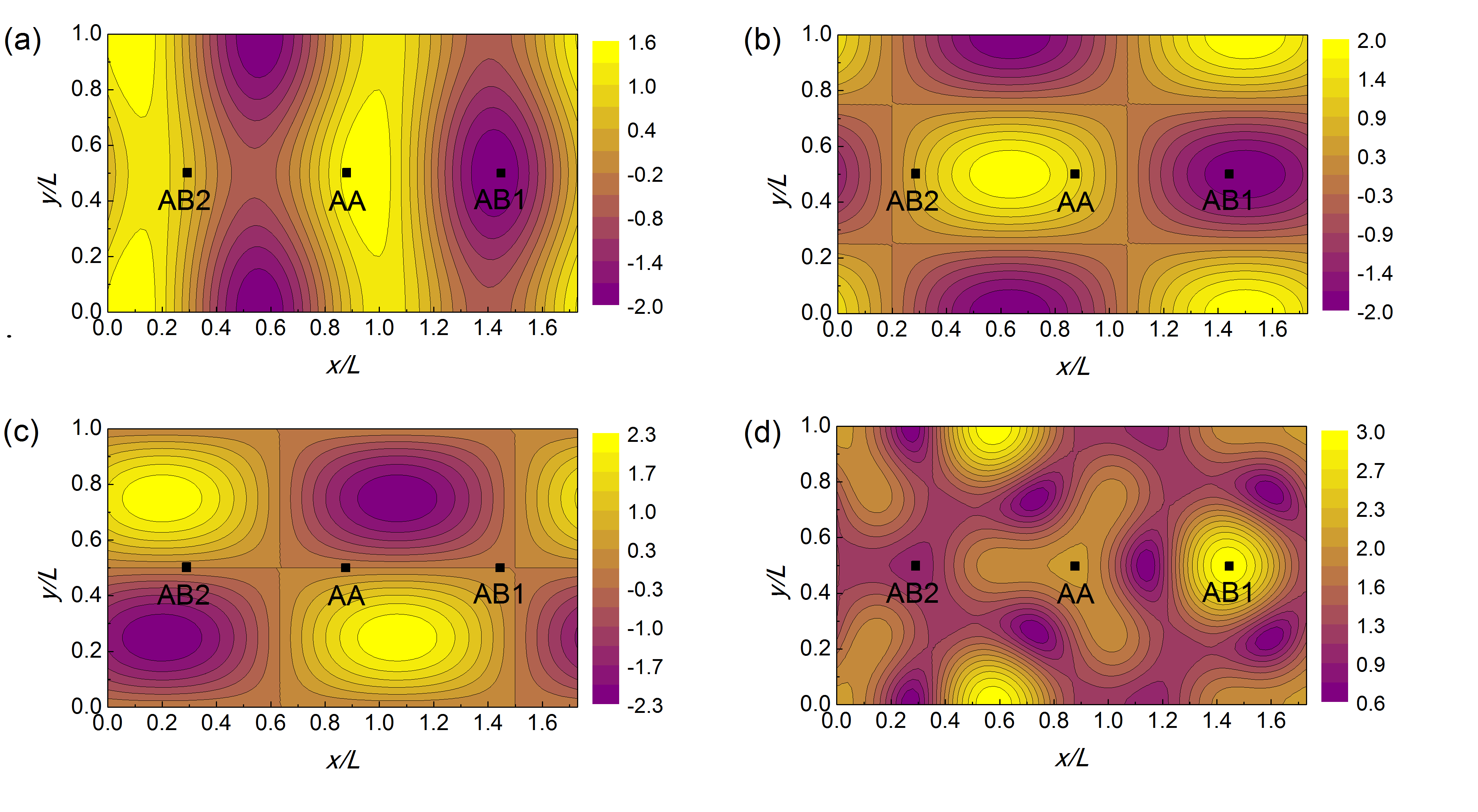}
	\caption{(Color online) Calculated tensile strains $\epsilon_x/\epsilon_0$ (a) and  $\epsilon_y/\epsilon_0$ (b), shear strain $\tau/\epsilon_0$ (c) and total strain $\epsilon/\epsilon_0$ (d) in free layers of the graphene/boron-nitride heterostructure as functions of coordinates $x/L$ and $y/L$ in the armchair and zigzag directions, respectively.  The unit $\epsilon_0$ equals $1.94 \cdot 10^{-3}$ for the graphene layer and $2.35 \cdot 10^{-3}$ for the boron nitride layer (see text). The perturbation in the period $L$ of the Moir\'{e} superstructure is neglected ($L \approx L_0$). The point $x = 0$ and $y = 0$ corresponds to the AA stacking.}
	\label{fig:strains}
\end{figure*}

One of the consequences of the perturbation approach used is that in the limit of $L \approx L_0$ the strain in each free layer does not depend on whether the second layer is free or fixed. This is clear from Eq. \ref{eq_9} as the coefficient $\alpha$, which determines the relative displacement $\Delta u$ of atoms in the layers, is inversely proportional to the reduced elastic constant $k$ of the system. In the case when both of the layers are free to relax in the plane, absolute displacements of atoms in the graphene and boron nitride layers correspond to the fractions $k_{\mathrm{BN}}/(k_{\mathrm{C}}+k_{\mathrm{BN}})$ and 
$k_{\mathrm{C}}/(k_{\mathrm{C}}+k_{\mathrm{BN}})$ of the relative displacement $\Delta u$, respectively. As a result, the same displacement as on the fixed substrate with $k = k_{\mathrm{C}}$ or $k = k_{\mathrm{BN}}$, respectively, is obtained. 

The distribution patterns of strain in free layers for $L \approx L_0$ are shown in Fig. \ref{fig:strains} in units of $\epsilon_0=-4\pi^2 \alpha U_1/L_0^2 = 1.94 \cdot 10^{-3}$ for the graphene layer and $2.35 \cdot 10^{-3}$ for the boron nitride layer. Such a strain distribution is qualitatively similar to the distribution of bond lengths observed in atomistic simulations \cite{Wijk2014, Leven2016, Argentero2017} and elastic energy maps obtained within continuum models \cite{Jung2015,Kumar2015}. As can be expected, the maximal strains of about $3\epsilon_0$ are observed in the regions with the AB1 stacking, where the layers tend to be commensurate in order to minimize the interlayer interaction energy (Fig. \ref{fig:strains}). These maximal strains of 0.6 -- 0.7\% are in agreement with variations of the bond length of about 1\% measured experimentally for the graphene layer on the boron nitride crystal \cite{Woods2014}. The minimal strains of about $0.6\epsilon_0$ are achieved for the stacking that is intermediate between the AA and AB1 stackings. Adjusting the period $L$ of the  Moir\'{e} superstructure allows to reduce the strain though the $\bar{\epsilon}$ term (see Eq. \ref{eq_6}). 

With account of Eq. \ref{eq_9}, the energy of the heterostructure relative to the incommensurate state with no in-plane deformation of the layers due to the interlayer interaction takes the form
\begin{equation} \label{eq_10}
	\begin{split}
		\Delta w  = \frac{k\delta^2}{(1-\nu)}\left(\frac{\Delta L}{L}\right)^2 - \frac{3(1-\nu^2)U_1^2}{(4-\nu)k\delta^2}\left(1+ \frac{\Delta L}{L_0}\right)^2 .
	\end{split}
\end{equation}

The optimal change of the period of the Moir\'{e} superstructure is then roughly
\begin{equation} \label{eq_11}
	\begin{split}
		\frac{\Delta L}{L_0} \approx \frac{3(1+\nu)(1-\nu)^2}{(4-\nu)}\left(\frac{U_1}{k\delta^2}\right)^2.
	\end{split}
\end{equation}

When one of the graphene or boron nitride layers is fixed, the correction to the period of the Moir\'{e} superstructure is rather small. For the fixed graphene layer, $\Delta L = 0.32$ nm, i.e. 2.3\% of the period $L_0$ of the Moir\'{e} superstructure without account of in-plane deformation of the layers due to the interlayer interaction. For the fixed boron nitride layer, $\Delta L = 0.21$ nm, i.e. 1.5\% of $L_0$.  In the latter case, the estimated period of the Moir\'{e} superstructure is $L = 14.4$ nm. The same as $L_0$, this is close to the experimentally measured period of the Moir\'{e} superstructure for the aligned graphene layer on the boron nitride crystal of about 14 nm \cite{Woods2014,Tang2013} and $15\pm1$ nm \cite{Yang2013}. However, these experimental data are insufficient to confirm the effect of the interlayer interaction. A more significant correction to the period of the Moir\'{e} superstructure of $\Delta L = 1.30$ nm, i.e. 9.2\% of $L_0$, with the resulting period of $L = 15.5$ nm, is expected when both of the graphene and boron nitride layers are free to relax in the plane, which can be the case for the graphene/boron-nitride bilayer on an incommensurate substrate, for example, on a rotated layer of graphene or boron nitride. Experimental measurements for such a system could help to distinguish the effect of the interlayer interaction. 

\subsection{Stacking dislocations in slightly incommensurate systems}
Let us now consider stacking dislocations in slightly incommensurate graphene/boron-nitride systems. Stretching graphene so that its lattice constant increases up to the one for boron nitride $a_{BN}$ results in formation of the commensurate structure with the minimal interlayer interaction energy. Futher stretching or compressing the graphene layer should lead to a competition in the elastic energy of the free boron nitride layer and interlayer interaction energy resolved through the transition to the slightly incommesurate state with stacking dislocations. In other words, a commensurate-incommensurate phase transition \cite{Pokrovsky1978} should take place at some critical strain associated with the difference in the average lattice constants of the graphene and boron nitride layers. 

As discussed in the beginning of section IV, a slightly incommensurate system can be realized by biaxial stretching of the graphene layer for bilayer geometry or by uniaxial stretching if the boron nitride layer is a ribbon of the width smaller than the period of the Moir\'{e} superstructure. In the latter case stretching should take place along the ribbon axis and we disregard the effect of the ribbon edges on the potential surface of interlayer interaction energy, which is reasonable if the ribbon width is much greater than the lattice constant and the edges are properly terminated, e.g. by hydrogen. We also assume that the density of stacking dislocations is low so that the interaction between the dislocations can be neglected. These assumptions allow us to use the formalism of the two-chain Frenkel-Kontorova model \cite{Bichoutskaia2006}, in which two layers are represented as chains of particles connected by harmonic springs and coupled by van der Waals interactions. The model has been already applied to study the commensurate-incommensurate phase transition in double-walled carbon nanotubes \cite{Bichoutskaia2006, Popov2009}, bilayer graphene \cite{Popov2011,Lebedeva2016} and boron nitride \cite{Lebedev2016,Lebedeva2016} as well as edge stacking dislocations in bilayer graphene \cite{Lebedeva2017a}.

Following the approach from papers \cite{Popov2011,Lebedeva2016,Lebedev2016}, it is first necessary to choose the adequate approximation of the dislocation path, i.e. the curve on the potential energy surface described by the dependence of relative displacement $\vec{u}$ of the layers on the coordinate in the direction perpendicular to the boundary between commensurate domains that minimizes the formation energy of dislocations. In the case of graphene/boron-nitride heterostructures, the minimum energy path between the AB1 minima passing through the SP stacking only slightly deviates from the straight line between adjacent energy minima in the zigzag direction. The barrier along this straight path  (corresponding to the SP' stacking) is 9.81 meV/atom (Fig.~\ref{fig:pes}c). This is only 4\% higher than the barrier along the minimum energy path through the SP stacking and within the accuracy of the DFT calculations with corrections for van der Waals interactions. The SP' stacking is obtained from the SP stacking by shifting the layers by only 0.29 \AA. Therefore, it can be safely assumed that the dislocation path in graphene/boron-nitride heterostructures lies along the straight line between adjacent energy minima AB1.

On the basis of Eq. \ref{eq_2}, the relative potential energy along the straight dislocation path is then approximated as
\begin{equation} \label{eq_approx}
	\begin{split}
		V(u) = V_\mathrm{max} \big(1-\cos(k_0 u)\big)/2,
	\end{split}
\end{equation}
where $k_0 = 2\pi/b$, $u$ is the relative displacement of the layers along the dislocation path, which changes from 0 to the magnitude of the Burgers vector $b = a_{BN}$, and $V_\mathrm{max}$ is the barrier per unit area. The barrier along this path calculated from Eq. \ref{eq_2} is $U_\mathrm{max}=2(2U_{1CN}-U_{1CB}) =9.81$ meV/atom, which is exactly equal to our DFT value. The variation of the barrier due to changes in the lattice constant from $a$ to $a_{BN}$ lies within 2\% and is neglected. Therefore, we estimate the barrier per unit area as $V_\mathrm{max} = 4U_\mathrm{max}/(\sqrt{3}a_{BN}^2)=0.0571$ J/m$^2$. 

The dislocation width depends on the angle $\beta$ between the Burgers vector and the normal to the boundary between commensurate domains \cite{Lebedev2016,Lebedeva2016} and is given by 
\begin{equation} \label{eq_width}
	\begin{split}
		l_\mathrm{D} (\beta) = b \sqrt{\frac{K(\beta)}{2V_\mathrm{max}}},
	\end{split}
\end{equation}
where $K(\beta) = E \cos^2{\beta} +G \sin^2{\beta}$ describes the dependence of the reduced elastic constant of the layers on fractions of tensile and shear character in the dislocation, $E$ and $G$ are the reduced tensile and shear elastic constants. When the boron nitride and graphene layers are of similar width, both of the layers participate in the formation of the dislocation and the reduced constants are given by $E =E_{C}E_{BN}/(E_{C}+E_{BN})$ and $G =G_{C}G_{BN}/(G_{C}+G_{BN})$, where $E_{C(BN)} = k_{C(BN)}/(1-\nu_{C(BN)}^2)$ and $G_{C(BN)} = k_{C(BN)}/2(1+\nu_{C(BN)})$, respectively. For the boron nitride ribbon on the wide graphene layer, the dislocation appears only in the boron nitride layer and, correspondingly, $E \approx E_{BN}$ and $G \approx G_{BN}$. 

\begin{figure*}
	\centering
	\includegraphics[width=\textwidth]{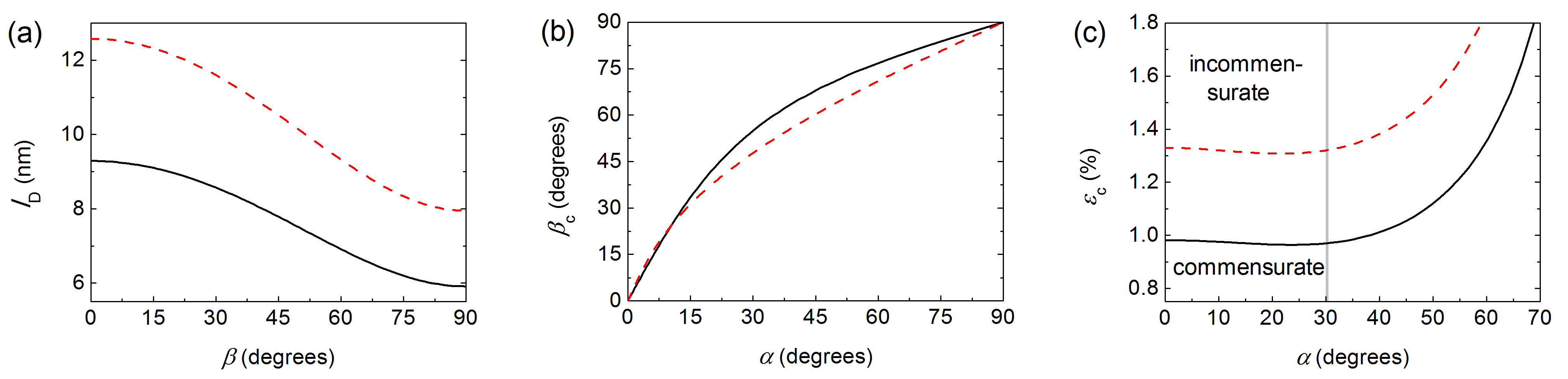}
	\caption{(Color online) Calculated properties of stacking dislocations in graphene and boron nitride layers of similar width with the lattice mismatch in one of the directions (black solid lines) or a boron nitride ribbon on a wide graphene layer with the lattice mismatch along the ribbon axis (red/gray dashed lines) : (a) dislocation width $l_\mathrm{D}$ (in nm) as a function of the angle  $\beta$ (in degrees) between the Burgers vector and the normal to the boundary between commensurate domains, (b) optimal angle $\beta_\mathrm{c}$ (in degrees) at the critical strain as a function of the angle $\alpha$ (in degrees) between the Burgers vector and the direction in which the lattice constants of the layers are slightly different, (c) critical strain $\epsilon_\mathrm{c}$ (in \%) for formation of dislocations with the angle $\beta=\beta_\mathrm{c}$ associated with the average lattice mismatch between the graphene and boron nitride layers as a function of angle $\alpha$ (in degrees). The absolute critical strain in the graphene layer is given by $\epsilon^\mathrm{C}_\mathrm{c,\pm} (\alpha) =\delta \pm \epsilon_\mathrm{c} (\alpha)$. The phase diagram for the commensurate-incommensurate phase transition corresponds to the left part of panel (c) with $\alpha \le 30^{\circ}$ and is separated from the rest of the figure by the vertical gray line.}
	\label{fig:disl}
\end{figure*}

As seen from Fig. \ref{fig:disl}a, the dislocation width $l_\mathrm{D}$ decreases with increasing the angle  $\beta$, i.e. changing the dislocation character from tensile to shear. The estimated dislocation widths $l_\mathrm{D}$ in the case of similar widths of the graphene and boron nitride layers are 6 -- 9 nm, which is much smaller then the width of full dislocations in bilayer boron nitride of 12 -- 16 nm (Ref. \onlinecite{Lebedev2016,Lebedeva2016}) evaluated within the same approach and only a little smaller than the corresponding results for partial dislocations in bilayer graphene of 8 -- 14 nm (Ref. \onlinecite{Popov2011,Lebedeva2016}). These values also lie within the range of experimental data on dislocation widths in bilayer graphene of 6 -- 11 nm (Ref. \onlinecite{Alden2013,Lin2013,Yankowitz2014}). While the barrier $V_\mathrm{max}$ along the dislocation path in graphene/boron-nitride heterostructures is several times greater than that in bilayer graphene, the length of this path is also greater (the dislocation path length is equal to the lattice constant $a_{BN}$ in the heterostructure and to the bond length $l_{C} \approx a_{BN}/\sqrt{3}$ in bilayer graphene) and the resulting dislocation widths are close. 

The dislocation energy $W_0$, i.e. the energy of the state with one stacking dislocation relative to the commensurate state with equal lattice constants of the layers per unit length of the boundary between commensurate domains, follows the same dependence on the angle $\beta$ as the dislocation width $W_0 (\beta) =2b\sqrt{2K(\beta)V_\mathrm{max}}/\pi$ (Refs. \onlinecite{Lebedev2016,Lebedeva2016}). For the system of two layers of similar width this energy ranges from 0.42 meV/\AA~for tensile dilsocations to 0.27 meV/\AA~for shear dislocations, which is about four times greater than for bilayer graphene \cite{Popov2011,Lebedeva2016} and comparable to the estimates for bilayer boron nitride \cite{Lebedev2016,Lebedeva2016}. For the boron nitride ribbon on the wide graphene layer, the corresponding values are 0.57 meV/\AA~and 0.36 meV/\AA, respectively.

Let us now consider the commensurate-incommensurate phase transition in the system in which the average lattice constant of graphene in one of the directions (along the ribbon axis in the case  of the boron nitride layer of the ribbon shape) is slightly smaller or larger than the average lattice constant of boron nitride. The commensurate phase should be observed for strains $ \epsilon^\mathrm{C}$ in the graphene layer within the interval $\epsilon^\mathrm{C}_\mathrm{c,-}=\delta - \epsilon_\mathrm{c} < \epsilon^\mathrm{C} < \epsilon^\mathrm{C}_\mathrm{c,+}=\delta + \epsilon_\mathrm{c}$, where $\epsilon_\mathrm{c}$ is the critical strain associated with the average lattice mismatch of the graphene and boron nitride layers at which the formation energy of stacking dislocations goes to zero. The commensurate-incommensurate phase transition in similar systems of pure graphene and boron nitride with the uniaxially stretched substrate layer was considered in our previous paper \cite{Lebedeva2016}. In that paper we derived the expressions for $\epsilon_\mathrm{c}$ and the optimal angle $\beta_c$ between the boundary separating commensurate domains and the Burgers vector which minimizes the formation energy of stacking dislocations at the critical strain (or at any strain in the substrate layer if the free layer is a ribbon). According to these derivations, the optimal angle $\beta_c$ between the boundary separating commensurate domains and the Burgers vector depends on the angle $\alpha$ between the Burgers vector and the direction in which the average lattice constants of the layers are slightly different  (Fig. \ref{fig:disl}b) as  
\begin{equation} \label{eq_beta}
\begin{split}
\tan{(\alpha - \beta_\mathrm{c})} = -\frac{(E-G)\sin{2\beta_\mathrm{c}}}{2K(\beta_\mathrm{c})}.
\end{split}
\end{equation}

In the case of the graphene/boron-nitride systems the critical strain $\epsilon_\mathrm{c}$ associated with the average lattice mismatch between the graphene and boron nitride layers \cite{Lebedeva2016} is given by  
\begin{equation} \label{eq_10}
\begin{split}
\epsilon_\mathrm{c} (\alpha)= \frac{W_0 (\beta_\mathrm{c})}{k_{BN} b \cos{\alpha} \cos{(\alpha - \beta_c)}},
\end{split}
\end{equation}
while the absolute critical strain in the graphene layer takes the values of $\epsilon^\mathrm{C}_\mathrm{c,\pm} (\alpha) =\delta \pm \epsilon_\mathrm{c}(\alpha)$. 

As seen from Fig. \ref{fig:disl}c, the critical strain $\epsilon_\mathrm{c}$ for $\alpha = 0^{\circ}$ is about 0.98\% when the layers are of similar width and 1.33\% for the boron nitride ribbon on the wide graphene layer. The dependence of the critical strain on the angle $\alpha$ is rather weak for angles within $30^{\circ}$. A shallow minimum in the critical  strain $\epsilon_\mathrm{c}$ of  about 0.96\% is reached at $\alpha = 23.5^{\circ}$ for the layers of similar width. For the boron nitride ribbon on the wide graphene layer, the minimal $\epsilon_\mathrm{c}$ is about 1.31\% and is reached at $\alpha = 21.3^{\circ}$. With increasing $\alpha$ beyond $30^{\circ}$ the critical strain $\epsilon_\mathrm{c}$ grows fast and tends to infinity at $\alpha \to 90^{\circ}$. Such a behavior is explained by the fact that the formation energy of stacking dislocations cannot be reduced by stretching the substrate layer in the direction perpendicular to the Burgers vector. 

As the Burgers vector is aligned along one of the zigzag directions, for a given geometry of the system, only six orientations of the Burgers vector are possible. Among these six types of possible dislocations, the dislocations with the angle $\alpha = \alpha_0$, where $0 \le \alpha_0 \le 30^{\circ}$ is the smallest angle between one of the zigzag directions and the direction in which the average lattice constants of the graphene and boron nitride layers are slightly different, are characterized by the lowest critical strain $\epsilon_\mathrm{c}$  and should appear first upon changing the lattice constant of the graphene layer relative to the boron nitride layer. Therefore, $\epsilon_\mathrm{c} (\alpha_0)$ corresponds to the critical strain associated with the average lattice mismatch of the layers at which the commensurate-incommensurate phase transition takes place in graphene/boron-nitride heterostructures (Fig. \ref{fig:disl}c). It should be mentioned that this critical strain for the layers of similar width is only slightly greater than the corresponding value for bilayer boron nitride of 0.75\% and almost three times greater than the critical strain for formation of the first dislocation in bilayer graphene of about 0.36\% (Ref. \onlinecite{Lebedeva2016}).

\section{Conclusions and discussion}
The potential energy surface of commensurate graphene and boron nitride layers has been calculated within the DFT approach based on the vdW-DF2 functional \cite{Lee2010}. It is shown that such an approach combined with the use of the experimental interlayer distance provides reliable characteristics of the potential energy surface that are in agreement with more accurate but much more expensive computational methods, such as RPA. The calculated barrier to relative motion of graphene and boron nitride layers of 9.5 meV/atom is several times greater than the values for two graphene or boron nitride layers. 

It has been checked that the approximation of the calculated potential energy surface by the first spatial Fourier components, an approach which is widely used in literature \cite{Zhou2015, Jung2015, Kumar2015} to analyze electronic properties, is highly accurate. Such an approximation allows to describe the potential energy surface of flat graphene/boron-nitride heterostructures using only two independent parameters. This means that a large number of physical properties of graphene/boron-nitride systems are interrelated and are expressed through these two parameters as well. In the present paper  we have used the approximation of the potential energy surface to estimate the shear mode frequency in the commensurate state, the barrier to relative rotation of graphene and boron nitride layers from the commensurate state to the fully incommensurate one, the width of stacking dislocations in slightly incommensurate systems and the period of the Moir\'{e} superstructure of the highly incommensurate ground state of the graphene/boron-nitride bilayer.

It is shown that the interlayer interaction in the highly incommensurate ground state of the graphene/boron-nitride systems with the layers aligned leads to modulation of atomic positions in the layers, which results in the increase of the period of Moir\'{e} superstructure by 9\% in the case of free graphene and boron nitride layers on an incommensurate substrate and about 2\% in the case when one of the layers is fixed. The maximal strains in the graphene and boron nitride layers are on the order of 0.6 -- 0.7\%, in agreement with the experimental observations \cite{Woods2014}.

Commensurate or slightly incommensurate graphene/boron-nitride systems can be realized by stretching the graphene layer or restricting the system geometry to flakes of the size smaller than the period of the Moir\'{e} superstructure. The shear mode frequencies for the commensurate graphene/boron-nitride heterostructure, a graphene flake on a boron nitride layer and a boron nitride flake on a graphene layer have been calculated. The estimated barrier to relative rotation of the layers from the commesurate to fully incommensurate state of 7.4 meV/atom is comparable to the previously reported values for pure graphene or boron nitride systems. 

The characteristrics of stacking dislocations have been studied for slightly incommensurate systems of boron nitride and graphene layers of similar width with the biaxially stretched graphene layer or a boron nitride ribbon of the width smaller than the period of the Moir\'{e} superstructure on a wide graphene layer stretched along the ribbon axis. The estimated dislocation widths for these two types of systems lie in the ranges of 6 -- 9 nm and  8 -- 13 nm, respectively, where the lower bound corresponds to shear dislocations and the higher one to tensile dislocations. It is suggested that changing the lattice constant of graphene with respect to boron nitride in one of the directions (in the system with the boron nitride ribbon along the ribbon axis) should result in observation of the commensurate-incommensurate phase transition. The estimated strain intervals for the graphene layer corresponding to the commensurate phase are about 0.8 -- 2.7\% in the case when the graphene and boron nitride layers are of similar width and 0.4  --  3.1\% for the boron nitride ribbon on the wide graphene layer.

Let us now discuss the possibility of experimental measurements of the calculated properties related to the interaction of graphene and boron nitride layers. The estimated barrier to relative rotation of the layers is relevant for dynamics of flakes on periodic substrates, i.e. for such processes as superlubricity \cite{Dienwiebel2004,Dienwiebel2005,Filippov2008,Xu2013} and diffusion \cite{Lebedeva2010,Lebedeva2011a}. In the latter case the diffusion coefficient of flakes is exponentially dependent on this barrier \cite{Lebedeva2010,Lebedeva2011a}. The shear mode frequency in bilayers can be measured by Raman scattering \cite{Tan2012} and coherent phonon spectroscopy \cite{Boschetto2013}, as demonstrated for graphene \cite{Tan2012,Boschetto2013}. The transmission electron microscopy \cite{Alden2013,Lin2013} and scanning tunneling microscopy \cite{Yankowitz2014} allow direct measurements of the dislocation width by analogy with papers \cite{Alden2013,Lin2013,Yankowitz2014} on graphene. The experiments for heterostructures with the stretched graphene layer should make it possible to observe the commesurate-incommensurate phase transition. The period of the Moir\'{e} pattern for graphene on boron nitride crystal has been already determined by scanning tunneling microscopy and atomic force measurements \cite{Woods2014,Tang2013,Yang2013}. The consideration of flat heterostructures with both free layers, which can be realized in the case of an incommensurate substrate, such as a rotated graphene or boron nitride layer, would allow to distinguish the change in the period of the Moir\'{e} pattern coming from the interlayer interaction. Such studies could provide a valuable experimental insight into the graphene/boron-nitride interlayer interaction due to the link between these properties and the potential energy surface established in the present paper. 

\begin{acknowledgements}
AL and AP acknowledge the Russian Foundation of Basic Research (16-52-00181) and computational time on the Multipurpose Computing Complex NRC ``Kurchatov Institute". IL acknowledges the financial support from Grupos Consolidados UPV/EHU del Gobierno Vasco (IT578-13) and EU-H2020 project ``MOSTOPHOS" (n. 646259). 
\end{acknowledgements}

\bibliography{rsc}

\begin{thebibliography}{101}%
\makeatletter
\providecommand \@ifxundefined [1]{%
 \@ifx{#1\undefined}
}%
\providecommand \@ifnum [1]{%
 \ifnum #1\expandafter \@firstoftwo
 \else \expandafter \@secondoftwo
 \fi
}%
\providecommand \@ifx [1]{%
 \ifx #1\expandafter \@firstoftwo
 \else \expandafter \@secondoftwo
 \fi
}%
\providecommand \natexlab [1]{#1}%
\providecommand \enquote  [1]{``#1''}%
\providecommand \bibnamefont  [1]{#1}%
\providecommand \bibfnamefont [1]{#1}%
\providecommand \citenamefont [1]{#1}%
\providecommand \href@noop [0]{\@secondoftwo}%
\providecommand \href [0]{\begingroup \@sanitize@url \@href}%
\providecommand \@href[1]{\@@startlink{#1}\@@href}%
\providecommand \@@href[1]{\endgroup#1\@@endlink}%
\providecommand \@sanitize@url [0]{\catcode `\\12\catcode `\$12\catcode
  `\&12\catcode `\#12\catcode `\^12\catcode `\_12\catcode `\%12\relax}%
\providecommand \@@startlink[1]{}%
\providecommand \@@endlink[0]{}%
\providecommand \url  [0]{\begingroup\@sanitize@url \@url }%
\providecommand \@url [1]{\endgroup\@href {#1}{\urlprefix }}%
\providecommand \urlprefix  [0]{URL }%
\providecommand \Eprint [0]{\href }%
\providecommand \doibase [0]{http://dx.doi.org/}%
\providecommand \selectlanguage [0]{\@gobble}%
\providecommand \bibinfo  [0]{\@secondoftwo}%
\providecommand \bibfield  [0]{\@secondoftwo}%
\providecommand \translation [1]{[#1]}%
\providecommand \BibitemOpen [0]{}%
\providecommand \bibitemStop [0]{}%
\providecommand \bibitemNoStop [0]{.\EOS\space}%
\providecommand \EOS [0]{\spacefactor3000\relax}%
\providecommand \BibitemShut  [1]{\csname bibitem#1\endcsname}%
\let\auto@bib@innerbib\@empty
%</preamble>
\bibitem [{\citenamefont {Geim}\ and\ \citenamefont
  {Grigorieva}(2013)}]{Geim2013}%
  \BibitemOpen
  \bibfield  {author} {\bibinfo {author} {\bibfnamefont {A.~K.}\ \bibnamefont
  {Geim}}\ and\ \bibinfo {author} {\bibfnamefont {I.~V.}\ \bibnamefont
  {Grigorieva}},\ }\href@noop {} {\bibfield  {journal} {\bibinfo  {journal}
  {Nature}\ }\textbf {\bibinfo {volume} {499}},\ \bibinfo {pages} {419}
  (\bibinfo {year} {2013})}\BibitemShut {NoStop}%
\bibitem [{\citenamefont {Novoselov}\ and\ \citenamefont {{Castro
  Neto}}(2012)}]{Novoselov2012}%
  \BibitemOpen
  \bibfield  {author} {\bibinfo {author} {\bibfnamefont {K.~S.}\ \bibnamefont
  {Novoselov}}\ and\ \bibinfo {author} {\bibfnamefont {A.~H.}\ \bibnamefont
  {{Castro Neto}}},\ }\href@noop {} {\bibfield  {journal} {\bibinfo  {journal}
  {Physica Scripta}\ }\textbf {\bibinfo {volume} {2012}},\ \bibinfo {pages}
  {014006} (\bibinfo {year} {2012})}\BibitemShut {NoStop}%
\bibitem [{\citenamefont {Dean}\ \emph {et~al.}(2010)\citenamefont {Dean},
  \citenamefont {Young}, \citenamefont {Meric}, \citenamefont {Lee},
  \citenamefont {Wang}, \citenamefont {Sorgenfrei}, \citenamefont {Watanabe},
  \citenamefont {Taniguchi}, \citenamefont {Kim}, \citenamefont {Shepard},\
  and\ \citenamefont {Hone}}]{Dean2010}%
  \BibitemOpen
  \bibfield  {author} {\bibinfo {author} {\bibfnamefont {C.~R.}\ \bibnamefont
  {Dean}}, \bibinfo {author} {\bibfnamefont {A.~F.}\ \bibnamefont {Young}},
  \bibinfo {author} {\bibfnamefont {I.}~\bibnamefont {Meric}}, \bibinfo
  {author} {\bibfnamefont {C.}~\bibnamefont {Lee}}, \bibinfo {author}
  {\bibfnamefont {L.}~\bibnamefont {Wang}}, \bibinfo {author} {\bibfnamefont
  {S.}~\bibnamefont {Sorgenfrei}}, \bibinfo {author} {\bibfnamefont
  {K.}~\bibnamefont {Watanabe}}, \bibinfo {author} {\bibfnamefont
  {T.}~\bibnamefont {Taniguchi}}, \bibinfo {author} {\bibfnamefont
  {P.}~\bibnamefont {Kim}}, \bibinfo {author} {\bibfnamefont {K.~L.}\
  \bibnamefont {Shepard}}, \ and\ \bibinfo {author} {\bibfnamefont
  {J.}~\bibnamefont {Hone}},\ }\href {\doibase 10.1038/nnano.2010.172}
  {\bibfield  {journal} {\bibinfo  {journal} {Nature Nanotechnology}\ }\textbf
  {\bibinfo {volume} {5}},\ \bibinfo {pages} {722} (\bibinfo {year}
  {2010})}\BibitemShut {NoStop}%
\bibitem [{\citenamefont {Xue}\ \emph {et~al.}(2011)\citenamefont {Xue},
  \citenamefont {Sanchez-Yamagishi}, \citenamefont {Bulmash}, \citenamefont
  {Jacquod}, \citenamefont {Deshpande}, \citenamefont {Watanabe}, \citenamefont
  {Taniguchi}, \citenamefont {Jarillo-Herrero},\ and\ \citenamefont
  {LeRoy}}]{Xue2011}%
  \BibitemOpen
  \bibfield  {author} {\bibinfo {author} {\bibfnamefont {J.}~\bibnamefont
  {Xue}}, \bibinfo {author} {\bibfnamefont {J.}~\bibnamefont
  {Sanchez-Yamagishi}}, \bibinfo {author} {\bibfnamefont {D.}~\bibnamefont
  {Bulmash}}, \bibinfo {author} {\bibfnamefont {P.}~\bibnamefont {Jacquod}},
  \bibinfo {author} {\bibfnamefont {A.}~\bibnamefont {Deshpande}}, \bibinfo
  {author} {\bibfnamefont {K.}~\bibnamefont {Watanabe}}, \bibinfo {author}
  {\bibfnamefont {T.}~\bibnamefont {Taniguchi}}, \bibinfo {author}
  {\bibfnamefont {P.}~\bibnamefont {Jarillo-Herrero}}, \ and\ \bibinfo {author}
  {\bibfnamefont {B.~J.}\ \bibnamefont {LeRoy}},\ }\href {\doibase
  10.1038/nmat2968} {\bibfield  {journal} {\bibinfo  {journal} {Nat. Mater.}\
  }\textbf {\bibinfo {volume} {10}},\ \bibinfo {pages} {282} (\bibinfo {year}
  {2011})}\BibitemShut {NoStop}%
\bibitem [{\citenamefont {Zomer}\ \emph {et~al.}(2011)\citenamefont {Zomer},
  \citenamefont {Dash}, \citenamefont {Tombros},\ and\ \citenamefont {van
  Wees}}]{Zomer2011}%
  \BibitemOpen
  \bibfield  {author} {\bibinfo {author} {\bibfnamefont {P.~J.}\ \bibnamefont
  {Zomer}}, \bibinfo {author} {\bibfnamefont {S.~P.}\ \bibnamefont {Dash}},
  \bibinfo {author} {\bibfnamefont {N.}~\bibnamefont {Tombros}}, \ and\
  \bibinfo {author} {\bibfnamefont {B.~J.}\ \bibnamefont {van Wees}},\ }\href
  {\doibase 10.1063/1.3665405} {\bibfield  {journal} {\bibinfo  {journal}
  {Applied Physics Letters}\ }\textbf {\bibinfo {volume} {99}},\ \bibinfo
  {pages} {232104} (\bibinfo {year} {2011})}\BibitemShut {NoStop}%
\bibitem [{\citenamefont {Decker}\ \emph {et~al.}(2011)\citenamefont {Decker},
  \citenamefont {Wang}, \citenamefont {Brar}, \citenamefont {Regan},
  \citenamefont {Tsai}, \citenamefont {Wu}, \citenamefont {Gannett},
  \citenamefont {Zettl},\ and\ \citenamefont {Crommie}}]{Decker2011}%
  \BibitemOpen
  \bibfield  {author} {\bibinfo {author} {\bibfnamefont {R.}~\bibnamefont
  {Decker}}, \bibinfo {author} {\bibfnamefont {Y.}~\bibnamefont {Wang}},
  \bibinfo {author} {\bibfnamefont {V.~W.}\ \bibnamefont {Brar}}, \bibinfo
  {author} {\bibfnamefont {W.}~\bibnamefont {Regan}}, \bibinfo {author}
  {\bibfnamefont {H.-Z.}\ \bibnamefont {Tsai}}, \bibinfo {author}
  {\bibfnamefont {Q.}~\bibnamefont {Wu}}, \bibinfo {author} {\bibfnamefont
  {W.}~\bibnamefont {Gannett}}, \bibinfo {author} {\bibfnamefont
  {A.}~\bibnamefont {Zettl}}, \ and\ \bibinfo {author} {\bibfnamefont {M.~F.}\
  \bibnamefont {Crommie}},\ }\href {\doibase 10.1021/nl2005115} {\bibfield
  {journal} {\bibinfo  {journal} {Nano Lett.}\ }\textbf {\bibinfo {volume}
  {11}},\ \bibinfo {pages} {2291} (\bibinfo {year} {2011})}\BibitemShut
  {NoStop}%
\bibitem [{\citenamefont {Dean}\ \emph {et~al.}(2012)\citenamefont {Dean},
  \citenamefont {Young}, \citenamefont {Wang}, \citenamefont {Meric},
  \citenamefont {Lee}, \citenamefont {Watanabe}, \citenamefont {Taniguchi},
  \citenamefont {Shepard}, \citenamefont {Kim},\ and\ \citenamefont
  {Hone}}]{Dean2012}%
  \BibitemOpen
  \bibfield  {author} {\bibinfo {author} {\bibfnamefont {C.}~\bibnamefont
  {Dean}}, \bibinfo {author} {\bibfnamefont {A.~F.}\ \bibnamefont {Young}},
  \bibinfo {author} {\bibfnamefont {L.}~\bibnamefont {Wang}}, \bibinfo {author}
  {\bibfnamefont {I.}~\bibnamefont {Meric}}, \bibinfo {author} {\bibfnamefont
  {G.-H.}\ \bibnamefont {Lee}}, \bibinfo {author} {\bibfnamefont
  {K.}~\bibnamefont {Watanabe}}, \bibinfo {author} {\bibfnamefont
  {T.}~\bibnamefont {Taniguchi}}, \bibinfo {author} {\bibfnamefont
  {K.}~\bibnamefont {Shepard}}, \bibinfo {author} {\bibfnamefont
  {P.}~\bibnamefont {Kim}}, \ and\ \bibinfo {author} {\bibfnamefont
  {J.}~\bibnamefont {Hone}},\ }\href {\doibase 10.1016/j.ssc.2012.04.021}
  {\bibfield  {journal} {\bibinfo  {journal} {Solid State Communications}\
  }\textbf {\bibinfo {volume} {152}},\ \bibinfo {pages} {1275} (\bibinfo {year}
  {2012})}\BibitemShut {NoStop}%
\bibitem [{\citenamefont {Ponomarenko}\ \emph {et~al.}(2011)\citenamefont
  {Ponomarenko}, \citenamefont {Geim}, \citenamefont {Zhukov}, \citenamefont
  {Jalil}, \citenamefont {Morozov}, \citenamefont {Novoselov}, \citenamefont
  {Grigorieva}, \citenamefont {Hill}, \citenamefont {Cheianov}, \citenamefont
  {Fal'ko}, \citenamefont {Watanabe}, \citenamefont {Taniguchi},\ and\
  \citenamefont {Gorbachev}}]{Ponomarenko2011}%
  \BibitemOpen
  \bibfield  {author} {\bibinfo {author} {\bibfnamefont {L.~A.}\ \bibnamefont
  {Ponomarenko}}, \bibinfo {author} {\bibfnamefont {A.~K.}\ \bibnamefont
  {Geim}}, \bibinfo {author} {\bibfnamefont {A.~A.}\ \bibnamefont {Zhukov}},
  \bibinfo {author} {\bibfnamefont {R.}~\bibnamefont {Jalil}}, \bibinfo
  {author} {\bibfnamefont {S.~V.}\ \bibnamefont {Morozov}}, \bibinfo {author}
  {\bibfnamefont {K.~S.}\ \bibnamefont {Novoselov}}, \bibinfo {author}
  {\bibfnamefont {I.~V.}\ \bibnamefont {Grigorieva}}, \bibinfo {author}
  {\bibfnamefont {E.~H.}\ \bibnamefont {Hill}}, \bibinfo {author}
  {\bibfnamefont {V.~V.}\ \bibnamefont {Cheianov}}, \bibinfo {author}
  {\bibfnamefont {V.~I.}\ \bibnamefont {Fal'ko}}, \bibinfo {author}
  {\bibfnamefont {K.}~\bibnamefont {Watanabe}}, \bibinfo {author}
  {\bibfnamefont {T.}~\bibnamefont {Taniguchi}}, \ and\ \bibinfo {author}
  {\bibfnamefont {R.~V.}\ \bibnamefont {Gorbachev}},\ }\href@noop {} {\bibfield
   {journal} {\bibinfo  {journal} {Nature Phys.}\ }\textbf {\bibinfo {volume}
  {7}},\ \bibinfo {pages} {958} (\bibinfo {year} {2011})}\BibitemShut {NoStop}%
\bibitem [{\citenamefont {Kim}\ \emph {et~al.}(2011)\citenamefont {Kim},
  \citenamefont {Yu}, \citenamefont {Song},\ and\ \citenamefont
  {Yu}}]{Kim2011}%
  \BibitemOpen
  \bibfield  {author} {\bibinfo {author} {\bibfnamefont {E.}~\bibnamefont
  {Kim}}, \bibinfo {author} {\bibfnamefont {T.}~\bibnamefont {Yu}}, \bibinfo
  {author} {\bibfnamefont {E.~S.}\ \bibnamefont {Song}}, \ and\ \bibinfo
  {author} {\bibfnamefont {B.}~\bibnamefont {Yu}},\ }\href {\doibase
  10.1063/1.3604012} {\bibfield  {journal} {\bibinfo  {journal} {Applied
  Physics Letters}\ }\textbf {\bibinfo {volume} {98}},\ \bibinfo {pages}
  {262103} (\bibinfo {year} {2011})}\BibitemShut {NoStop}%
\bibitem [{\citenamefont {Jain}\ \emph {et~al.}(2013)\citenamefont {Jain},
  \citenamefont {Bansal}, \citenamefont {Durcan}, \citenamefont {Xu},\ and\
  \citenamefont {Yu}}]{Jain2013}%
  \BibitemOpen
  \bibfield  {author} {\bibinfo {author} {\bibfnamefont {N.}~\bibnamefont
  {Jain}}, \bibinfo {author} {\bibfnamefont {T.}~\bibnamefont {Bansal}},
  \bibinfo {author} {\bibfnamefont {C.~A.}\ \bibnamefont {Durcan}}, \bibinfo
  {author} {\bibfnamefont {Y.}~\bibnamefont {Xu}}, \ and\ \bibinfo {author}
  {\bibfnamefont {B.}~\bibnamefont {Yu}},\ }\href {\doibase
  10.1016/j.carbon.2012.11.054} {\bibfield  {journal} {\bibinfo  {journal}
  {Carbon}\ }\textbf {\bibinfo {volume} {54}},\ \bibinfo {pages} {396}
  (\bibinfo {year} {2013})}\BibitemShut {NoStop}%
\bibitem [{\citenamefont {Lee}\ \emph {et~al.}(2011)\citenamefont {Lee},
  \citenamefont {Yu}, \citenamefont {Lee}, \citenamefont {Dean}, \citenamefont
  {Shepard}, \citenamefont {Kim},\ and\ \citenamefont {Hone}}]{Lee2011}%
  \BibitemOpen
  \bibfield  {author} {\bibinfo {author} {\bibfnamefont {G.-H.}\ \bibnamefont
  {Lee}}, \bibinfo {author} {\bibfnamefont {Y.-J.}\ \bibnamefont {Yu}},
  \bibinfo {author} {\bibfnamefont {C.}~\bibnamefont {Lee}}, \bibinfo {author}
  {\bibfnamefont {C.}~\bibnamefont {Dean}}, \bibinfo {author} {\bibfnamefont
  {K.~L.}\ \bibnamefont {Shepard}}, \bibinfo {author} {\bibfnamefont
  {P.}~\bibnamefont {Kim}}, \ and\ \bibinfo {author} {\bibfnamefont
  {J.}~\bibnamefont {Hone}},\ }\href {\doibase 10.1063/1.3662043} {\bibfield
  {journal} {\bibinfo  {journal} {Applied Physics Letters}\ }\textbf {\bibinfo
  {volume} {99}},\ \bibinfo {pages} {243114} (\bibinfo {year}
  {2011})}\BibitemShut {NoStop}%
\bibitem [{\citenamefont {Britnell}\ \emph
  {et~al.}(2012{\natexlab{a}})\citenamefont {Britnell}, \citenamefont
  {Gorbachev}, \citenamefont {Jalil}, \citenamefont {Belle}, \citenamefont
  {Schedin}, \citenamefont {Katsnelson}, \citenamefont {Eaves}, \citenamefont
  {Morozov}, \citenamefont {Mayorov}, \citenamefont {Peres}, \citenamefont
  {{Castro Neto}}, \citenamefont {Leist}, \citenamefont {Geim}, \citenamefont
  {Ponomarenko},\ and\ \citenamefont {Novoselov}}]{Britnell2012a}%
  \BibitemOpen
  \bibfield  {author} {\bibinfo {author} {\bibfnamefont {L.}~\bibnamefont
  {Britnell}}, \bibinfo {author} {\bibfnamefont {R.~V.}\ \bibnamefont
  {Gorbachev}}, \bibinfo {author} {\bibfnamefont {R.}~\bibnamefont {Jalil}},
  \bibinfo {author} {\bibfnamefont {B.~D.}\ \bibnamefont {Belle}}, \bibinfo
  {author} {\bibfnamefont {F.}~\bibnamefont {Schedin}}, \bibinfo {author}
  {\bibfnamefont {M.~I.}\ \bibnamefont {Katsnelson}}, \bibinfo {author}
  {\bibfnamefont {L.}~\bibnamefont {Eaves}}, \bibinfo {author} {\bibfnamefont
  {S.~V.}\ \bibnamefont {Morozov}}, \bibinfo {author} {\bibfnamefont {A.~S.}\
  \bibnamefont {Mayorov}}, \bibinfo {author} {\bibfnamefont {N.~M.~R.}\
  \bibnamefont {Peres}}, \bibinfo {author} {\bibfnamefont {A.~H.}\ \bibnamefont
  {{Castro Neto}}}, \bibinfo {author} {\bibfnamefont {J.}~\bibnamefont
  {Leist}}, \bibinfo {author} {\bibfnamefont {A.~K.}\ \bibnamefont {Geim}},
  \bibinfo {author} {\bibfnamefont {L.~A.}\ \bibnamefont {Ponomarenko}}, \ and\
  \bibinfo {author} {\bibfnamefont {K.~S.}\ \bibnamefont {Novoselov}},\ }\href
  {\doibase 10.1021/nl3002205} {\bibfield  {journal} {\bibinfo  {journal} {Nano
  Lett.}\ }\textbf {\bibinfo {volume} {12}},\ \bibinfo {pages} {1707} (\bibinfo
  {year} {2012}{\natexlab{a}})}\BibitemShut {NoStop}%
\bibitem [{\citenamefont {Britnell}\ \emph
  {et~al.}(2012{\natexlab{b}})\citenamefont {Britnell}, \citenamefont
  {Gorbachev}, \citenamefont {Jalil}, \citenamefont {Belle}, \citenamefont
  {Schedin}, \citenamefont {Mishchenko}, \citenamefont {Georgiou},
  \citenamefont {Katsnelson}, \citenamefont {Eaves}, \citenamefont {Morozov},
  \citenamefont {Peres}, \citenamefont {Leist}, \citenamefont {Geim},
  \citenamefont {Novoselov},\ and\ \citenamefont {Ponomarenko}}]{Britnell2012}%
  \BibitemOpen
  \bibfield  {author} {\bibinfo {author} {\bibfnamefont {L.}~\bibnamefont
  {Britnell}}, \bibinfo {author} {\bibfnamefont {R.~V.}\ \bibnamefont
  {Gorbachev}}, \bibinfo {author} {\bibfnamefont {R.}~\bibnamefont {Jalil}},
  \bibinfo {author} {\bibfnamefont {B.~D.}\ \bibnamefont {Belle}}, \bibinfo
  {author} {\bibfnamefont {F.}~\bibnamefont {Schedin}}, \bibinfo {author}
  {\bibfnamefont {A.}~\bibnamefont {Mishchenko}}, \bibinfo {author}
  {\bibfnamefont {T.}~\bibnamefont {Georgiou}}, \bibinfo {author}
  {\bibfnamefont {M.~I.}\ \bibnamefont {Katsnelson}}, \bibinfo {author}
  {\bibfnamefont {L.}~\bibnamefont {Eaves}}, \bibinfo {author} {\bibfnamefont
  {S.~V.}\ \bibnamefont {Morozov}}, \bibinfo {author} {\bibfnamefont
  {N.~M.~R.}\ \bibnamefont {Peres}}, \bibinfo {author} {\bibfnamefont
  {J.}~\bibnamefont {Leist}}, \bibinfo {author} {\bibfnamefont {A.~K.}\
  \bibnamefont {Geim}}, \bibinfo {author} {\bibfnamefont {K.~S.}\ \bibnamefont
  {Novoselov}}, \ and\ \bibinfo {author} {\bibfnamefont {L.~A.}\ \bibnamefont
  {Ponomarenko}},\ }\href {\doibase 10.1126/science.128461} {\bibfield
  {journal} {\bibinfo  {journal} {Science}\ }\textbf {\bibinfo {volume}
  {335}},\ \bibinfo {pages} {947} (\bibinfo {year}
  {2012}{\natexlab{b}})}\BibitemShut {NoStop}%
\bibitem [{\citenamefont {\"{O}z\c{c}elik}\ and\ \citenamefont
  {Ciraci}(2013)}]{Ozcelik2013}%
  \BibitemOpen
  \bibfield  {author} {\bibinfo {author} {\bibfnamefont {V.~O.}\ \bibnamefont
  {\"{O}z\c{c}elik}}\ and\ \bibinfo {author} {\bibfnamefont {S.}~\bibnamefont
  {Ciraci}},\ }\href {\doibase 10.1021/jp403706e} {\bibfield  {journal}
  {\bibinfo  {journal} {The Journal of Physical Chemistry C}\ }\textbf
  {\bibinfo {volume} {117}},\ \bibinfo {pages} {15327} (\bibinfo {year}
  {2013})}\BibitemShut {NoStop}%
\bibitem [{\citenamefont {Shi}\ \emph {et~al.}(2014)\citenamefont {Shi},
  \citenamefont {Hanlumyuang}, \citenamefont {Liu}, \citenamefont {Gong},
  \citenamefont {Gao}, \citenamefont {Li}, \citenamefont {Kono}, \citenamefont
  {Lou}, \citenamefont {Vajtai}, \citenamefont {Sharma},\ and\ \citenamefont
  {Ajayan}}]{Shi2014}%
  \BibitemOpen
  \bibfield  {author} {\bibinfo {author} {\bibfnamefont {G.}~\bibnamefont
  {Shi}}, \bibinfo {author} {\bibfnamefont {Y.}~\bibnamefont {Hanlumyuang}},
  \bibinfo {author} {\bibfnamefont {Z.}~\bibnamefont {Liu}}, \bibinfo {author}
  {\bibfnamefont {Y.}~\bibnamefont {Gong}}, \bibinfo {author} {\bibfnamefont
  {W.}~\bibnamefont {Gao}}, \bibinfo {author} {\bibfnamefont {B.}~\bibnamefont
  {Li}}, \bibinfo {author} {\bibfnamefont {J.}~\bibnamefont {Kono}}, \bibinfo
  {author} {\bibfnamefont {J.}~\bibnamefont {Lou}}, \bibinfo {author}
  {\bibfnamefont {R.}~\bibnamefont {Vajtai}}, \bibinfo {author} {\bibfnamefont
  {P.}~\bibnamefont {Sharma}}, \ and\ \bibinfo {author} {\bibfnamefont {P.~M.}\
  \bibnamefont {Ajayan}},\ }\href {\doibase 10.1021/nl4037824} {\bibfield
  {journal} {\bibinfo  {journal} {Nano Letters}\ }\textbf {\bibinfo {volume}
  {14}},\ \bibinfo {pages} {1739} (\bibinfo {year} {2014})}\BibitemShut
  {NoStop}%
\bibitem [{\citenamefont {Li}\ \emph {et~al.}(2012)\citenamefont {Li},
  \citenamefont {Sun}, \citenamefont {Chen}, \citenamefont {Zhou},
  \citenamefont {Wang}, \citenamefont {Ding},\ and\ \citenamefont
  {Zhang}}]{Li2012}%
  \BibitemOpen
  \bibfield  {author} {\bibinfo {author} {\bibfnamefont {Y.-J.}\ \bibnamefont
  {Li}}, \bibinfo {author} {\bibfnamefont {Q.-Q.}\ \bibnamefont {Sun}},
  \bibinfo {author} {\bibfnamefont {L.}~\bibnamefont {Chen}}, \bibinfo {author}
  {\bibfnamefont {P.}~\bibnamefont {Zhou}}, \bibinfo {author} {\bibfnamefont
  {P.-F.}\ \bibnamefont {Wang}}, \bibinfo {author} {\bibfnamefont {S.-J.}\
  \bibnamefont {Ding}}, \ and\ \bibinfo {author} {\bibfnamefont {D.~W.}\
  \bibnamefont {Zhang}},\ }\href {\doibase 10.1063/1.3701267} {\bibfield
  {journal} {\bibinfo  {journal} {AIP Advances}\ }\textbf {\bibinfo {volume}
  {2}},\ \bibinfo {pages} {012191} (\bibinfo {year} {2012})}\BibitemShut
  {NoStop}%
\bibitem [{\citenamefont {Gorbachev}\ \emph {et~al.}(2012)\citenamefont
  {Gorbachev}, \citenamefont {Geim}, \citenamefont {Katsnelson}, \citenamefont
  {Novoselov}, \citenamefont {Tudorovskiy}, \citenamefont {Grigorieva},
  \citenamefont {MacDonald}, \citenamefont {Morozov}, \citenamefont {Watanabe},
  \citenamefont {Taniguchi},\ and\ \citenamefont
  {Ponomarenko}}]{Gorbachev2012}%
  \BibitemOpen
  \bibfield  {author} {\bibinfo {author} {\bibfnamefont {R.~V.}\ \bibnamefont
  {Gorbachev}}, \bibinfo {author} {\bibfnamefont {A.~K.}\ \bibnamefont {Geim}},
  \bibinfo {author} {\bibfnamefont {M.~I.}\ \bibnamefont {Katsnelson}},
  \bibinfo {author} {\bibfnamefont {K.~S.}\ \bibnamefont {Novoselov}}, \bibinfo
  {author} {\bibfnamefont {T.}~\bibnamefont {Tudorovskiy}}, \bibinfo {author}
  {\bibfnamefont {I.~V.}\ \bibnamefont {Grigorieva}}, \bibinfo {author}
  {\bibfnamefont {A.~H.}\ \bibnamefont {MacDonald}}, \bibinfo {author}
  {\bibfnamefont {S.~V.}\ \bibnamefont {Morozov}}, \bibinfo {author}
  {\bibfnamefont {K.}~\bibnamefont {Watanabe}}, \bibinfo {author}
  {\bibfnamefont {T.}~\bibnamefont {Taniguchi}}, \ and\ \bibinfo {author}
  {\bibfnamefont {L.~A.}\ \bibnamefont {Ponomarenko}},\ }\href {\doibase
  10.1038/nphys2441} {\bibfield  {journal} {\bibinfo  {journal} {Nature
  Physics}\ }\textbf {\bibinfo {volume} {8}},\ \bibinfo {pages} {896} (\bibinfo
  {year} {2012})}\BibitemShut {NoStop}%
\bibitem [{\citenamefont {Perali}\ \emph {et~al.}(2013)\citenamefont {Perali},
  \citenamefont {Neilson},\ and\ \citenamefont {Hamilton}}]{Perali2013}%
  \BibitemOpen
  \bibfield  {author} {\bibinfo {author} {\bibfnamefont {A.}~\bibnamefont
  {Perali}}, \bibinfo {author} {\bibfnamefont {D.}~\bibnamefont {Neilson}}, \
  and\ \bibinfo {author} {\bibfnamefont {A.~R.}\ \bibnamefont {Hamilton}},\
  }\href {\doibase 10.1103/PhysRevLett.110.146803} {\bibfield  {journal}
  {\bibinfo  {journal} {Phys. Rev. Lett.}\ }\textbf {\bibinfo {volume} {110}},\
  \bibinfo {pages} {146803} (\bibinfo {year} {2013})}\BibitemShut {NoStop}%
\bibitem [{\citenamefont {Woods}\ \emph {et~al.}(2014)\citenamefont {Woods},
  \citenamefont {Britnell}, \citenamefont {Eckmann}, \citenamefont {Ma},
  \citenamefont {Lu}, \citenamefont {Guo}, \citenamefont {Lin}, \citenamefont
  {Yu}, \citenamefont {Cao}, \citenamefont {Gorbachev}, \citenamefont
  {Kretinin}, \citenamefont {Park}, \citenamefont {Ponomarenko}, \citenamefont
  {Katsnelson}, \citenamefont {Gornostyrev}, \citenamefont {Watanabe},
  \citenamefont {Taniguchi}, \citenamefont {Casiraghi}, \citenamefont {Gao},
  \citenamefont {Geim},\ and\ \citenamefont {Novoselov}}]{Woods2014}%
  \BibitemOpen
  \bibfield  {author} {\bibinfo {author} {\bibfnamefont {C.~R.}\ \bibnamefont
  {Woods}}, \bibinfo {author} {\bibfnamefont {L.}~\bibnamefont {Britnell}},
  \bibinfo {author} {\bibfnamefont {A.}~\bibnamefont {Eckmann}}, \bibinfo
  {author} {\bibfnamefont {R.~S.}\ \bibnamefont {Ma}}, \bibinfo {author}
  {\bibfnamefont {J.~C.}\ \bibnamefont {Lu}}, \bibinfo {author} {\bibfnamefont
  {H.~M.}\ \bibnamefont {Guo}}, \bibinfo {author} {\bibfnamefont
  {X.}~\bibnamefont {Lin}}, \bibinfo {author} {\bibfnamefont {G.~L.}\
  \bibnamefont {Yu}}, \bibinfo {author} {\bibfnamefont {Y.}~\bibnamefont
  {Cao}}, \bibinfo {author} {\bibfnamefont {R.~V.}\ \bibnamefont {Gorbachev}},
  \bibinfo {author} {\bibfnamefont {A.~V.}\ \bibnamefont {Kretinin}}, \bibinfo
  {author} {\bibfnamefont {J.}~\bibnamefont {Park}}, \bibinfo {author}
  {\bibfnamefont {L.~A.}\ \bibnamefont {Ponomarenko}}, \bibinfo {author}
  {\bibfnamefont {M.~I.}\ \bibnamefont {Katsnelson}}, \bibinfo {author}
  {\bibfnamefont {Y.~N.}\ \bibnamefont {Gornostyrev}}, \bibinfo {author}
  {\bibfnamefont {K.}~\bibnamefont {Watanabe}}, \bibinfo {author}
  {\bibfnamefont {T.}~\bibnamefont {Taniguchi}}, \bibinfo {author}
  {\bibfnamefont {C.}~\bibnamefont {Casiraghi}}, \bibinfo {author}
  {\bibfnamefont {H.-J.}\ \bibnamefont {Gao}}, \bibinfo {author} {\bibfnamefont
  {A.~K.}\ \bibnamefont {Geim}}, \ and\ \bibinfo {author} {\bibfnamefont
  {K.~S.}\ \bibnamefont {Novoselov}},\ }\href {\doibase 10.1038/nphys2954}
  {\bibfield  {journal} {\bibinfo  {journal} {Nature Physics}\ }\textbf
  {\bibinfo {volume} {10}},\ \bibinfo {pages} {451} (\bibinfo {year}
  {2014})}\BibitemShut {NoStop}%
\bibitem [{\citenamefont {Tang}\ \emph {et~al.}(2013)\citenamefont {Tang},
  \citenamefont {Wang}, \citenamefont {Zhang}, \citenamefont {Li},
  \citenamefont {Xie}, \citenamefont {Liu}, \citenamefont {Liu}, \citenamefont
  {Li}, \citenamefont {Huang}, \citenamefont {Xie},\ and\ \citenamefont
  {Jiang}}]{Tang2013}%
  \BibitemOpen
  \bibfield  {author} {\bibinfo {author} {\bibfnamefont {S.}~\bibnamefont
  {Tang}}, \bibinfo {author} {\bibfnamefont {H.}~\bibnamefont {Wang}}, \bibinfo
  {author} {\bibfnamefont {Y.}~\bibnamefont {Zhang}}, \bibinfo {author}
  {\bibfnamefont {A.}~\bibnamefont {Li}}, \bibinfo {author} {\bibfnamefont
  {H.}~\bibnamefont {Xie}}, \bibinfo {author} {\bibfnamefont {X.}~\bibnamefont
  {Liu}}, \bibinfo {author} {\bibfnamefont {L.}~\bibnamefont {Liu}}, \bibinfo
  {author} {\bibfnamefont {T.}~\bibnamefont {Li}}, \bibinfo {author}
  {\bibfnamefont {F.}~\bibnamefont {Huang}}, \bibinfo {author} {\bibfnamefont
  {X.}~\bibnamefont {Xie}}, \ and\ \bibinfo {author} {\bibfnamefont
  {M.}~\bibnamefont {Jiang}},\ }\href@noop {} {\bibfield  {journal} {\bibinfo
  {journal} {Scientific Reports}\ }\textbf {\bibinfo {volume} {3}},\ \bibinfo
  {pages} {2666} (\bibinfo {year} {2013})}\BibitemShut {NoStop}%
\bibitem [{\citenamefont {Yang}\ \emph {et~al.}(2013)\citenamefont {Yang},
  \citenamefont {Chen}, \citenamefont {Shi}, \citenamefont {Liu}, \citenamefont
  {Zhang}, \citenamefont {Xie}, \citenamefont {Cheng}, \citenamefont {Wang},
  \citenamefont {Yang}, \citenamefont {Shi}, \citenamefont {Watanabe},
  \citenamefont {Taniguchi}, \citenamefont {Yao}, \citenamefont {Zhang},\ and\
  \citenamefont {Zhang}}]{Yang2013}%
  \BibitemOpen
  \bibfield  {author} {\bibinfo {author} {\bibfnamefont {W.}~\bibnamefont
  {Yang}}, \bibinfo {author} {\bibfnamefont {G.}~\bibnamefont {Chen}}, \bibinfo
  {author} {\bibfnamefont {Z.}~\bibnamefont {Shi}}, \bibinfo {author}
  {\bibfnamefont {C.-C.}\ \bibnamefont {Liu}}, \bibinfo {author} {\bibfnamefont
  {L.}~\bibnamefont {Zhang}}, \bibinfo {author} {\bibfnamefont
  {G.}~\bibnamefont {Xie}}, \bibinfo {author} {\bibfnamefont {M.}~\bibnamefont
  {Cheng}}, \bibinfo {author} {\bibfnamefont {D.}~\bibnamefont {Wang}},
  \bibinfo {author} {\bibfnamefont {R.}~\bibnamefont {Yang}}, \bibinfo {author}
  {\bibfnamefont {D.}~\bibnamefont {Shi}}, \bibinfo {author} {\bibfnamefont
  {K.}~\bibnamefont {Watanabe}}, \bibinfo {author} {\bibfnamefont
  {T.}~\bibnamefont {Taniguchi}}, \bibinfo {author} {\bibfnamefont
  {Y.}~\bibnamefont {Yao}}, \bibinfo {author} {\bibfnamefont {Y.}~\bibnamefont
  {Zhang}}, \ and\ \bibinfo {author} {\bibfnamefont {G.}~\bibnamefont
  {Zhang}},\ }\href@noop {} {\bibfield  {journal} {\bibinfo  {journal} {Nature
  Materials}\ }\textbf {\bibinfo {volume} {12}},\ \bibinfo {pages} {792}
  (\bibinfo {year} {2013})}\BibitemShut {NoStop}%
\bibitem [{\citenamefont {Yankowitz}\ \emph {et~al.}(2012)\citenamefont
  {Yankowitz}, \citenamefont {Xue}, \citenamefont {Cormode}, \citenamefont
  {Sanchez-Yamagishi}, \citenamefont {Watanabe}, \citenamefont {Taniguchi},
  \citenamefont {Jarillo-Herrero}, \citenamefont {Jacquod},\ and\ \citenamefont
  {LeRoy}}]{Yankowitz2012}%
  \BibitemOpen
  \bibfield  {author} {\bibinfo {author} {\bibfnamefont {M.}~\bibnamefont
  {Yankowitz}}, \bibinfo {author} {\bibfnamefont {J.}~\bibnamefont {Xue}},
  \bibinfo {author} {\bibfnamefont {D.}~\bibnamefont {Cormode}}, \bibinfo
  {author} {\bibfnamefont {J.~D.}\ \bibnamefont {Sanchez-Yamagishi}}, \bibinfo
  {author} {\bibfnamefont {K.}~\bibnamefont {Watanabe}}, \bibinfo {author}
  {\bibfnamefont {T.}~\bibnamefont {Taniguchi}}, \bibinfo {author}
  {\bibfnamefont {P.}~\bibnamefont {Jarillo-Herrero}}, \bibinfo {author}
  {\bibfnamefont {P.}~\bibnamefont {Jacquod}}, \ and\ \bibinfo {author}
  {\bibfnamefont {B.~J.}\ \bibnamefont {LeRoy}},\ }\href {\doibase
  10.1038/nphys2272} {\bibfield  {journal} {\bibinfo  {journal} {Nature
  Physics}\ }\textbf {\bibinfo {volume} {8}},\ \bibinfo {pages} {382} (\bibinfo
  {year} {2012})}\BibitemShut {NoStop}%
\bibitem [{\citenamefont {Dean}\ \emph {et~al.}(2013)\citenamefont {Dean},
  \citenamefont {Wang}, \citenamefont {Maher}, \citenamefont {Forsythe},
  \citenamefont {Ghahari}, \citenamefont {Gao}, \citenamefont {Katoch},
  \citenamefont {Ishigami}, \citenamefont {Moon}, \citenamefont {Koshino},
  \citenamefont {Taniguchi}, \citenamefont {Watanabe}, \citenamefont {Shepard},
  \citenamefont {Hone},\ and\ \citenamefont {Kim}}]{Dean2013}%
  \BibitemOpen
  \bibfield  {author} {\bibinfo {author} {\bibfnamefont {C.~R.}\ \bibnamefont
  {Dean}}, \bibinfo {author} {\bibfnamefont {L.}~\bibnamefont {Wang}}, \bibinfo
  {author} {\bibfnamefont {P.}~\bibnamefont {Maher}}, \bibinfo {author}
  {\bibfnamefont {C.}~\bibnamefont {Forsythe}}, \bibinfo {author}
  {\bibfnamefont {F.}~\bibnamefont {Ghahari}}, \bibinfo {author} {\bibfnamefont
  {Y.}~\bibnamefont {Gao}}, \bibinfo {author} {\bibfnamefont {J.}~\bibnamefont
  {Katoch}}, \bibinfo {author} {\bibfnamefont {M.}~\bibnamefont {Ishigami}},
  \bibinfo {author} {\bibfnamefont {P.}~\bibnamefont {Moon}}, \bibinfo {author}
  {\bibfnamefont {M.}~\bibnamefont {Koshino}}, \bibinfo {author} {\bibfnamefont
  {T.}~\bibnamefont {Taniguchi}}, \bibinfo {author} {\bibfnamefont
  {K.}~\bibnamefont {Watanabe}}, \bibinfo {author} {\bibfnamefont {K.~L.}\
  \bibnamefont {Shepard}}, \bibinfo {author} {\bibfnamefont {J.}~\bibnamefont
  {Hone}}, \ and\ \bibinfo {author} {\bibfnamefont {P.}~\bibnamefont {Kim}},\
  }\href@noop {} {\bibfield  {journal} {\bibinfo  {journal} {Nature}\ }\textbf
  {\bibinfo {volume} {497}},\ \bibinfo {pages} {598} (\bibinfo {year}
  {2013})}\BibitemShut {NoStop}%
\bibitem [{\citenamefont {Hunt}\ \emph {et~al.}(2013)\citenamefont {Hunt},
  \citenamefont {Sanchez-Yamagishi}, \citenamefont {Young}, \citenamefont
  {Yankowitz}, \citenamefont {LeRoy}, \citenamefont {Watanabe}, \citenamefont
  {Taniguchi}, \citenamefont {Moon}, \citenamefont {Koshino}, \citenamefont
  {Jarillo-Herrero},\ and\ \citenamefont {Ashoori}}]{Hunt2013}%
  \BibitemOpen
  \bibfield  {author} {\bibinfo {author} {\bibfnamefont {B.}~\bibnamefont
  {Hunt}}, \bibinfo {author} {\bibfnamefont {J.~D.}\ \bibnamefont
  {Sanchez-Yamagishi}}, \bibinfo {author} {\bibfnamefont {A.~F.}\ \bibnamefont
  {Young}}, \bibinfo {author} {\bibfnamefont {M.}~\bibnamefont {Yankowitz}},
  \bibinfo {author} {\bibfnamefont {B.~J.}\ \bibnamefont {LeRoy}}, \bibinfo
  {author} {\bibfnamefont {K.}~\bibnamefont {Watanabe}}, \bibinfo {author}
  {\bibfnamefont {T.}~\bibnamefont {Taniguchi}}, \bibinfo {author}
  {\bibfnamefont {P.}~\bibnamefont {Moon}}, \bibinfo {author} {\bibfnamefont
  {M.}~\bibnamefont {Koshino}}, \bibinfo {author} {\bibfnamefont
  {P.}~\bibnamefont {Jarillo-Herrero}}, \ and\ \bibinfo {author} {\bibfnamefont
  {R.~C.}\ \bibnamefont {Ashoori}},\ }\href {\doibase 10.1126/science.1237240}
  {\bibfield  {journal} {\bibinfo  {journal} {Science}\ }\textbf {\bibinfo
  {volume} {340}},\ \bibinfo {pages} {1427} (\bibinfo {year}
  {2013})}\BibitemShut {NoStop}%
\bibitem [{\citenamefont {Ponomarenko}\ \emph {et~al.}(2013)\citenamefont
  {Ponomarenko}, \citenamefont {Gorbachev}, \citenamefont {Yu}, \citenamefont
  {Elias}, \citenamefont {Jalil}, \citenamefont {Patel}, \citenamefont
  {Mishchenko}, \citenamefont {Mayorov}, \citenamefont {Woods}, \citenamefont
  {Wallbank}, \citenamefont {Mucha-Kruczynski}, \citenamefont {Piot},
  \citenamefont {Potemski}, \citenamefont {Grigorieva}, \citenamefont
  {Novoselov}, \citenamefont {Guinea}, \citenamefont {Fal'ko},\ and\
  \citenamefont {Geim}}]{Ponomarenko2013}%
  \BibitemOpen
  \bibfield  {author} {\bibinfo {author} {\bibfnamefont {L.~A.}\ \bibnamefont
  {Ponomarenko}}, \bibinfo {author} {\bibfnamefont {R.~V.}\ \bibnamefont
  {Gorbachev}}, \bibinfo {author} {\bibfnamefont {G.~L.}\ \bibnamefont {Yu}},
  \bibinfo {author} {\bibfnamefont {D.~C.}\ \bibnamefont {Elias}}, \bibinfo
  {author} {\bibfnamefont {R.}~\bibnamefont {Jalil}}, \bibinfo {author}
  {\bibfnamefont {A.~A.}\ \bibnamefont {Patel}}, \bibinfo {author}
  {\bibfnamefont {A.}~\bibnamefont {Mishchenko}}, \bibinfo {author}
  {\bibfnamefont {A.~S.}\ \bibnamefont {Mayorov}}, \bibinfo {author}
  {\bibfnamefont {C.~R.}\ \bibnamefont {Woods}}, \bibinfo {author}
  {\bibfnamefont {J.~R.}\ \bibnamefont {Wallbank}}, \bibinfo {author}
  {\bibfnamefont {M.}~\bibnamefont {Mucha-Kruczynski}}, \bibinfo {author}
  {\bibfnamefont {B.~A.}\ \bibnamefont {Piot}}, \bibinfo {author}
  {\bibfnamefont {M.}~\bibnamefont {Potemski}}, \bibinfo {author}
  {\bibfnamefont {I.~V.}\ \bibnamefont {Grigorieva}}, \bibinfo {author}
  {\bibfnamefont {K.~S.}\ \bibnamefont {Novoselov}}, \bibinfo {author}
  {\bibfnamefont {F.}~\bibnamefont {Guinea}}, \bibinfo {author} {\bibfnamefont
  {V.~I.}\ \bibnamefont {Fal'ko}}, \ and\ \bibinfo {author} {\bibfnamefont
  {A.~K.}\ \bibnamefont {Geim}},\ }\href@noop {} {\bibfield  {journal}
  {\bibinfo  {journal} {Nature}\ }\textbf {\bibinfo {volume} {497}},\ \bibinfo
  {pages} {594} (\bibinfo {year} {2013})}\BibitemShut {NoStop}%
\bibitem [{\citenamefont {Chen}\ \emph {et~al.}(2014)\citenamefont {Chen},
  \citenamefont {Shi}, \citenamefont {Yang}, \citenamefont {Lu}, \citenamefont
  {Lai}, \citenamefont {Yan}, \citenamefont {Wang}, \citenamefont {Zhang},\
  and\ \citenamefont {Li}}]{Chen2014}%
  \BibitemOpen
  \bibfield  {author} {\bibinfo {author} {\bibfnamefont {Z.-G.}\ \bibnamefont
  {Chen}}, \bibinfo {author} {\bibfnamefont {Z.}~\bibnamefont {Shi}}, \bibinfo
  {author} {\bibfnamefont {W.}~\bibnamefont {Yang}}, \bibinfo {author}
  {\bibfnamefont {X.}~\bibnamefont {Lu}}, \bibinfo {author} {\bibfnamefont
  {Y.}~\bibnamefont {Lai}}, \bibinfo {author} {\bibfnamefont {H.}~\bibnamefont
  {Yan}}, \bibinfo {author} {\bibfnamefont {F.}~\bibnamefont {Wang}}, \bibinfo
  {author} {\bibfnamefont {G.}~\bibnamefont {Zhang}}, \ and\ \bibinfo {author}
  {\bibfnamefont {Z.}~\bibnamefont {Li}},\ }\href@noop {} {\bibfield  {journal}
  {\bibinfo  {journal} {Nature Communications}\ }\textbf {\bibinfo {volume}
  {5}},\ \bibinfo {pages} {4461} (\bibinfo {year} {2014})}\BibitemShut
  {NoStop}%
\bibitem [{\citenamefont {Mucha-Kruczy\ifmmode~\acute{n}\else \'{n}\fi{}ski}\
  \emph {et~al.}(2016)\citenamefont {Mucha-Kruczy\ifmmode~\acute{n}\else
  \'{n}\fi{}ski}, \citenamefont {Wallbank},\ and\ \citenamefont
  {Fal'ko}}]{Mucha2016}%
  \BibitemOpen
  \bibfield  {author} {\bibinfo {author} {\bibfnamefont {M.}~\bibnamefont
  {Mucha-Kruczy\ifmmode~\acute{n}\else \'{n}\fi{}ski}}, \bibinfo {author}
  {\bibfnamefont {J.~R.}\ \bibnamefont {Wallbank}}, \ and\ \bibinfo {author}
  {\bibfnamefont {V.~I.}\ \bibnamefont {Fal'ko}},\ }\href {\doibase
  10.1103/PhysRevB.93.085409} {\bibfield  {journal} {\bibinfo  {journal} {Phys.
  Rev. B}\ }\textbf {\bibinfo {volume} {93}},\ \bibinfo {pages} {085409}
  (\bibinfo {year} {2016})}\BibitemShut {NoStop}%
\bibitem [{\citenamefont {Kumar}\ \emph {et~al.}(2015)\citenamefont {Kumar},
  \citenamefont {Er}, \citenamefont {Dong}, \citenamefont {Li},\ and\
  \citenamefont {Shenoy}}]{Kumar2015}%
  \BibitemOpen
  \bibfield  {author} {\bibinfo {author} {\bibfnamefont {H.}~\bibnamefont
  {Kumar}}, \bibinfo {author} {\bibfnamefont {D.}~\bibnamefont {Er}}, \bibinfo
  {author} {\bibfnamefont {L.}~\bibnamefont {Dong}}, \bibinfo {author}
  {\bibfnamefont {J.}~\bibnamefont {Li}}, \ and\ \bibinfo {author}
  {\bibfnamefont {V.~B.}\ \bibnamefont {Shenoy}},\ }\href@noop {} {\bibfield
  {journal} {\bibinfo  {journal} {Scientific Reports}\ }\textbf {\bibinfo
  {volume} {5}},\ \bibinfo {pages} {10872} (\bibinfo {year}
  {2015})}\BibitemShut {NoStop}%
\bibitem [{\citenamefont {Slotman}\ \emph {et~al.}(2015)\citenamefont
  {Slotman}, \citenamefont {van Wijk}, \citenamefont {Zhao}, \citenamefont
  {Fasolino}, \citenamefont {Katsnelson},\ and\ \citenamefont
  {Yuan}}]{Slotman2015}%
  \BibitemOpen
  \bibfield  {author} {\bibinfo {author} {\bibfnamefont {G.~J.}\ \bibnamefont
  {Slotman}}, \bibinfo {author} {\bibfnamefont {M.~M.}\ \bibnamefont {van
  Wijk}}, \bibinfo {author} {\bibfnamefont {P.-L.}\ \bibnamefont {Zhao}},
  \bibinfo {author} {\bibfnamefont {A.}~\bibnamefont {Fasolino}}, \bibinfo
  {author} {\bibfnamefont {M.~I.}\ \bibnamefont {Katsnelson}}, \ and\ \bibinfo
  {author} {\bibfnamefont {S.}~\bibnamefont {Yuan}},\ }\href {\doibase
  10.1103/PhysRevLett.115.186801} {\bibfield  {journal} {\bibinfo  {journal}
  {Phys. Rev. Lett.}\ }\textbf {\bibinfo {volume} {115}},\ \bibinfo {pages}
  {186801} (\bibinfo {year} {2015})}\BibitemShut {NoStop}%
\bibitem [{\citenamefont {Eckmann}\ \emph {et~al.}(2013)\citenamefont
  {Eckmann}, \citenamefont {Park}, \citenamefont {Yang}, \citenamefont {Elias},
  \citenamefont {Mayorov}, \citenamefont {Yu}, \citenamefont {Jalil},
  \citenamefont {Novoselov}, \citenamefont {Gorbachev}, \citenamefont
  {Lazzeri}, \citenamefont {Geim},\ and\ \citenamefont
  {Casiraghi}}]{Eckmann2013}%
  \BibitemOpen
  \bibfield  {author} {\bibinfo {author} {\bibfnamefont {A.}~\bibnamefont
  {Eckmann}}, \bibinfo {author} {\bibfnamefont {J.}~\bibnamefont {Park}},
  \bibinfo {author} {\bibfnamefont {H.}~\bibnamefont {Yang}}, \bibinfo {author}
  {\bibfnamefont {D.}~\bibnamefont {Elias}}, \bibinfo {author} {\bibfnamefont
  {A.~S.}\ \bibnamefont {Mayorov}}, \bibinfo {author} {\bibfnamefont
  {G.}~\bibnamefont {Yu}}, \bibinfo {author} {\bibfnamefont {R.}~\bibnamefont
  {Jalil}}, \bibinfo {author} {\bibfnamefont {K.~S.}\ \bibnamefont
  {Novoselov}}, \bibinfo {author} {\bibfnamefont {R.~V.}\ \bibnamefont
  {Gorbachev}}, \bibinfo {author} {\bibfnamefont {M.}~\bibnamefont {Lazzeri}},
  \bibinfo {author} {\bibfnamefont {A.~K.}\ \bibnamefont {Geim}}, \ and\
  \bibinfo {author} {\bibfnamefont {C.}~\bibnamefont {Casiraghi}},\ }\href
  {\doibase 10.1021/nl402679b} {\bibfield  {journal} {\bibinfo  {journal} {Nano
  Lett.}\ }\textbf {\bibinfo {volume} {13}},\ \bibinfo {pages} {5242} (\bibinfo
  {year} {2013})}\BibitemShut {NoStop}%
\bibitem [{\citenamefont {Jung}\ \emph {et~al.}(2015)\citenamefont {Jung},
  \citenamefont {DaSilva}, \citenamefont {MacDonald},\ and\ \citenamefont
  {Adam}}]{Jung2015}%
  \BibitemOpen
  \bibfield  {author} {\bibinfo {author} {\bibfnamefont {J.}~\bibnamefont
  {Jung}}, \bibinfo {author} {\bibfnamefont {A.~M.}\ \bibnamefont {DaSilva}},
  \bibinfo {author} {\bibfnamefont {A.~H.}\ \bibnamefont {MacDonald}}, \ and\
  \bibinfo {author} {\bibfnamefont {S.}~\bibnamefont {Adam}},\ }\href {\doibase
  10.1038/ncomms7308} {\bibfield  {journal} {\bibinfo  {journal} {Nature
  Communications}\ }\textbf {\bibinfo {volume} {6}},\ \bibinfo {pages} {6308}
  (\bibinfo {year} {2015})}\BibitemShut {NoStop}%
\bibitem [{\citenamefont {Sachs}\ \emph {et~al.}(2011)\citenamefont {Sachs},
  \citenamefont {Wehling}, \citenamefont {Katsnelson},\ and\ \citenamefont
  {Lichtenstein}}]{Sachs2011}%
  \BibitemOpen
  \bibfield  {author} {\bibinfo {author} {\bibfnamefont {B.}~\bibnamefont
  {Sachs}}, \bibinfo {author} {\bibfnamefont {T.~O.}\ \bibnamefont {Wehling}},
  \bibinfo {author} {\bibfnamefont {M.~I.}\ \bibnamefont {Katsnelson}}, \ and\
  \bibinfo {author} {\bibfnamefont {A.~I.}\ \bibnamefont {Lichtenstein}},\
  }\href {\doibase 10.1103/PhysRevB.84.195414} {\bibfield  {journal} {\bibinfo
  {journal} {Phys. Rev. B}\ }\textbf {\bibinfo {volume} {84}},\ \bibinfo
  {pages} {195414} (\bibinfo {year} {2011})}\BibitemShut {NoStop}%
\bibitem [{\citenamefont {Argentero}\ \emph {et~al.}(2017)\citenamefont
  {Argentero}, \citenamefont {Mittelberger}, \citenamefont {Monazam},
  \citenamefont {Cao}, \citenamefont {Pennycook}, \citenamefont {Mangler},
  \citenamefont {Kramberger}, \citenamefont {Kotakoski}, \citenamefont {Geim},\
  and\ \citenamefont {Meyer}}]{Argentero2017}%
  \BibitemOpen
  \bibfield  {author} {\bibinfo {author} {\bibfnamefont {G.}~\bibnamefont
  {Argentero}}, \bibinfo {author} {\bibfnamefont {A.}~\bibnamefont
  {Mittelberger}}, \bibinfo {author} {\bibfnamefont {M.~R.~A.}\ \bibnamefont
  {Monazam}}, \bibinfo {author} {\bibfnamefont {Y.}~\bibnamefont {Cao}},
  \bibinfo {author} {\bibfnamefont {T.~J.}\ \bibnamefont {Pennycook}}, \bibinfo
  {author} {\bibfnamefont {C.}~\bibnamefont {Mangler}}, \bibinfo {author}
  {\bibfnamefont {C.}~\bibnamefont {Kramberger}}, \bibinfo {author}
  {\bibfnamefont {J.}~\bibnamefont {Kotakoski}}, \bibinfo {author}
  {\bibfnamefont {A.~K.}\ \bibnamefont {Geim}}, \ and\ \bibinfo {author}
  {\bibfnamefont {J.~C.}\ \bibnamefont {Meyer}},\ }\href {\doibase
  10.1021/acs.nanolett.6b04360} {\bibfield  {journal} {\bibinfo  {journal}
  {Nano Lett.}\ }\textbf {\bibinfo {volume} {17}},\ \bibinfo {pages} {1409}
  (\bibinfo {year} {2017})}\BibitemShut {NoStop}%
\bibitem [{\citenamefont {Woods}\ \emph {et~al.}(2016)\citenamefont {Woods},
  \citenamefont {Withers}, \citenamefont {Zhu}, \citenamefont {Cao},
  \citenamefont {Yu}, \citenamefont {Kozikov}, \citenamefont {Ben~Shalom},
  \citenamefont {Morozov}, \citenamefont {van Wijk}, \citenamefont {Fasolino},
  \citenamefont {Katsnelson}, \citenamefont {Watanabe}, \citenamefont
  {Taniguchi}, \citenamefont {Geim}, \citenamefont {Mishchenko},\ and\
  \citenamefont {Novoselov}}]{Woods2016}%
  \BibitemOpen
  \bibfield  {author} {\bibinfo {author} {\bibfnamefont {C.~R.}\ \bibnamefont
  {Woods}}, \bibinfo {author} {\bibfnamefont {F.}~\bibnamefont {Withers}},
  \bibinfo {author} {\bibfnamefont {M.~J.}\ \bibnamefont {Zhu}}, \bibinfo
  {author} {\bibfnamefont {Y.}~\bibnamefont {Cao}}, \bibinfo {author}
  {\bibfnamefont {G.}~\bibnamefont {Yu}}, \bibinfo {author} {\bibfnamefont
  {A.}~\bibnamefont {Kozikov}}, \bibinfo {author} {\bibfnamefont
  {M.}~\bibnamefont {Ben~Shalom}}, \bibinfo {author} {\bibfnamefont {S.~V.}\
  \bibnamefont {Morozov}}, \bibinfo {author} {\bibfnamefont {M.~M.}\
  \bibnamefont {van Wijk}}, \bibinfo {author} {\bibfnamefont {A.}~\bibnamefont
  {Fasolino}}, \bibinfo {author} {\bibfnamefont {M.~I.}\ \bibnamefont
  {Katsnelson}}, \bibinfo {author} {\bibfnamefont {K.}~\bibnamefont
  {Watanabe}}, \bibinfo {author} {\bibfnamefont {T.}~\bibnamefont {Taniguchi}},
  \bibinfo {author} {\bibfnamefont {A.~K.}\ \bibnamefont {Geim}}, \bibinfo
  {author} {\bibfnamefont {A.}~\bibnamefont {Mishchenko}}, \ and\ \bibinfo
  {author} {\bibfnamefont {K.~S.}\ \bibnamefont {Novoselov}},\ }\href@noop {}
  {\bibfield  {journal} {\bibinfo  {journal} {Nature Communications}\ }\textbf
  {\bibinfo {volume} {7}},\ \bibinfo {pages} {10800} (\bibinfo {year}
  {2016})}\BibitemShut {NoStop}%
\bibitem [{\citenamefont {Leven}\ \emph {et~al.}(2013)\citenamefont {Leven},
  \citenamefont {Krepel}, \citenamefont {Shemesh},\ and\ \citenamefont
  {Hod}}]{Leven2013}%
  \BibitemOpen
  \bibfield  {author} {\bibinfo {author} {\bibfnamefont {I.}~\bibnamefont
  {Leven}}, \bibinfo {author} {\bibfnamefont {D.}~\bibnamefont {Krepel}},
  \bibinfo {author} {\bibfnamefont {O.}~\bibnamefont {Shemesh}}, \ and\
  \bibinfo {author} {\bibfnamefont {O.}~\bibnamefont {Hod}},\ }\href {\doibase
  10.1021/jz301758c} {\bibfield  {journal} {\bibinfo  {journal} {J. Phys. Chem.
  Lett.}\ }\textbf {\bibinfo {volume} {4}},\ \bibinfo {pages} {115} (\bibinfo
  {year} {2013})}\BibitemShut {NoStop}%
\bibitem [{\citenamefont {Fan}\ \emph {et~al.}(2011)\citenamefont {Fan},
  \citenamefont {Zhao}, \citenamefont {Wang}, \citenamefont {Zhang},\ and\
  \citenamefont {Zhang}}]{Fan2011}%
  \BibitemOpen
  \bibfield  {author} {\bibinfo {author} {\bibfnamefont {Y.}~\bibnamefont
  {Fan}}, \bibinfo {author} {\bibfnamefont {M.}~\bibnamefont {Zhao}}, \bibinfo
  {author} {\bibfnamefont {Z.}~\bibnamefont {Wang}}, \bibinfo {author}
  {\bibfnamefont {X.}~\bibnamefont {Zhang}}, \ and\ \bibinfo {author}
  {\bibfnamefont {H.}~\bibnamefont {Zhang}},\ }\href {\doibase
  10.1063/1.3556640} {\bibfield  {journal} {\bibinfo  {journal} {Applied
  Physics Letters}\ }\textbf {\bibinfo {volume} {98}},\ \bibinfo {pages}
  {083103} (\bibinfo {year} {2011})}\BibitemShut {NoStop}%
\bibitem [{\citenamefont {Giovannetti}\ \emph {et~al.}(2007)\citenamefont
  {Giovannetti}, \citenamefont {Khomyakov}, \citenamefont {Brocks},
  \citenamefont {Kelly},\ and\ \citenamefont {van~den
  Brink}}]{Giovannetti2007}%
  \BibitemOpen
  \bibfield  {author} {\bibinfo {author} {\bibfnamefont {G.}~\bibnamefont
  {Giovannetti}}, \bibinfo {author} {\bibfnamefont {P.~A.}\ \bibnamefont
  {Khomyakov}}, \bibinfo {author} {\bibfnamefont {G.}~\bibnamefont {Brocks}},
  \bibinfo {author} {\bibfnamefont {P.~J.}\ \bibnamefont {Kelly}}, \ and\
  \bibinfo {author} {\bibfnamefont {J.}~\bibnamefont {van~den Brink}},\ }\href
  {\doibase 10.1103/PhysRevB.76.073103} {\bibfield  {journal} {\bibinfo
  {journal} {Phys. Rev. B}\ }\textbf {\bibinfo {volume} {76}},\ \bibinfo
  {pages} {073103} (\bibinfo {year} {2007})}\BibitemShut {NoStop}%
\bibitem [{\citenamefont {Leven}\ \emph {et~al.}(2016)\citenamefont {Leven},
  \citenamefont {Maaravi}, \citenamefont {Azuri}, \citenamefont {Kronik},\ and\
  \citenamefont {Hod}}]{Leven2016}%
  \BibitemOpen
  \bibfield  {author} {\bibinfo {author} {\bibfnamefont {I.}~\bibnamefont
  {Leven}}, \bibinfo {author} {\bibfnamefont {T.}~\bibnamefont {Maaravi}},
  \bibinfo {author} {\bibfnamefont {I.}~\bibnamefont {Azuri}}, \bibinfo
  {author} {\bibfnamefont {L.}~\bibnamefont {Kronik}}, \ and\ \bibinfo {author}
  {\bibfnamefont {O.}~\bibnamefont {Hod}},\ }\href {\doibase
  10.1021/acs.jctc.6b00147} {\bibfield  {journal} {\bibinfo  {journal} {J.
  Chem. Theor. Comp.}\ }\textbf {\bibinfo {volume} {12}},\ \bibinfo {pages}
  {2896} (\bibinfo {year} {2016})}\BibitemShut {NoStop}%
\bibitem [{\citenamefont {Zhao}\ \emph {et~al.}(2014)\citenamefont {Zhao},
  \citenamefont {Li},\ and\ \citenamefont {Zhao}}]{Zhao2014}%
  \BibitemOpen
  \bibfield  {author} {\bibinfo {author} {\bibfnamefont {X.}~\bibnamefont
  {Zhao}}, \bibinfo {author} {\bibfnamefont {L.}~\bibnamefont {Li}}, \ and\
  \bibinfo {author} {\bibfnamefont {M.}~\bibnamefont {Zhao}},\ }\href {\doibase
  10.1088/0953-8984/26/9/095002} {\bibfield  {journal} {\bibinfo  {journal} {J.
  Phys. Condens. Matter}\ }\textbf {\bibinfo {volume} {26}},\ \bibinfo {pages}
  {095002} (\bibinfo {year} {2014})}\BibitemShut {NoStop}%
\bibitem [{\citenamefont {Zhou}\ \emph {et~al.}(2015)\citenamefont {Zhou},
  \citenamefont {Han}, \citenamefont {Dai}, \citenamefont {Sun},\ and\
  \citenamefont {Srolovitz}}]{Zhou2015}%
  \BibitemOpen
  \bibfield  {author} {\bibinfo {author} {\bibfnamefont {S.}~\bibnamefont
  {Zhou}}, \bibinfo {author} {\bibfnamefont {J.}~\bibnamefont {Han}}, \bibinfo
  {author} {\bibfnamefont {S.}~\bibnamefont {Dai}}, \bibinfo {author}
  {\bibfnamefont {J.}~\bibnamefont {Sun}}, \ and\ \bibinfo {author}
  {\bibfnamefont {D.~J.}\ \bibnamefont {Srolovitz}},\ }\href {\doibase
  10.1103/PhysRevB.92.155438} {\bibfield  {journal} {\bibinfo  {journal} {Phys.
  Rev. B}\ }\textbf {\bibinfo {volume} {92}},\ \bibinfo {pages} {155438}
  (\bibinfo {year} {2015})}\BibitemShut {NoStop}%
\bibitem [{\citenamefont {Lebedeva}\ \emph
  {et~al.}(2011{\natexlab{a}})\citenamefont {Lebedeva}, \citenamefont
  {Knizhnik}, \citenamefont {Popov}, \citenamefont {Lozovik},\ and\
  \citenamefont {Potapkin}}]{Lebedeva2011}%
  \BibitemOpen
  \bibfield  {author} {\bibinfo {author} {\bibfnamefont {I.~V.}\ \bibnamefont
  {Lebedeva}}, \bibinfo {author} {\bibfnamefont {A.~A.}\ \bibnamefont
  {Knizhnik}}, \bibinfo {author} {\bibfnamefont {A.~M.}\ \bibnamefont {Popov}},
  \bibinfo {author} {\bibfnamefont {Y.~E.}\ \bibnamefont {Lozovik}}, \ and\
  \bibinfo {author} {\bibfnamefont {B.~V.}\ \bibnamefont {Potapkin}},\ }\href
  {\doibase 10.1039/c0cp02614j} {\bibfield  {journal} {\bibinfo  {journal}
  {Phys. Chem. Chem. Phys.}\ }\textbf {\bibinfo {volume} {13}},\ \bibinfo
  {pages} {5687} (\bibinfo {year} {2011}{\natexlab{a}})}\BibitemShut {NoStop}%
\bibitem [{\citenamefont {Reguzzoni}\ \emph {et~al.}(2012)\citenamefont
  {Reguzzoni}, \citenamefont {Fasolino}, \citenamefont {Molinari},\ and\
  \citenamefont {Righi}}]{Reguzzoni2012}%
  \BibitemOpen
  \bibfield  {author} {\bibinfo {author} {\bibfnamefont {M.}~\bibnamefont
  {Reguzzoni}}, \bibinfo {author} {\bibfnamefont {A.}~\bibnamefont {Fasolino}},
  \bibinfo {author} {\bibfnamefont {E.}~\bibnamefont {Molinari}}, \ and\
  \bibinfo {author} {\bibfnamefont {M.~C.}\ \bibnamefont {Righi}},\ }\href
  {\doibase 10.1103/PhysRevB.86.245434} {\bibfield  {journal} {\bibinfo
  {journal} {Phys. Rev. B}\ }\textbf {\bibinfo {volume} {86}},\ \bibinfo
  {pages} {245434} (\bibinfo {year} {2012})}\BibitemShut {NoStop}%
\bibitem [{\citenamefont {Lee}\ \emph {et~al.}(2010)\citenamefont {Lee},
  \citenamefont {Murray}, \citenamefont {Kong}, \citenamefont {Lundqvist},\
  and\ \citenamefont {Langreth}}]{Lee2010}%
  \BibitemOpen
  \bibfield  {author} {\bibinfo {author} {\bibfnamefont {K.}~\bibnamefont
  {Lee}}, \bibinfo {author} {\bibfnamefont {E.~D.}\ \bibnamefont {Murray}},
  \bibinfo {author} {\bibfnamefont {L.}~\bibnamefont {Kong}}, \bibinfo {author}
  {\bibfnamefont {B.~I.}\ \bibnamefont {Lundqvist}}, \ and\ \bibinfo {author}
  {\bibfnamefont {D.~C.}\ \bibnamefont {Langreth}},\ }\href {\doibase
  10.1103/PhysRevB.82.081101} {\bibfield  {journal} {\bibinfo  {journal} {Phys.
  Rev. B}\ }\textbf {\bibinfo {volume} {82}},\ \bibinfo {pages} {081101}
  (\bibinfo {year} {2010})}\BibitemShut {NoStop}%
\bibitem [{\citenamefont {van Wijk}\ \emph {et~al.}(2014)\citenamefont {van
  Wijk}, \citenamefont {Schuring}, \citenamefont {Katsnelson},\ and\
  \citenamefont {Fasolino}}]{Wijk2014}%
  \BibitemOpen
  \bibfield  {author} {\bibinfo {author} {\bibfnamefont {M.~M.}\ \bibnamefont
  {van Wijk}}, \bibinfo {author} {\bibfnamefont {A.}~\bibnamefont {Schuring}},
  \bibinfo {author} {\bibfnamefont {M.~I.}\ \bibnamefont {Katsnelson}}, \ and\
  \bibinfo {author} {\bibfnamefont {A.}~\bibnamefont {Fasolino}},\ }\href
  {\doibase 10.1103/PhysRevLett.113.135504} {\bibfield  {journal} {\bibinfo
  {journal} {Phys. Rev. Lett.}\ }\textbf {\bibinfo {volume} {113}},\ \bibinfo
  {pages} {135504} (\bibinfo {year} {2014})}\BibitemShut {NoStop}%
\bibitem [{\citenamefont {Neek-Amal}\ and\ \citenamefont
  {Peeters}(2014)}]{Neek-Amal2014}%
  \BibitemOpen
  \bibfield  {author} {\bibinfo {author} {\bibfnamefont {M.}~\bibnamefont
  {Neek-Amal}}\ and\ \bibinfo {author} {\bibfnamefont {F.~M.}\ \bibnamefont
  {Peeters}},\ }\href {\doibase 10.1063/1.4863661} {\bibfield  {journal}
  {\bibinfo  {journal} {Applied Physics Letters}\ }\textbf {\bibinfo {volume}
  {104}},\ \bibinfo {pages} {041909} (\bibinfo {year} {2014})}\BibitemShut
  {NoStop}%
\bibitem [{\citenamefont {Popov}\ \emph
  {et~al.}(2012{\natexlab{a}})\citenamefont {Popov}, \citenamefont {Lebedeva},
  \citenamefont {Knizhnik}, \citenamefont {Lozovik},\ and\ \citenamefont
  {Potapkin}}]{Popov2012}%
  \BibitemOpen
  \bibfield  {author} {\bibinfo {author} {\bibfnamefont {A.~M.}\ \bibnamefont
  {Popov}}, \bibinfo {author} {\bibfnamefont {I.~V.}\ \bibnamefont {Lebedeva}},
  \bibinfo {author} {\bibfnamefont {A.~A.}\ \bibnamefont {Knizhnik}}, \bibinfo
  {author} {\bibfnamefont {Y.~E.}\ \bibnamefont {Lozovik}}, \ and\ \bibinfo
  {author} {\bibfnamefont {B.~V.}\ \bibnamefont {Potapkin}},\ }\href {\doibase
  10.1016/j.cplett.2012.03.082} {\bibfield  {journal} {\bibinfo  {journal}
  {Chem. Phys. Lett.}\ }\textbf {\bibinfo {volume} {536}},\ \bibinfo {pages}
  {82} (\bibinfo {year} {2012}{\natexlab{a}})}\BibitemShut {NoStop}%
\bibitem [{\citenamefont {Lebedev}\ \emph {et~al.}(2016)\citenamefont
  {Lebedev}, \citenamefont {Lebedeva}, \citenamefont {Knizhnik},\ and\
  \citenamefont {Popov}}]{Lebedev2016}%
  \BibitemOpen
  \bibfield  {author} {\bibinfo {author} {\bibfnamefont {A.~V.}\ \bibnamefont
  {Lebedev}}, \bibinfo {author} {\bibfnamefont {I.~V.}\ \bibnamefont
  {Lebedeva}}, \bibinfo {author} {\bibfnamefont {A.~A.}\ \bibnamefont
  {Knizhnik}}, \ and\ \bibinfo {author} {\bibfnamefont {A.~M.}\ \bibnamefont
  {Popov}},\ }\href {\doibase 10.1039/C5RA20882C} {\bibfield  {journal}
  {\bibinfo  {journal} {RSC Advances}\ }\textbf {\bibinfo {volume} {6}},\
  \bibinfo {pages} {6423} (\bibinfo {year} {2016})}\BibitemShut {NoStop}%
\bibitem [{\citenamefont {Kresse}\ and\ \citenamefont
  {Furthm\"{u}ller}(1996)}]{Kresse1996}%
  \BibitemOpen
  \bibfield  {author} {\bibinfo {author} {\bibfnamefont {G.}~\bibnamefont
  {Kresse}}\ and\ \bibinfo {author} {\bibfnamefont {J.}~\bibnamefont
  {Furthm\"{u}ller}},\ }\href {\doibase 10.1103/PhysRevB.54.11169} {\bibfield
  {journal} {\bibinfo  {journal} {Phys. Rev. B}\ }\textbf {\bibinfo {volume}
  {54}},\ \bibinfo {pages} {11169} (\bibinfo {year} {1996})}\BibitemShut
  {NoStop}%
\bibitem [{\citenamefont {Kresse}\ and\ \citenamefont
  {Joubert}(1999)}]{Kresse1999}%
  \BibitemOpen
  \bibfield  {author} {\bibinfo {author} {\bibfnamefont {G.}~\bibnamefont
  {Kresse}}\ and\ \bibinfo {author} {\bibfnamefont {D.}~\bibnamefont
  {Joubert}},\ }\href {\doibase 10.1103/PhysRevB.59.1758} {\bibfield  {journal}
  {\bibinfo  {journal} {Phys. Rev. B}\ }\textbf {\bibinfo {volume} {59}},\
  \bibinfo {pages} {1758} (\bibinfo {year} {1999})}\BibitemShut {NoStop}%
\bibitem [{\citenamefont {Monkhorst}\ and\ \citenamefont
  {Pack}(1976)}]{Monkhorst1976}%
  \BibitemOpen
  \bibfield  {author} {\bibinfo {author} {\bibfnamefont {H.~J.}\ \bibnamefont
  {Monkhorst}}\ and\ \bibinfo {author} {\bibfnamefont {J.~D.}\ \bibnamefont
  {Pack}},\ }\href {\doibase 10.1103/PhysRevB.13.5188} {\bibfield  {journal}
  {\bibinfo  {journal} {Phys. Rev. B}\ }\textbf {\bibinfo {volume} {13}},\
  \bibinfo {pages} {5188} (\bibinfo {year} {1976})}\BibitemShut {NoStop}%
\bibitem [{\citenamefont {Pease}(1950)}]{Pease1950}%
  \BibitemOpen
  \bibfield  {author} {\bibinfo {author} {\bibfnamefont {R.~S.}\ \bibnamefont
  {Pease}},\ }\href {\doibase 10.1038/165722b0} {\bibfield  {journal} {\bibinfo
   {journal} {Nature}\ }\textbf {\bibinfo {volume} {165}},\ \bibinfo {pages}
  {722} (\bibinfo {year} {1950})}\BibitemShut {NoStop}%
\bibitem [{\citenamefont {Pease}(1952)}]{Pease1952}%
  \BibitemOpen
  \bibfield  {author} {\bibinfo {author} {\bibfnamefont {R.~S.}\ \bibnamefont
  {Pease}},\ }\href {\doibase 10.1107/S0365110X52001064} {\bibfield  {journal}
  {\bibinfo  {journal} {Acta Crystallographica}\ }\textbf {\bibinfo {volume}
  {5}},\ \bibinfo {pages} {356} (\bibinfo {year} {1952})}\BibitemShut {NoStop}%
\bibitem [{\citenamefont {Lynch}\ and\ \citenamefont
  {Drickamer}(1966)}]{Lynch1966}%
  \BibitemOpen
  \bibfield  {author} {\bibinfo {author} {\bibfnamefont {R.~W.}\ \bibnamefont
  {Lynch}}\ and\ \bibinfo {author} {\bibfnamefont {H.~G.}\ \bibnamefont
  {Drickamer}},\ }\href {\doibase 10.1063/1.1726442} {\bibfield  {journal}
  {\bibinfo  {journal} {J. Chem. Phys.}\ }\textbf {\bibinfo {volume} {44}},\
  \bibinfo {pages} {181} (\bibinfo {year} {1966})}\BibitemShut {NoStop}%
\bibitem [{\citenamefont {Solozhenko}\ \emph {et~al.}(1995)\citenamefont
  {Solozhenko}, \citenamefont {Will},\ and\ \citenamefont
  {Elf}}]{Solozhenko1995}%
  \BibitemOpen
  \bibfield  {author} {\bibinfo {author} {\bibfnamefont {V.~L.}\ \bibnamefont
  {Solozhenko}}, \bibinfo {author} {\bibfnamefont {G.}~\bibnamefont {Will}}, \
  and\ \bibinfo {author} {\bibfnamefont {F.}~\bibnamefont {Elf}},\ }\href
  {\doibase 10.1016/0038-1098(95)00381-9} {\bibfield  {journal} {\bibinfo
  {journal} {Solid State Communications}\ }\textbf {\bibinfo {volume} {96}},\
  \bibinfo {pages} {1} (\bibinfo {year} {1995})}\BibitemShut {NoStop}%
\bibitem [{\citenamefont {Solozhenko}\ and\ \citenamefont
  {Peun}(1997)}]{Solozhenko1997}%
  \BibitemOpen
  \bibfield  {author} {\bibinfo {author} {\bibfnamefont {V.~L.}\ \bibnamefont
  {Solozhenko}}\ and\ \bibinfo {author} {\bibfnamefont {T.}~\bibnamefont
  {Peun}},\ }\href {\doibase 10.1016/S0022-3697(97)00037-1} {\bibfield
  {journal} {\bibinfo  {journal} {J. Phys. Chem. Solids}\ }\textbf {\bibinfo
  {volume} {58}},\ \bibinfo {pages} {1321} (\bibinfo {year}
  {1997})}\BibitemShut {NoStop}%
\bibitem [{\citenamefont {Solozhenko}\ \emph {et~al.}(2001)\citenamefont
  {Solozhenko}, \citenamefont {Lazarenko}, \citenamefont {Petitet},\ and\
  \citenamefont {Kanaev}}]{Solozhenko2001}%
  \BibitemOpen
  \bibfield  {author} {\bibinfo {author} {\bibfnamefont {V.~L.}\ \bibnamefont
  {Solozhenko}}, \bibinfo {author} {\bibfnamefont {A.~G.}\ \bibnamefont
  {Lazarenko}}, \bibinfo {author} {\bibfnamefont {J.-P.}\ \bibnamefont
  {Petitet}}, \ and\ \bibinfo {author} {\bibfnamefont {A.~V.}\ \bibnamefont
  {Kanaev}},\ }\href {\doibase 10.1016/S0022-3697(01)00030-0} {\bibfield
  {journal} {\bibinfo  {journal} {J. Phys. Chem. Solids}\ }\textbf {\bibinfo
  {volume} {62}},\ \bibinfo {pages} {1331} (\bibinfo {year}
  {2001})}\BibitemShut {NoStop}%
\bibitem [{\citenamefont {Paszkowicz}\ \emph {et~al.}(2002)\citenamefont
  {Paszkowicz}, \citenamefont {Pelka}, \citenamefont {Knapp}, \citenamefont
  {Szyszko},\ and\ \citenamefont {Podsiadlo}}]{Paszkowicz2002}%
  \BibitemOpen
  \bibfield  {author} {\bibinfo {author} {\bibfnamefont {W.}~\bibnamefont
  {Paszkowicz}}, \bibinfo {author} {\bibfnamefont {J.~B.}\ \bibnamefont
  {Pelka}}, \bibinfo {author} {\bibfnamefont {M.}~\bibnamefont {Knapp}},
  \bibinfo {author} {\bibfnamefont {T.}~\bibnamefont {Szyszko}}, \ and\
  \bibinfo {author} {\bibfnamefont {S.}~\bibnamefont {Podsiadlo}},\ }\href
  {\doibase 10.1007/s003390100999} {\bibfield  {journal} {\bibinfo  {journal}
  {Appl. Phys. A}\ }\textbf {\bibinfo {volume} {75}},\ \bibinfo {pages} {431}
  (\bibinfo {year} {2002})}\BibitemShut {NoStop}%
\bibitem [{\citenamefont {Bosak}\ \emph {et~al.}(2006)\citenamefont {Bosak},
  \citenamefont {Serrano}, \citenamefont {Krisch}, \citenamefont {Watanabe},
  \citenamefont {Taniguchi},\ and\ \citenamefont {Kanda}}]{Bosak2006}%
  \BibitemOpen
  \bibfield  {author} {\bibinfo {author} {\bibfnamefont {A.}~\bibnamefont
  {Bosak}}, \bibinfo {author} {\bibfnamefont {J.}~\bibnamefont {Serrano}},
  \bibinfo {author} {\bibfnamefont {M.}~\bibnamefont {Krisch}}, \bibinfo
  {author} {\bibfnamefont {K.}~\bibnamefont {Watanabe}}, \bibinfo {author}
  {\bibfnamefont {T.}~\bibnamefont {Taniguchi}}, \ and\ \bibinfo {author}
  {\bibfnamefont {H.}~\bibnamefont {Kanda}},\ }\href {\doibase
  10.1103/PhysRevB.73.041402} {\bibfield  {journal} {\bibinfo  {journal} {Phys.
  Rev. B}\ }\textbf {\bibinfo {volume} {73}},\ \bibinfo {pages} {041402}
  (\bibinfo {year} {2006})}\BibitemShut {NoStop}%
\bibitem [{\citenamefont {Fuchizaki}\ \emph {et~al.}(2008)\citenamefont
  {Fuchizaki}, \citenamefont {Nakamichi}, \citenamefont {Saitoh},\ and\
  \citenamefont {Katayama}}]{Fuchizaki2008}%
  \BibitemOpen
  \bibfield  {author} {\bibinfo {author} {\bibfnamefont {K.}~\bibnamefont
  {Fuchizaki}}, \bibinfo {author} {\bibfnamefont {T.}~\bibnamefont
  {Nakamichi}}, \bibinfo {author} {\bibfnamefont {H.}~\bibnamefont {Saitoh}}, \
  and\ \bibinfo {author} {\bibfnamefont {Y.}~\bibnamefont {Katayama}},\ }\href
  {\doibase 10.1016/j.ssc.2008.09.031} {\bibfield  {journal} {\bibinfo
  {journal} {Solid State Communications}\ }\textbf {\bibinfo {volume} {148}},\
  \bibinfo {pages} {390} (\bibinfo {year} {2008})}\BibitemShut {NoStop}%
\bibitem [{\citenamefont {Bernal}(1924)}]{Bernal1924}%
  \BibitemOpen
  \bibfield  {author} {\bibinfo {author} {\bibfnamefont {J.~D.}\ \bibnamefont
  {Bernal}},\ }\href {\doibase 10.1098/rspa.1924.0101} {\bibfield  {journal}
  {\bibinfo  {journal} {Proc. R. Soc. London, Ser. A}\ }\textbf {\bibinfo
  {volume} {106}},\ \bibinfo {pages} {749} (\bibinfo {year}
  {1924})}\BibitemShut {NoStop}%
\bibitem [{\citenamefont {Baskin}\ and\ \citenamefont
  {Meyer}(1955)}]{Baskin1955}%
  \BibitemOpen
  \bibfield  {author} {\bibinfo {author} {\bibfnamefont {V.}~\bibnamefont
  {Baskin}}\ and\ \bibinfo {author} {\bibfnamefont {L.}~\bibnamefont {Meyer}},\
  }\href {\doibase 10.1103/PhysRev.100.544} {\bibfield  {journal} {\bibinfo
  {journal} {Phys. Rev.}\ }\textbf {\bibinfo {volume} {100}},\ \bibinfo {pages}
  {544} (\bibinfo {year} {1955})}\BibitemShut {NoStop}%
\bibitem [{\citenamefont {Ludsteck}(1972)}]{Ludsteck1972}%
  \BibitemOpen
  \bibfield  {author} {\bibinfo {author} {\bibfnamefont {V.~A.}\ \bibnamefont
  {Ludsteck}},\ }\href {\doibase 10.1107/S0567739472000130} {\bibfield
  {journal} {\bibinfo  {journal} {Acta Crystallographica, Section A}\ }\textbf
  {\bibinfo {volume} {28}},\ \bibinfo {pages} {59} (\bibinfo {year}
  {1972})}\BibitemShut {NoStop}%
\bibitem [{\citenamefont {Trucano}\ and\ \citenamefont
  {Chen}(1975)}]{Trucano1975}%
  \BibitemOpen
  \bibfield  {author} {\bibinfo {author} {\bibfnamefont {P.}~\bibnamefont
  {Trucano}}\ and\ \bibinfo {author} {\bibfnamefont {R.}~\bibnamefont {Chen}},\
  }\href {\doibase 10.1038/258136a0} {\bibfield  {journal} {\bibinfo  {journal}
  {Nature}\ }\textbf {\bibinfo {volume} {258}},\ \bibinfo {pages} {136}
  (\bibinfo {year} {1975})}\BibitemShut {NoStop}%
\bibitem [{\citenamefont {Zhao}\ and\ \citenamefont {Spain}(1989)}]{Zhao1989}%
  \BibitemOpen
  \bibfield  {author} {\bibinfo {author} {\bibfnamefont {Y.~X.}\ \bibnamefont
  {Zhao}}\ and\ \bibinfo {author} {\bibfnamefont {I.~L.}\ \bibnamefont
  {Spain}},\ }\href {\doibase 10.1103/PhysRevB.40.993} {\bibfield  {journal}
  {\bibinfo  {journal} {Phys. Rev. B}\ }\textbf {\bibinfo {volume} {40}},\
  \bibinfo {pages} {993} (\bibinfo {year} {1989})}\BibitemShut {NoStop}%
\bibitem [{\citenamefont {Bosak}\ \emph {et~al.}(2007)\citenamefont {Bosak},
  \citenamefont {Krisch}, \citenamefont {Mohr}, \citenamefont {Maultzsch},\
  and\ \citenamefont {Thomsen}}]{Bosak2007}%
  \BibitemOpen
  \bibfield  {author} {\bibinfo {author} {\bibfnamefont {A.}~\bibnamefont
  {Bosak}}, \bibinfo {author} {\bibfnamefont {M.}~\bibnamefont {Krisch}},
  \bibinfo {author} {\bibfnamefont {M.}~\bibnamefont {Mohr}}, \bibinfo {author}
  {\bibfnamefont {J.}~\bibnamefont {Maultzsch}}, \ and\ \bibinfo {author}
  {\bibfnamefont {C.}~\bibnamefont {Thomsen}},\ }\href {\doibase
  10.1103/PhysRevB.75.153408} {\bibfield  {journal} {\bibinfo  {journal} {Phys.
  Rev. B}\ }\textbf {\bibinfo {volume} {75}},\ \bibinfo {pages} {153408}
  (\bibinfo {year} {2007})}\BibitemShut {NoStop}%
\bibitem [{\citenamefont {Lebedeva}\ \emph
  {et~al.}(2017{\natexlab{a}})\citenamefont {Lebedeva}, \citenamefont
  {Lebedev}, \citenamefont {Popov},\ and\ \citenamefont
  {Knizhnik}}]{Lebedeva2017}%
  \BibitemOpen
  \bibfield  {author} {\bibinfo {author} {\bibfnamefont {I.~V.}\ \bibnamefont
  {Lebedeva}}, \bibinfo {author} {\bibfnamefont {A.~V.}\ \bibnamefont
  {Lebedev}}, \bibinfo {author} {\bibfnamefont {A.~M.}\ \bibnamefont {Popov}},
  \ and\ \bibinfo {author} {\bibfnamefont {A.~A.}\ \bibnamefont {Knizhnik}},\
  }\href {\doibase 10.1016/j.commatsci.2016.11.011} {\bibfield  {journal}
  {\bibinfo  {journal} {Computational Materials Science}\ }\textbf {\bibinfo
  {volume} {128}},\ \bibinfo {pages} {45} (\bibinfo {year}
  {2017}{\natexlab{a}})}\BibitemShut {NoStop}%
\bibitem [{\citenamefont {Lebedeva}\ \emph {et~al.}(2016)\citenamefont
  {Lebedeva}, \citenamefont {Lebedev}, \citenamefont {Popov},\ and\
  \citenamefont {Knizhnik}}]{Lebedeva2016}%
  \BibitemOpen
  \bibfield  {author} {\bibinfo {author} {\bibfnamefont {I.~V.}\ \bibnamefont
  {Lebedeva}}, \bibinfo {author} {\bibfnamefont {A.~V.}\ \bibnamefont
  {Lebedev}}, \bibinfo {author} {\bibfnamefont {A.~M.}\ \bibnamefont {Popov}},
  \ and\ \bibinfo {author} {\bibfnamefont {A.~A.}\ \bibnamefont {Knizhnik}},\
  }\href {\doibase 10.1103/PhysRevB.93.235414} {\bibfield  {journal} {\bibinfo
  {journal} {Phys. Rev. B}\ }\textbf {\bibinfo {volume} {93}},\ \bibinfo
  {pages} {235414} (\bibinfo {year} {2016})}\BibitemShut {NoStop}%
\bibitem [{\citenamefont {Yoo}\ \emph {et~al.}(1997)\citenamefont {Yoo},
  \citenamefont {Akella}, \citenamefont {Cynn},\ and\ \citenamefont
  {Nicol}}]{Yoo1997}%
  \BibitemOpen
  \bibfield  {author} {\bibinfo {author} {\bibfnamefont {C.~S.}\ \bibnamefont
  {Yoo}}, \bibinfo {author} {\bibfnamefont {J.}~\bibnamefont {Akella}},
  \bibinfo {author} {\bibfnamefont {H.}~\bibnamefont {Cynn}}, \ and\ \bibinfo
  {author} {\bibfnamefont {M.}~\bibnamefont {Nicol}},\ }\href {\doibase
  10.1103/PhysRevB.56.140} {\bibfield  {journal} {\bibinfo  {journal} {Phys.
  Rev. B}\ }\textbf {\bibinfo {volume} {56}},\ \bibinfo {pages} {140} (\bibinfo
  {year} {1997})}\BibitemShut {NoStop}%
\bibitem [{\citenamefont {Yamamura}\ \emph {et~al.}(1997)\citenamefont
  {Yamamura}, \citenamefont {Takata},\ and\ \citenamefont
  {Sakata}}]{Yamamura1997}%
  \BibitemOpen
  \bibfield  {author} {\bibinfo {author} {\bibfnamefont {S.}~\bibnamefont
  {Yamamura}}, \bibinfo {author} {\bibfnamefont {M.}~\bibnamefont {Takata}}, \
  and\ \bibinfo {author} {\bibfnamefont {M.}~\bibnamefont {Sakata}},\ }\href
  {\doibase 10.1016/S0022-3697(96)00134-5} {\bibfield  {journal} {\bibinfo
  {journal} {Journal of Physics and Chemistry of Solids}\ }\textbf {\bibinfo
  {volume} {58}},\ \bibinfo {pages} {177} (\bibinfo {year} {1997})}\BibitemShut
  {NoStop}%
\bibitem [{\citenamefont {Kolmogorov}\ and\ \citenamefont
  {Crespi}(2005)}]{Kolmogorov2005}%
  \BibitemOpen
  \bibfield  {author} {\bibinfo {author} {\bibfnamefont {A.~N.}\ \bibnamefont
  {Kolmogorov}}\ and\ \bibinfo {author} {\bibfnamefont {V.~H.}\ \bibnamefont
  {Crespi}},\ }\href {\doibase 10.1103/PhysRevB.71.235415} {\bibfield
  {journal} {\bibinfo  {journal} {Phys. Rev. B}\ }\textbf {\bibinfo {volume}
  {71}},\ \bibinfo {pages} {235415} (\bibinfo {year} {2005})}\BibitemShut
  {NoStop}%
\bibitem [{\citenamefont {Aoki}\ and\ \citenamefont
  {Amawashi}(2007)}]{Aoki2007}%
  \BibitemOpen
  \bibfield  {author} {\bibinfo {author} {\bibfnamefont {M.}~\bibnamefont
  {Aoki}}\ and\ \bibinfo {author} {\bibfnamefont {H.}~\bibnamefont
  {Amawashi}},\ }\href {\doibase 10.1016/j.ssc.2007.02.013} {\bibfield
  {journal} {\bibinfo  {journal} {Solid State Communications}\ }\textbf
  {\bibinfo {volume} {142}},\ \bibinfo {pages} {123} (\bibinfo {year}
  {2007})}\BibitemShut {NoStop}%
\bibitem [{\citenamefont {Ershova}\ \emph {et~al.}(2010)\citenamefont
  {Ershova}, \citenamefont {Lillestolen},\ and\ \citenamefont
  {Bichoutskaia}}]{Ershova2010}%
  \BibitemOpen
  \bibfield  {author} {\bibinfo {author} {\bibfnamefont {O.~V.}\ \bibnamefont
  {Ershova}}, \bibinfo {author} {\bibfnamefont {T.~C.}\ \bibnamefont
  {Lillestolen}}, \ and\ \bibinfo {author} {\bibfnamefont {E.}~\bibnamefont
  {Bichoutskaia}},\ }\href {\doibase 10.1039/C000370K} {\bibfield  {journal}
  {\bibinfo  {journal} {Phys. Chem. Chem. Phys.}\ }\textbf {\bibinfo {volume}
  {12}},\ \bibinfo {pages} {6483} (\bibinfo {year} {2010})}\BibitemShut
  {NoStop}%
\bibitem [{\citenamefont {Alden}\ \emph {et~al.}(2013)\citenamefont {Alden},
  \citenamefont {Tsen}, \citenamefont {Huang}, \citenamefont {Hovden},
  \citenamefont {Brown}, \citenamefont {Park}, \citenamefont {Muller},\ and\
  \citenamefont {McEuen}}]{Alden2013}%
  \BibitemOpen
  \bibfield  {author} {\bibinfo {author} {\bibfnamefont {J.~S.}\ \bibnamefont
  {Alden}}, \bibinfo {author} {\bibfnamefont {A.~W.}\ \bibnamefont {Tsen}},
  \bibinfo {author} {\bibfnamefont {P.~Y.}\ \bibnamefont {Huang}}, \bibinfo
  {author} {\bibfnamefont {R.}~\bibnamefont {Hovden}}, \bibinfo {author}
  {\bibfnamefont {L.}~\bibnamefont {Brown}}, \bibinfo {author} {\bibfnamefont
  {J.}~\bibnamefont {Park}}, \bibinfo {author} {\bibfnamefont {D.~A.}\
  \bibnamefont {Muller}}, \ and\ \bibinfo {author} {\bibfnamefont {P.~L.}\
  \bibnamefont {McEuen}},\ }\href {\doibase 10.1073/pnas.1309394110} {\bibfield
   {journal} {\bibinfo  {journal} {PNAS}\ }\textbf {\bibinfo {volume} {110}},\
  \bibinfo {pages} {11256} (\bibinfo {year} {2013})}\BibitemShut {NoStop}%
\bibitem [{\citenamefont {Constantinescu}\ \emph {et~al.}(2013)\citenamefont
  {Constantinescu}, \citenamefont {Kuc},\ and\ \citenamefont
  {Heine}}]{Constantinescu2013}%
  \BibitemOpen
  \bibfield  {author} {\bibinfo {author} {\bibfnamefont {G.}~\bibnamefont
  {Constantinescu}}, \bibinfo {author} {\bibfnamefont {A.}~\bibnamefont {Kuc}},
  \ and\ \bibinfo {author} {\bibfnamefont {T.}~\bibnamefont {Heine}},\ }\href
  {\doibase 10.1103/PhysRevLett.111.036104} {\bibfield  {journal} {\bibinfo
  {journal} {Phys. Rev. Lett.}\ }\textbf {\bibinfo {volume} {111}},\ \bibinfo
  {pages} {036104} (\bibinfo {year} {2013})}\BibitemShut {NoStop}%
\bibitem [{\citenamefont {Liu}\ \emph {et~al.}(2012)\citenamefont {Liu},
  \citenamefont {Liu}, \citenamefont {Cheng}, \citenamefont {Li}, \citenamefont
  {Wang},\ and\ \citenamefont {Zheng}}]{Liu2012}%
  \BibitemOpen
  \bibfield  {author} {\bibinfo {author} {\bibfnamefont {Z.}~\bibnamefont
  {Liu}}, \bibinfo {author} {\bibfnamefont {J.~Z.}\ \bibnamefont {Liu}},
  \bibinfo {author} {\bibfnamefont {Y.}~\bibnamefont {Cheng}}, \bibinfo
  {author} {\bibfnamefont {Z.}~\bibnamefont {Li}}, \bibinfo {author}
  {\bibfnamefont {L.}~\bibnamefont {Wang}}, \ and\ \bibinfo {author}
  {\bibfnamefont {Q.}~\bibnamefont {Zheng}},\ }\href {\doibase
  10.1103/PhysRevB.85.205418} {\bibfield  {journal} {\bibinfo  {journal} {Phys.
  Rev. B}\ }\textbf {\bibinfo {volume} {85}},\ \bibinfo {pages} {205418}
  (\bibinfo {year} {2012})}\BibitemShut {NoStop}%
\bibitem [{\citenamefont {Zacharia}\ \emph {et~al.}(2004)\citenamefont
  {Zacharia}, \citenamefont {Ulbricht},\ and\ \citenamefont
  {Hertel}}]{Zacharia2004}%
  \BibitemOpen
  \bibfield  {author} {\bibinfo {author} {\bibfnamefont {R.}~\bibnamefont
  {Zacharia}}, \bibinfo {author} {\bibfnamefont {H.}~\bibnamefont {Ulbricht}},
  \ and\ \bibinfo {author} {\bibfnamefont {T.}~\bibnamefont {Hertel}},\ }\href
  {\doibase 10.1103/PhysRevB.69.155406} {\bibfield  {journal} {\bibinfo
  {journal} {Phys. Rev. B}\ }\textbf {\bibinfo {volume} {69}},\ \bibinfo
  {pages} {155406} (\bibinfo {year} {2004})}\BibitemShut {NoStop}%
\bibitem [{\citenamefont {Benedict}\ \emph {et~al.}(1998)\citenamefont
  {Benedict}, \citenamefont {Chopra}, \citenamefont {Cohen}, \citenamefont
  {Zettl}, \citenamefont {Louie},\ and\ \citenamefont {Crespi}}]{Benedict1998}%
  \BibitemOpen
  \bibfield  {author} {\bibinfo {author} {\bibfnamefont {L.~X.}\ \bibnamefont
  {Benedict}}, \bibinfo {author} {\bibfnamefont {N.~G.}\ \bibnamefont
  {Chopra}}, \bibinfo {author} {\bibfnamefont {M.~L.}\ \bibnamefont {Cohen}},
  \bibinfo {author} {\bibfnamefont {A.}~\bibnamefont {Zettl}}, \bibinfo
  {author} {\bibfnamefont {S.~G.}\ \bibnamefont {Louie}}, \ and\ \bibinfo
  {author} {\bibfnamefont {V.~H.}\ \bibnamefont {Crespi}},\ }\href {\doibase
  10.1016/S0009-2614(97)01466-8} {\bibfield  {journal} {\bibinfo  {journal}
  {Chem. Phys. Lett.}\ }\textbf {\bibinfo {volume} {286}},\ \bibinfo {pages}
  {490} (\bibinfo {year} {1998})}\BibitemShut {NoStop}%
\bibitem [{\citenamefont {Girifalco}\ and\ \citenamefont
  {Lad}(1956)}]{Girifalco1956}%
  \BibitemOpen
  \bibfield  {author} {\bibinfo {author} {\bibfnamefont {L.~A.}\ \bibnamefont
  {Girifalco}}\ and\ \bibinfo {author} {\bibfnamefont {R.~A.}\ \bibnamefont
  {Lad}},\ }\href {\doibase 10.1063/1.1743030} {\bibfield  {journal} {\bibinfo
  {journal} {J. Chem. Phys.}\ }\textbf {\bibinfo {volume} {25}},\ \bibinfo
  {pages} {693} (\bibinfo {year} {1956})}\BibitemShut {NoStop}%
\bibitem [{\citenamefont {Lebedeva}\ \emph {et~al.}(2012)\citenamefont
  {Lebedeva}, \citenamefont {Knizhnik}, \citenamefont {Popov}, \citenamefont
  {Lozovik},\ and\ \citenamefont {Potapkin}}]{Lebedeva2012}%
  \BibitemOpen
  \bibfield  {author} {\bibinfo {author} {\bibfnamefont {I.~V.}\ \bibnamefont
  {Lebedeva}}, \bibinfo {author} {\bibfnamefont {A.~A.}\ \bibnamefont
  {Knizhnik}}, \bibinfo {author} {\bibfnamefont {A.~M.}\ \bibnamefont {Popov}},
  \bibinfo {author} {\bibfnamefont {Y.~E.}\ \bibnamefont {Lozovik}}, \ and\
  \bibinfo {author} {\bibfnamefont {B.~V.}\ \bibnamefont {Potapkin}},\ }\href
  {\doibase 10.1016/j.physe.2011.07.018} {\bibfield  {journal} {\bibinfo
  {journal} {Physica E}\ }\textbf {\bibinfo {volume} {44}},\ \bibinfo {pages}
  {949} (\bibinfo {year} {2012})}\BibitemShut {NoStop}%
\bibitem [{\citenamefont {Belikov}\ \emph {et~al.}(2004)\citenamefont
  {Belikov}, \citenamefont {Lozovik}, \citenamefont {Nikolaev},\ and\
  \citenamefont {Popov}}]{Belikov2004}%
  \BibitemOpen
  \bibfield  {author} {\bibinfo {author} {\bibfnamefont {A.~V.}\ \bibnamefont
  {Belikov}}, \bibinfo {author} {\bibfnamefont {Y.~E.}\ \bibnamefont
  {Lozovik}}, \bibinfo {author} {\bibfnamefont {A.~G.}\ \bibnamefont
  {Nikolaev}}, \ and\ \bibinfo {author} {\bibfnamefont {A.~M.}\ \bibnamefont
  {Popov}},\ }\href@noop {} {\bibfield  {journal} {\bibinfo  {journal} {Chem.
  Phys. Lett.}\ }\textbf {\bibinfo {volume} {385}},\ \bibinfo {pages} {72}
  (\bibinfo {year} {2004})}\BibitemShut {NoStop}%
\bibitem [{\citenamefont {Bichoutskaia}\ \emph {et~al.}(2005)\citenamefont
  {Bichoutskaia}, \citenamefont {Popov}, \citenamefont {El-Barbary},
  \citenamefont {Heggie},\ and\ \citenamefont {Lozovik}}]{Bichoutskaia2005}%
  \BibitemOpen
  \bibfield  {author} {\bibinfo {author} {\bibfnamefont {E.}~\bibnamefont
  {Bichoutskaia}}, \bibinfo {author} {\bibfnamefont {A.~M.}\ \bibnamefont
  {Popov}}, \bibinfo {author} {\bibfnamefont {A.}~\bibnamefont {El-Barbary}},
  \bibinfo {author} {\bibfnamefont {M.~I.}\ \bibnamefont {Heggie}}, \ and\
  \bibinfo {author} {\bibfnamefont {Y.~E.}\ \bibnamefont {Lozovik}},\
  }\href@noop {} {\bibfield  {journal} {\bibinfo  {journal} {Phys. Rev. B}\
  }\textbf {\bibinfo {volume} {71}},\ \bibinfo {pages} {113403} (\bibinfo
  {year} {2005})}\BibitemShut {NoStop}%
\bibitem [{\citenamefont {Bichoutskaia}\ \emph {et~al.}(2009)\citenamefont
  {Bichoutskaia}, \citenamefont {Popov}, \citenamefont {Lozovik}, \citenamefont
  {Ershova}, \citenamefont {Lebedeva},\ and\ \citenamefont
  {Knizhnik}}]{Bichoutskaia2009}%
  \BibitemOpen
  \bibfield  {author} {\bibinfo {author} {\bibfnamefont {E.}~\bibnamefont
  {Bichoutskaia}}, \bibinfo {author} {\bibfnamefont {A.~M.}\ \bibnamefont
  {Popov}}, \bibinfo {author} {\bibfnamefont {Y.~E.}\ \bibnamefont {Lozovik}},
  \bibinfo {author} {\bibfnamefont {O.~V.}\ \bibnamefont {Ershova}}, \bibinfo
  {author} {\bibfnamefont {I.~V.}\ \bibnamefont {Lebedeva}}, \ and\ \bibinfo
  {author} {\bibfnamefont {A.~A.}\ \bibnamefont {Knizhnik}},\ }\href@noop {}
  {\bibfield  {journal} {\bibinfo  {journal} {Phys. Rev. B}\ }\textbf {\bibinfo
  {volume} {80}},\ \bibinfo {pages} {165427} (\bibinfo {year}
  {2009})}\BibitemShut {NoStop}%
\bibitem [{\citenamefont {Popov}\ \emph {et~al.}(2009)\citenamefont {Popov},
  \citenamefont {Lozovik}, \citenamefont {Sobennikov},\ and\ \citenamefont
  {Knizhnik}}]{Popov2009}%
  \BibitemOpen
  \bibfield  {author} {\bibinfo {author} {\bibfnamefont {A.~M.}\ \bibnamefont
  {Popov}}, \bibinfo {author} {\bibfnamefont {Y.~E.}\ \bibnamefont {Lozovik}},
  \bibinfo {author} {\bibfnamefont {A.~S.}\ \bibnamefont {Sobennikov}}, \ and\
  \bibinfo {author} {\bibfnamefont {A.~A.}\ \bibnamefont {Knizhnik}},\ }\href
  {\doibase 10.1134/S1063776109040104} {\bibfield  {journal} {\bibinfo
  {journal} {JETP}\ }\textbf {\bibinfo {volume} {108}},\ \bibinfo {pages} {621}
  (\bibinfo {year} {2009})}\BibitemShut {NoStop}%
\bibitem [{\citenamefont {Popov}\ \emph
  {et~al.}(2012{\natexlab{b}})\citenamefont {Popov}, \citenamefont {Lebedeva},\
  and\ \citenamefont {Knizhnik}}]{Popov2012a}%
  \BibitemOpen
  \bibfield  {author} {\bibinfo {author} {\bibfnamefont {A.~M.}\ \bibnamefont
  {Popov}}, \bibinfo {author} {\bibfnamefont {I.~V.}\ \bibnamefont {Lebedeva}},
  \ and\ \bibinfo {author} {\bibfnamefont {A.~A.}\ \bibnamefont {Knizhnik}},\
  }\href@noop {} {\bibfield  {journal} {\bibinfo  {journal} {Appl. Phys.
  Lett.}\ }\textbf {\bibinfo {volume} {100}},\ \bibinfo {pages} {173101}
  (\bibinfo {year} {2012}{\natexlab{b}})}\BibitemShut {NoStop}%
\bibitem [{\citenamefont {Popov}\ \emph {et~al.}(2013)\citenamefont {Popov},
  \citenamefont {Lebedeva}, \citenamefont {Knizhnik}, \citenamefont {Lozovik},\
  and\ \citenamefont {Potapkin}}]{Popov2013}%
  \BibitemOpen
  \bibfield  {author} {\bibinfo {author} {\bibfnamefont {A.~M.}\ \bibnamefont
  {Popov}}, \bibinfo {author} {\bibfnamefont {I.~V.}\ \bibnamefont {Lebedeva}},
  \bibinfo {author} {\bibfnamefont {A.~A.}\ \bibnamefont {Knizhnik}}, \bibinfo
  {author} {\bibfnamefont {Y.~E.}\ \bibnamefont {Lozovik}}, \ and\ \bibinfo
  {author} {\bibfnamefont {B.~V.}\ \bibnamefont {Potapkin}},\ }\href@noop {}
  {\bibfield  {journal} {\bibinfo  {journal} {J. Chem. Phys.}\ }\textbf
  {\bibinfo {volume} {138}},\ \bibinfo {pages} {024703} (\bibinfo {year}
  {2013})}\BibitemShut {NoStop}%
\bibitem [{\citenamefont {Verhoeven}\ \emph {et~al.}(2004)\citenamefont
  {Verhoeven}, \citenamefont {Dienwiebel},\ and\ \citenamefont
  {Frenken}}]{Verhoeven2004}%
  \BibitemOpen
  \bibfield  {author} {\bibinfo {author} {\bibfnamefont {G.~S.}\ \bibnamefont
  {Verhoeven}}, \bibinfo {author} {\bibfnamefont {M.}~\bibnamefont
  {Dienwiebel}}, \ and\ \bibinfo {author} {\bibfnamefont {J.~W.~M.}\
  \bibnamefont {Frenken}},\ }\href@noop {} {\bibfield  {journal} {\bibinfo
  {journal} {Phys. Rev. B}\ }\textbf {\bibinfo {volume} {70}},\ \bibinfo
  {pages} {165418} (\bibinfo {year} {2004})}\BibitemShut {NoStop}%
\bibitem [{\citenamefont {Porovski\u{i}}\ and\ \citenamefont
  {Talapov}(1978)}]{Pokrovsky1978}%
  \BibitemOpen
  \bibfield  {author} {\bibinfo {author} {\bibfnamefont {V.~L.}\ \bibnamefont
  {Porovski\u{i}}}\ and\ \bibinfo {author} {\bibfnamefont {A.~L.}\ \bibnamefont
  {Talapov}},\ }\href@noop {} {\bibfield  {journal} {\bibinfo  {journal}
  {Soviet Physics JETP}\ }\textbf {\bibinfo {volume} {48}},\ \bibinfo {pages}
  {579} (\bibinfo {year} {1978})}\BibitemShut {NoStop}%
\bibitem [{\citenamefont {Chung}\ \emph {et~al.}(1987)\citenamefont {Chung},
  \citenamefont {Kara}, \citenamefont {Larese}, \citenamefont {Leung},\ and\
  \citenamefont {Frankl}}]{Chung1987}%
  \BibitemOpen
  \bibfield  {author} {\bibinfo {author} {\bibfnamefont {S.}~\bibnamefont
  {Chung}}, \bibinfo {author} {\bibfnamefont {A.}~\bibnamefont {Kara}},
  \bibinfo {author} {\bibfnamefont {J.~Z.}\ \bibnamefont {Larese}}, \bibinfo
  {author} {\bibfnamefont {W.~Y.}\ \bibnamefont {Leung}}, \ and\ \bibinfo
  {author} {\bibfnamefont {D.~R.}\ \bibnamefont {Frankl}},\ }\href {\doibase
  10.1103/PhysRevB.35.4870} {\bibfield  {journal} {\bibinfo  {journal} {Phys.
  Rev. B}\ }\textbf {\bibinfo {volume} {35}},\ \bibinfo {pages} {4870}
  (\bibinfo {year} {1987})}\BibitemShut {NoStop}%
\bibitem [{\citenamefont {Dienwiebel}\ \emph {et~al.}(2004)\citenamefont
  {Dienwiebel}, \citenamefont {Verhoeven}, \citenamefont {Pradeep},
  \citenamefont {Frenken}, \citenamefont {Heimberg},\ and\ \citenamefont
  {Zandbergen}}]{Dienwiebel2004}%
  \BibitemOpen
  \bibfield  {author} {\bibinfo {author} {\bibfnamefont {M.}~\bibnamefont
  {Dienwiebel}}, \bibinfo {author} {\bibfnamefont {G.~S.}\ \bibnamefont
  {Verhoeven}}, \bibinfo {author} {\bibfnamefont {N.}~\bibnamefont {Pradeep}},
  \bibinfo {author} {\bibfnamefont {J.~W.~M.}\ \bibnamefont {Frenken}},
  \bibinfo {author} {\bibfnamefont {J.~A.}\ \bibnamefont {Heimberg}}, \ and\
  \bibinfo {author} {\bibfnamefont {H.~W.}\ \bibnamefont {Zandbergen}},\ }\href
  {\doibase 10.1103/PhysRevLett.92.126101} {\bibfield  {journal} {\bibinfo
  {journal} {Phys. Rev. Lett.}\ }\textbf {\bibinfo {volume} {92}},\ \bibinfo
  {pages} {126101} (\bibinfo {year} {2004})}\BibitemShut {NoStop}%
\bibitem [{\citenamefont {Dienwiebel}\ \emph {et~al.}(2005)\citenamefont
  {Dienwiebel}, \citenamefont {Pradeep}, \citenamefont {Verhoeven},
  \citenamefont {Zandbergen},\ and\ \citenamefont {Frenken}}]{Dienwiebel2005}%
  \BibitemOpen
  \bibfield  {author} {\bibinfo {author} {\bibfnamefont {M.}~\bibnamefont
  {Dienwiebel}}, \bibinfo {author} {\bibfnamefont {N.}~\bibnamefont {Pradeep}},
  \bibinfo {author} {\bibfnamefont {G.~S.}\ \bibnamefont {Verhoeven}}, \bibinfo
  {author} {\bibfnamefont {H.~W.}\ \bibnamefont {Zandbergen}}, \ and\ \bibinfo
  {author} {\bibfnamefont {J.~W.}\ \bibnamefont {Frenken}},\ }\href {\doibase
  10.1016/j.susc.2004.12.011} {\bibfield  {journal} {\bibinfo  {journal} {Surf.
  Sci.}\ }\textbf {\bibinfo {volume} {576}},\ \bibinfo {pages} {197} (\bibinfo
  {year} {2005})}\BibitemShut {NoStop}%
\bibitem [{\citenamefont {Filippov}\ \emph {et~al.}(2008)\citenamefont
  {Filippov}, \citenamefont {Dienwiebel}, \citenamefont {Frenken},
  \citenamefont {Klafter},\ and\ \citenamefont {Urbakh}}]{Filippov2008}%
  \BibitemOpen
  \bibfield  {author} {\bibinfo {author} {\bibfnamefont {A.~E.}\ \bibnamefont
  {Filippov}}, \bibinfo {author} {\bibfnamefont {M.}~\bibnamefont
  {Dienwiebel}}, \bibinfo {author} {\bibfnamefont {J.~W.~M.}\ \bibnamefont
  {Frenken}}, \bibinfo {author} {\bibfnamefont {J.}~\bibnamefont {Klafter}}, \
  and\ \bibinfo {author} {\bibfnamefont {M.}~\bibnamefont {Urbakh}},\ }\href
  {\doibase 10.1103/PhysRevLett.100.046102} {\bibfield  {journal} {\bibinfo
  {journal} {Phys. Rev. Lett.}\ }\textbf {\bibinfo {volume} {100}},\ \bibinfo
  {pages} {046102} (\bibinfo {year} {2008})}\BibitemShut {NoStop}%
\bibitem [{\citenamefont {Xu}\ \emph {et~al.}(2013)\citenamefont {Xu},
  \citenamefont {Li}, \citenamefont {Yakobson},\ and\ \citenamefont
  {Ding}}]{Xu2013}%
  \BibitemOpen
  \bibfield  {author} {\bibinfo {author} {\bibfnamefont {Z.}~\bibnamefont
  {Xu}}, \bibinfo {author} {\bibfnamefont {X.}~\bibnamefont {Li}}, \bibinfo
  {author} {\bibfnamefont {B.~I.}\ \bibnamefont {Yakobson}}, \ and\ \bibinfo
  {author} {\bibfnamefont {F.}~\bibnamefont {Ding}},\ }\href {\doibase
  10.1039/C3NR01854G} {\bibfield  {journal} {\bibinfo  {journal} {Nanoscale}\
  }\textbf {\bibinfo {volume} {5}},\ \bibinfo {pages} {6736} (\bibinfo {year}
  {2013})}\BibitemShut {NoStop}%
\bibitem [{\citenamefont {Lebedeva}\ \emph {et~al.}(2010)\citenamefont
  {Lebedeva}, \citenamefont {Knizhnik}, \citenamefont {Popov}, \citenamefont
  {Ershova}, \citenamefont {Lozovik},\ and\ \citenamefont
  {Potapkin}}]{Lebedeva2010}%
  \BibitemOpen
  \bibfield  {author} {\bibinfo {author} {\bibfnamefont {I.~V.}\ \bibnamefont
  {Lebedeva}}, \bibinfo {author} {\bibfnamefont {A.~A.}\ \bibnamefont
  {Knizhnik}}, \bibinfo {author} {\bibfnamefont {A.~M.}\ \bibnamefont {Popov}},
  \bibinfo {author} {\bibfnamefont {O.~V.}\ \bibnamefont {Ershova}}, \bibinfo
  {author} {\bibfnamefont {Y.~E.}\ \bibnamefont {Lozovik}}, \ and\ \bibinfo
  {author} {\bibfnamefont {B.~V.}\ \bibnamefont {Potapkin}},\ }\href {\doibase
  10.1103/PhysRevB.82.155460} {\bibfield  {journal} {\bibinfo  {journal} {Phys.
  Rev. B}\ }\textbf {\bibinfo {volume} {82}},\ \bibinfo {pages} {155460}
  (\bibinfo {year} {2010})}\BibitemShut {NoStop}%
\bibitem [{\citenamefont {Lebedeva}\ \emph
  {et~al.}(2011{\natexlab{b}})\citenamefont {Lebedeva}, \citenamefont
  {Knizhnik}, \citenamefont {Popov}, \citenamefont {Ershova}, \citenamefont
  {Lozovik},\ and\ \citenamefont {Potapkin}}]{Lebedeva2011a}%
  \BibitemOpen
  \bibfield  {author} {\bibinfo {author} {\bibfnamefont {I.~V.}\ \bibnamefont
  {Lebedeva}}, \bibinfo {author} {\bibfnamefont {A.~A.}\ \bibnamefont
  {Knizhnik}}, \bibinfo {author} {\bibfnamefont {A.~M.}\ \bibnamefont {Popov}},
  \bibinfo {author} {\bibfnamefont {O.~V.}\ \bibnamefont {Ershova}}, \bibinfo
  {author} {\bibfnamefont {Y.~E.}\ \bibnamefont {Lozovik}}, \ and\ \bibinfo
  {author} {\bibfnamefont {B.~V.}\ \bibnamefont {Potapkin}},\ }\href {\doibase
  10.1063/1.3557819} {\bibfield  {journal} {\bibinfo  {journal} {J. Chem.
  Phys.}\ }\textbf {\bibinfo {volume} {134}},\ \bibinfo {pages} {104505}
  (\bibinfo {year} {2011}{\natexlab{b}})}\BibitemShut {NoStop}%
\bibitem [{\citenamefont {Boschetto}\ \emph {et~al.}(2013)\citenamefont
  {Boschetto}, \citenamefont {Malard}, \citenamefont {Lui}, \citenamefont
  {Mak}, \citenamefont {Li}, \citenamefont {Yan},\ and\ \citenamefont
  {Heinz}}]{Boschetto2013}%
  \BibitemOpen
  \bibfield  {author} {\bibinfo {author} {\bibfnamefont {D.}~\bibnamefont
  {Boschetto}}, \bibinfo {author} {\bibfnamefont {L.}~\bibnamefont {Malard}},
  \bibinfo {author} {\bibfnamefont {C.~H.}\ \bibnamefont {Lui}}, \bibinfo
  {author} {\bibfnamefont {K.~F.}\ \bibnamefont {Mak}}, \bibinfo {author}
  {\bibfnamefont {Z.}~\bibnamefont {Li}}, \bibinfo {author} {\bibfnamefont
  {H.}~\bibnamefont {Yan}}, \ and\ \bibinfo {author} {\bibfnamefont {T.~F.}\
  \bibnamefont {Heinz}},\ }\href {\doibase 10.1021/nl401713h} {\bibfield
  {journal} {\bibinfo  {journal} {Nano Lett.}\ }\textbf {\bibinfo {volume}
  {13}},\ \bibinfo {pages} {4620} (\bibinfo {year} {2013})}\BibitemShut
  {NoStop}%
\bibitem [{\citenamefont {Tan}\ \emph {et~al.}(2012)\citenamefont {Tan},
  \citenamefont {Han}, \citenamefont {Zhao}, \citenamefont {Wu}, \citenamefont
  {Chang}, \citenamefont {Wang}, \citenamefont {Wang}, \citenamefont {Bonini},
  \citenamefont {Marzari}, \citenamefont {Pugno}, \citenamefont {Savini},
  \citenamefont {Lombardo},\ and\ \citenamefont {Ferrari}}]{Tan2012}%
  \BibitemOpen
  \bibfield  {author} {\bibinfo {author} {\bibfnamefont {P.~H.}\ \bibnamefont
  {Tan}}, \bibinfo {author} {\bibfnamefont {W.~P.}\ \bibnamefont {Han}},
  \bibinfo {author} {\bibfnamefont {W.~J.}\ \bibnamefont {Zhao}}, \bibinfo
  {author} {\bibfnamefont {Z.~H.}\ \bibnamefont {Wu}}, \bibinfo {author}
  {\bibfnamefont {K.}~\bibnamefont {Chang}}, \bibinfo {author} {\bibfnamefont
  {H.}~\bibnamefont {Wang}}, \bibinfo {author} {\bibfnamefont {Y.~F.}\
  \bibnamefont {Wang}}, \bibinfo {author} {\bibfnamefont {N.}~\bibnamefont
  {Bonini}}, \bibinfo {author} {\bibfnamefont {N.}~\bibnamefont {Marzari}},
  \bibinfo {author} {\bibfnamefont {N.}~\bibnamefont {Pugno}}, \bibinfo
  {author} {\bibfnamefont {G.}~\bibnamefont {Savini}}, \bibinfo {author}
  {\bibfnamefont {A.}~\bibnamefont {Lombardo}}, \ and\ \bibinfo {author}
  {\bibfnamefont {A.~C.}\ \bibnamefont {Ferrari}},\ }\href {\doibase
  10.1038/nmat3245} {\bibfield  {journal} {\bibinfo  {journal} {Nat. Mater.}\
  }\textbf {\bibinfo {volume} {11}},\ \bibinfo {pages} {294} (\bibinfo {year}
  {2012})}\BibitemShut {NoStop}%
\bibitem [{\citenamefont {Bichoutskaia}\ \emph {et~al.}(2006)\citenamefont
  {Bichoutskaia}, \citenamefont {Heggie}, \citenamefont {Lozovik},\ and\
  \citenamefont {Popov}}]{Bichoutskaia2006}%
  \BibitemOpen
  \bibfield  {author} {\bibinfo {author} {\bibfnamefont {E.}~\bibnamefont
  {Bichoutskaia}}, \bibinfo {author} {\bibfnamefont {M.~I.}\ \bibnamefont
  {Heggie}}, \bibinfo {author} {\bibfnamefont {Y.~E.}\ \bibnamefont {Lozovik}},
  \ and\ \bibinfo {author} {\bibfnamefont {A.~M.}\ \bibnamefont {Popov}},\
  }\href {\doibase 10.1080/15363830600663412} {\bibfield  {journal} {\bibinfo
  {journal} {Fullerenes, Nanotubes, Carbon Nanostruct.}\ }\textbf {\bibinfo
  {volume} {14}},\ \bibinfo {pages} {131} (\bibinfo {year} {2006})}\BibitemShut
  {NoStop}%
\bibitem [{\citenamefont {Popov}\ \emph {et~al.}(2011)\citenamefont {Popov},
  \citenamefont {Lebedeva}, \citenamefont {Knizhnik}, \citenamefont {Lozovik},\
  and\ \citenamefont {Potapkin}}]{Popov2011}%
  \BibitemOpen
  \bibfield  {author} {\bibinfo {author} {\bibfnamefont {A.~M.}\ \bibnamefont
  {Popov}}, \bibinfo {author} {\bibfnamefont {I.~V.}\ \bibnamefont {Lebedeva}},
  \bibinfo {author} {\bibfnamefont {A.~A.}\ \bibnamefont {Knizhnik}}, \bibinfo
  {author} {\bibfnamefont {Y.~E.}\ \bibnamefont {Lozovik}}, \ and\ \bibinfo
  {author} {\bibfnamefont {B.~V.}\ \bibnamefont {Potapkin}},\ }\href {\doibase
  10.1103/PhysRevB.84.045404} {\bibfield  {journal} {\bibinfo  {journal} {Phys.
  Rev. B}\ }\textbf {\bibinfo {volume} {84}},\ \bibinfo {pages} {045404}
  (\bibinfo {year} {2011})}\BibitemShut {NoStop}%
\bibitem [{\citenamefont {Lebedeva}\ \emph
  {et~al.}(2017{\natexlab{b}})\citenamefont {Lebedeva}, \citenamefont
  {Knizhnik},\ and\ \citenamefont {Popov}}]{Lebedeva2017a}%
  \BibitemOpen
  \bibfield  {author} {\bibinfo {author} {\bibfnamefont {I.~V.}\ \bibnamefont
  {Lebedeva}}, \bibinfo {author} {\bibfnamefont {A.~A.}\ \bibnamefont
  {Knizhnik}}, \ and\ \bibinfo {author} {\bibfnamefont {A.~M.}\ \bibnamefont
  {Popov}},\ }\href {\doibase 10.1016/j.physe.2017.03.008} {\bibfield
  {journal} {\bibinfo  {journal} {Physica E: Low-dimensional Systems and
  Nanostructures}\ }\textbf {\bibinfo {volume} {90}},\ \bibinfo {pages} {49}
  (\bibinfo {year} {2017}{\natexlab{b}})}\BibitemShut {NoStop}%
\bibitem [{\citenamefont {Lin}\ \emph {et~al.}(2013)\citenamefont {Lin},
  \citenamefont {Fang}, \citenamefont {Zhou}, \citenamefont {Lupini},
  \citenamefont {Idrobo}, \citenamefont {Kong}, \citenamefont {Pennycook},\
  and\ \citenamefont {Pantelides}}]{Lin2013}%
  \BibitemOpen
  \bibfield  {author} {\bibinfo {author} {\bibfnamefont {J.}~\bibnamefont
  {Lin}}, \bibinfo {author} {\bibfnamefont {W.}~\bibnamefont {Fang}}, \bibinfo
  {author} {\bibfnamefont {W.}~\bibnamefont {Zhou}}, \bibinfo {author}
  {\bibfnamefont {A.~R.}\ \bibnamefont {Lupini}}, \bibinfo {author}
  {\bibfnamefont {J.~C.}\ \bibnamefont {Idrobo}}, \bibinfo {author}
  {\bibfnamefont {J.}~\bibnamefont {Kong}}, \bibinfo {author} {\bibfnamefont
  {S.~J.}\ \bibnamefont {Pennycook}}, \ and\ \bibinfo {author} {\bibfnamefont
  {S.~T.}\ \bibnamefont {Pantelides}},\ }\href {\doibase 10.1021/nl4013979}
  {\bibfield  {journal} {\bibinfo  {journal} {Nano Letters}\ }\textbf {\bibinfo
  {volume} {13}},\ \bibinfo {pages} {3262} (\bibinfo {year}
  {2013})}\BibitemShut {NoStop}%
\bibitem [{\citenamefont {Yankowitz}\ \emph {et~al.}(2014)\citenamefont
  {Yankowitz}, \citenamefont {Wang}, \citenamefont {Birdwell}, \citenamefont
  {Chen}, \citenamefont {Watanabe}, \citenamefont {Taniguchi}, \citenamefont
  {Jacquod}, \citenamefont {San-Jose}, \citenamefont {Jarillo-Herrero},\ and\
  \citenamefont {LeRoy}}]{Yankowitz2014}%
  \BibitemOpen
  \bibfield  {author} {\bibinfo {author} {\bibfnamefont {M.}~\bibnamefont
  {Yankowitz}}, \bibinfo {author} {\bibfnamefont {J.~I.-J.}\ \bibnamefont
  {Wang}}, \bibinfo {author} {\bibfnamefont {A.~G.}\ \bibnamefont {Birdwell}},
  \bibinfo {author} {\bibfnamefont {Y.-A.}\ \bibnamefont {Chen}}, \bibinfo
  {author} {\bibfnamefont {K.}~\bibnamefont {Watanabe}}, \bibinfo {author}
  {\bibfnamefont {T.}~\bibnamefont {Taniguchi}}, \bibinfo {author}
  {\bibfnamefont {P.}~\bibnamefont {Jacquod}}, \bibinfo {author} {\bibfnamefont
  {P.}~\bibnamefont {San-Jose}}, \bibinfo {author} {\bibfnamefont
  {P.}~\bibnamefont {Jarillo-Herrero}}, \ and\ \bibinfo {author} {\bibfnamefont
  {B.~J.}\ \bibnamefont {LeRoy}},\ }\href {\doibase 10.1038/nmat3965}
  {\bibfield  {journal} {\bibinfo  {journal} {Nat. Mater.}\ }\textbf {\bibinfo
  {volume} {13}},\ \bibinfo {pages} {786} (\bibinfo {year} {2014})}\BibitemShut
  {NoStop}%
\end{thebibliography}%

\end{document}